\documentclass[pre,showpacs,twocolumn,nofootinbib,superscriptaddress]{revtex4-1}

\usepackage[dvips]{graphicx}
\usepackage{amsmath,amssymb}
\usepackage{etoolbox}
\usepackage{amsfonts}
\usepackage{color}
\usepackage{url}
\usepackage{bm}
\usepackage{placeins}
\usepackage[usenames,dvipsnames]{xcolor}

\begin{document}

\newcommand{\Eq}[1]{Eq.~\eqref{#1}}
\newcommand{\Eqs}[1]{Eqs.~\eqref{#1}}
\newcommand{\be}{\begin{equation}}
\newcommand{\ee}{\end{equation}}


\newcommand{\diff}{\text{d}}
\newcommand{\ra}{\rangle}
\newcommand{\la}{\langle}

\newcommand{\bn}{{\bm n}}
\newcommand{\trN}{\text{Tr}_N \negmedspace}
\newcommand{\BES}{\mathcal{S}}

\newcommand{\ri}{\mathrm{i}}
\newcommand{\ze}{\zeta}
\newcommand{\re}{\mathrm{e}}
\newcommand{\rd}{\mathrm{d}}
\newcommand{\rs}{\mathrm{s}}
\newcommand{\rb}{\mathrm{b}}
\newcommand{\no}{\hat{n}}
\renewcommand{\ao}{\hat{a}}
\newcommand{\aod}{\hat{a}^\dag}
\newcommand{\Ho}{\hat{H}}
\newcommand{\No}{\hat{N}}
\newcommand{\Uo}{\hat{U}}
\newcommand{\tr}{\mathrm{tr}}
\newcommand{\bA}{{\bm A}}
\newcommand{\br}{{\bm r}}
\newcommand{\bF}{{\bm F}}
\newcommand{\bp}{{\bm p}}
\newcommand{\bx}{{\bm x}}
\newcommand{\hc}{\mathrm{h.c.}}
\renewcommand{\epsilon}{\varepsilon}

\newcommand{\avgo}[1]{\bar #1}
\newcommand{\zetao}{\hat\zeta}
\newcommand{\f}[2]{\zeta_{#1#2}}

\newcommand{\fixme}[1]{{\color{green} \bf [FixMe: #1]}}

\title{Non-equilibrium steady states of ideal bosonic and fermionic quantum gases}

\date{\today}

\author{Daniel Vorberg}
\email{dv@pks.mpg.de}
\affiliation{Max-Planck-Institut f\"ur Physik komplexer Systeme, N\"othnitzer Str.\ 38, 01187 Dresden, Germany}

\author{Waltraut Wustmann}
\affiliation{Max-Planck-Institut f\"ur Physik komplexer Systeme, N\"othnitzer Str.\ 38, 01187 Dresden, Germany}
\affiliation{Technische Universit\"at Dresden, Institut f\"ur Theoretische Physik and Center for Dynamics, 01062 Dresden, Germany}

\author{Henning Schomerus}
\affiliation{Max-Planck-Institut f\"ur Physik komplexer Systeme, N\"othnitzer Str.\ 38, 01187 Dresden, Germany}
\affiliation{Department of Physics, Lancaster University, Lancaster, LA1 4YB, United Kingdom}

\author{Roland Ketzmerick}
\affiliation{Max-Planck-Institut f\"ur Physik komplexer Systeme, N\"othnitzer Str.\ 38, 01187 Dresden, Germany}
\affiliation{Technische Universit\"at Dresden, Institut f\"ur Theoretische Physik and Center for Dynamics, 01062 Dresden, Germany}

\author{Andr\'e Eckardt}
\email{eckardt@pks.mpg.de}
\affiliation{Max-Planck-Institut f\"ur Physik komplexer Systeme, N\"othnitzer Str.\ 38, 01187 Dresden, Germany}

\pacs{05.30.Jp, 05.70.Ln, 67.10.Ba, 67.85.Jk}

\begin{abstract}
We investigate non-equilibrium steady states of driven-dissipative ideal quantum gases of both bosons and 
fermions. We focus on systems of sharp particle number that are driven out of equilibrium either by the coupling to 
several heat baths of different temperature or by time-periodic driving in combination with the coupling to a heat bath. Within the framework of (Floquet-)Born-Markov theory, several analytical and numerical methods are 
described in detail. This includes a mean-field theory in terms of occupation numbers, an augmented mean-field theory 
taking into account also non-trivial two-particle correlations, and quantum-jump-type Monte-Carlo simulations. 
For the case of the ideal Fermi gas, these methods are applied to simple lattice models and the possibility of achieving 
exotic states via bath engineering is pointed out.  
The largest part of this work is devoted to bosonic quantum gases and the phenomenon of Bose selection, a non-equilibrium 
generalization of Bose condensation, where multiple single-particle states are selected to acquire a large occupation 
[Phys.\ Rev.\ Lett.\ {\bf111}, 240405 (2013)]. In this context, among others, we provide a theory for transitions where 
the set of selected states changes, describe an efficient algorithm for finding the set of selected states, 
investigate beyond-mean-field effects, and identify the dominant mechanisms for heat transport in the Bose selected state.

\end{abstract}

\maketitle

\section{Introduction}
\label{sec:Introduction}

There is a huge current interest in non-equilibrium phenomena of many-body systems beyond the hydrodynamic description 
of systems retaining approximate local equilibrium. Recent work concerns several paradigmatic scenarios, like the 
dynamics away from equilibrium in response to a slow or an abrupt parameter variation \cite{CampisiEtAl11,Dziarmaga10,
PolkovnikovEtAl11}, the possible relaxation towards equilibrium \cite{Dziarmaga10,PolkovnikovEtAl11} versus many-body 
localization \cite{BaskoEtAl06,HuseEtAl13}, and the control of many-body physics by means of strong periodic forcing
\cite{EckardtEtAl05,OkaAoki09,ZenesiniEtAl09,StruckEtAl11,JotzuEtAl14,AidelsburgerEtAl15}. Also the possibility to achieve transient light-induced superconductivity 
above the equilibrium critical temperature attracted enormous interest \cite{FaustiEtAl11}. 

Another fundamental scenario of many-body dynamics are driven-dissipative quantum systems and their 
non-equilibrium steady states \cite{LepriEtAl03,ZiaSchmittmann07,DubiVentra11,ZimbovskayaPederson11,Prosen11,LiEtAl12,
MuellerEtAl12,DharEtAl12,LutzRenzoniEtAl13,Prosen14}. These include, for example, time-periodically driven open many-body 
systems \cite{VorbergEtAl13,ShiraiEtAl15,ChenEtAl15,IadecolaEtA15,SeetharamEtAl15} and photonic many-body systems
\cite{HartmannEtAl08,SchoelkopfGirvin08,KlaersEtAl10,HoukEtAl12,CarusottoCiuti13,ByrnesEtAl14}. In contrast to 
equilibrium states, which depend on a few thermodynamic parameters like temperature and chemical potential only, such
non-equilibrium steady states depend on the very details of the environment. On the one hand, this makes their 
theoretical treatment challenging. On the other hand, it offers also interesting opportunities to engineer the state and 
the properties of a many-body system beyond the constraints of thermal equilibrium in a robust and controlled fashion.

\begin{figure}[t]
\centering
\includegraphics[width=1\linewidth]{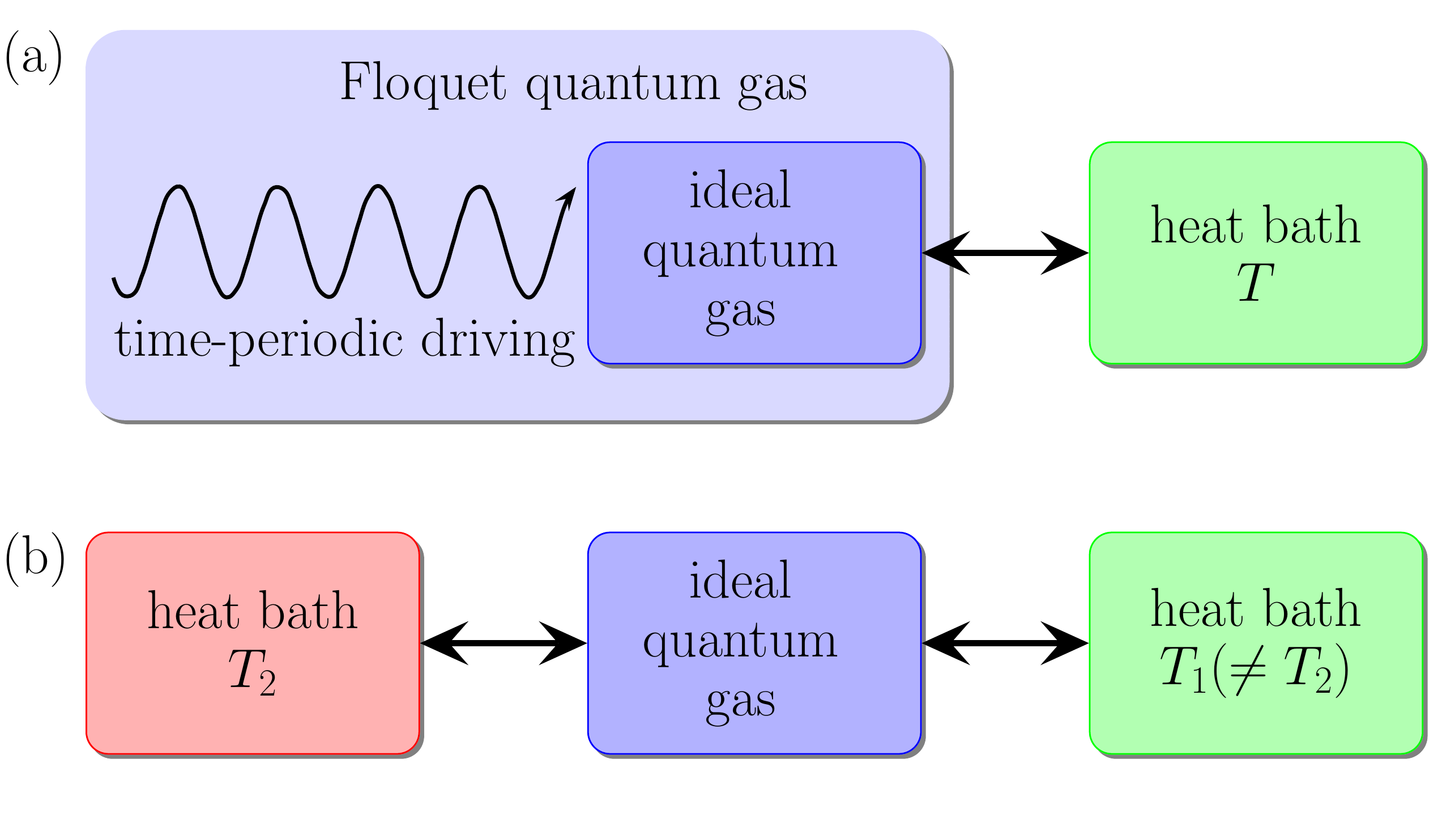}
\caption{(color online) Two paradigmatic examples of driven-dissipative ideal quantum gases possessing non-equilibrium steady states. 
(a) Periodically driven system weakly coupled to a heat bath. 
(b) Autonomous system weakly coupled to two heat baths of different temperature.
}
\label{fig:sys_config}
\end{figure}

In this context, it was recently pointed out that already an ideal Bose gas of $N$ particles can exhibit intriguing 
behavior, when it is driven into a steady state far from equilibrium, e.g., by coupling it to two heat baths of different 
temperature or by time-periodic driving in the presence of a heat bath (see Fig.~\ref{fig:sys_config}). In the quantum 
degenerate regime of large densities, the Bose gas undergoes a generalized form of Bose condensation, where multiple 
single-particle states can be selected to acquire large occupations \cite{VorbergEtAl13}. Namely, the single-particle 
states unambiguously separate into two groups: one that is called \emph{Bose selected}, whose occupations increase 
linearly when the total particle number is increased at fixed system size, and another one whose occupations saturate. This phenomenon is a consequence of the 
bosonic quantum statistics. It includes standard Bose condensation into a single quantum state, fragmented Bose 
condensation into a small number of single-particle states each acquiring a macroscopic occupation, and the case where a 
fraction of all single-particle states acquires large, but individually non-extensive occupations. The 
properties of the system, like its coherence or its heat conductivity, sensitively depend on which of these scenarios 
occurs. 

The physics of driven-dissipative ideal Bose gases is intimately related also to collective effects in classical systems 
and processes, where bunching phenomena have been identified as analog of Bose condensation. This includes the dynamics 
of networks and economic models \cite{BianconiBarabasi01,BurdaEtAl02}, classical transport and traffic
\cite{Chowdhury00,EvansHanney05,EvansEtAl06,SchwarzkopfEtAl08,HirschbergEtAl09,KimEtAl10,ThompsonEtAl10,
SchadschneiderEtAl10,GrosskinskyEtAl11}, chemical reactions \cite{vanKampen}, as well as population dynamics and 
evolutionary game theory \cite{KnebelEtAl13}. These connections have recently been discussed nicely by Knebel \emph{et al.\ }\cite{KnebelEtAl15}.

In this paper, we investigate non-equilibrium steady states of driven-dissipative ideal quantum gases of both bosons and 
fermions. We focus on systems of sharp particle number that exchange energy with the environment. These quantum gases are 
driven out of equilibrium either by the coupling to several heat baths of different temperature or by time-periodic 
driving in combination with the coupling to a heat bath (see Fig.~\ref{fig:sys_config}). 
We treat the problem using (Floquet-)Born-Markov theory \cite{BreuerPetruccione,BluemelEtAl91,KohlerEtAl97,BreuerEtAl00,
HoneEtAl09}, which is valid in the limit of weak system-bath coupling. In section~\ref{sec:quantum_gas} this theoretical 
framework is reviewed and applied to the problem of the ideal quantum gas. Morever, several model systems are introduced. 
In order to treat the resulting many-body master equation, we then describe analytical and numerical methods for 
computing the steady state (Section~\ref{sec:methods}). This includes a standard mean-field description in terms of
single-particle occupation numbers. We, moreover, derive an augmented mean-field theory taking into account also
non-trivial two-particle correlations, and explain how to apply quantum-jump-type Monte-Carlo simulations to the problem. 
These methods are then applied to both the ideal Bose gas (Section~\ref{sec:bosons}) and the ideal Fermi gas
(Section~\ref{sec:fermions}). 

Our treatment of the fermionic case in Section~\ref{sec:fermions} is rather brief and demonstrates the application of our theory to simple lattice models and the possibility to achieve exotic states via bath engineering. These results can be 
relevant, e.g., for the problem of realizing Floquet topological insulators with periodically forced electronic systems 
(graphene \cite{OkaAoki09} or semiconductor heterostructures \cite{LindnerEtAl11}).

The largest part of this paper is devoted to bosonic quantum gases and the phenomenon of Bose selection discussed in 
Section~\ref{sec:bosons}. Here we first review equilibrium Bose condensation (Sec.~\ref{sec:equilibrium}) and Bose 
selection in non-equilibrium steady states (Secs.~\ref{sec:driven_dissipative_bose_gas} to
\ref{sec:proof} give a detailed discussion of the results of Reference \cite{VorbergEtAl13}). After that, we derive a 
theory for transitions where the set of selected states changes (Sec.~\ref{sec:transition}), present an efficient 
algorithm for finding the set of selected states (Sec.~\ref{sec:algorithm}), discuss the possibility of approaching a 
preasymptotic state at intermediate densities before the true asymptotic state is reached at large densities 
(Sec.~\ref{sec:preasymptotic}), investigate the properties of systems described by non-fully connected rate matrices 
(Sec.~\ref{sec:ZeroRates}), study the role of fluctuations and beyond mean-field effects (Sec.~\ref{sec:beyond_mf}),
and identify the dominant mechanisms for heat transport in the Bose selected state Sec.~\ref{sec:heat_transport}.

\section{General framework and models}
\label{sec:quantum_gas}
In this section we set up the master equations for an ideal quantum gas of $N$ indistinguishable, noninteracting 
particles, weakly coupled to one or several heat baths. We cover both the case of an autonomous system with
time-independent Hamiltonian $\hat H$ and the case of a Floquet system with time-periodic Hamiltonian $\hat H(t)=\hat H(t+\tau)$.
This captures the non-equilibrium situations depicted in Fig.~\ref{fig:sys_config}.
In the case of the periodically driven system, we encounter the Floquet states 
$|\phi_i(t)\ra=\re^{-\ri  \epsilon_i t/\hbar} |i(t)\ra$, which are quasistationary (i.e.\ time-periodic) solutions of the 
dynamics generated by $\Ho(t)$ \cite{Shirley65,Zeldovich67,Sambe73}.
Here, $|i(t)\ra=|i(t+\tau)\ra$ denotes time-periodic Floquet modes while $\epsilon_i$ are the quasienergies, which are 
defined modulo the energy quantum $\hbar\omega$ with angular driving frequency $\omega=2\pi/\tau$. We start with the 
single-particle equations. In Sec.~\ref{sec:ideal_gas_rate_equation}, we will then generalize to the many-body case.

\subsection{Single-particle master equation}
\label{sec:RateEquation}
We consider the time evolution of the density operator $\hat{\rho}$ in a single-particle system. In the weak-coupling 
limit, where the full rotating-wave approximation is valid, this time evolution is governed by a master equation of 
Lindblad type \cite{BreuerPetruccione}, which in the interaction picture reads
\begin{equation}\label{eq:EOMrho_Lindblad}
\frac{\rd\hat{\rho}(t)}{\rd t}=\mathcal{D}[\hat\rho]= \sum_{i,j} R_{ij}\left(\hat L_{ij} \hat\rho(t)
\hat L_{ij}^{\dagger}-\frac{1}{2}\{\hat\rho(t), \hat L_{ij}^{\dagger}\hat L_{ij}\}\right).
\end{equation}
Here $\{A,B\} = AB+BA$ denotes the anticommutator. The indices enumerate the energy eigenstates of the autonomous system, 
or the Floquet states of the periodically driven system. In practice, we will restrict the number $M$ of participating single-particle states to be finite.
The dissipation causes transitions from eigenstate $|j\ra$ to eigenstate $|i\ra$ according to the
jump operator $\hat L_{ij}=| i\ra\la j|$, where $R_{ij}$ is the corresponding transition rate.
This description is valid in the  weak-coupling limit, where the level broadening $\hbar R_{ij}$ due to the transitions 
is much smaller than the (typical) energy separation of neighboring (quasi)energy levels in the spectrum of the system. The characteristic time 
scale $\tau_S$ of the unitary dynamics is then much smaller than the time scale $\tau_R$ of the dissipative relaxation,
$\tau_S\ll \tau_R$, which allows to employ the full rotating-wave approximation leading to Eq.~(\ref{eq:EOMrho_Lindblad}) \cite{BluemelEtAl91,KohlerEtAl97,BreuerEtAl00,HoneEtAl09}.

Since the resulting Lindblad equation \eqref{eq:EOMrho_Lindblad} is diagonal in the basis of states $|i\ra$, the dynamics 
of the occupation probabilities $p_i=\la i|\hat\rho |i\ra$ decouples from the off-diagonal elements of the density 
operator, which decay as one approaches the steady state. The dynamics of the diagonal elements are described by the 
Pauli master equation
\begin{equation}
\label{eq:spme}
{\dot p}_i(t) =
\sum_{j} \left[ R_{ij} p_j(t)- R_{ji}p_i(t)\right].
\end{equation}
The terms of the sum correspond to the net probability flux from states $j$ to state $i$. 
The uniqueness of the steady state $\hat\rho=\sum_i p_i|i\ra\la i|$, obtained by requiring $\dot p_i=0$, is guaranteed by 
the Frobenius-Perron theorem, which holds if every state is connected with all the other states by a sequence of 
transitions with non-vanishing rates \cite{Schnakenberg76}.

For the weak coupling to the environment considered here, the rates $R_{ij}$ in Eq.~\eqref{eq:spme} can in general be 
determined in the Born-Markov (Floquet-Born-Markov) approximation for 
autonomous (time-periodically driven) systems. We will consider that a bath is given by a collection of harmonic 
oscillators $\alpha$ with angular frequency $\omega_\alpha$ and annihilation operator $\hat b_\alpha$, described by 
the bath Hamiltonian $\Ho_B=\sum_\alpha\hbar\omega_\alpha \hat b^\dag_{\alpha} \hat b_\alpha$. The bath is in thermal 
equilibrium with  temperature $T$ and coupled to the system via the Hamiltonian
$\Ho_{SB}=\hat{v}\sum_\alpha c_{\alpha} (\hat b^\dag_\alpha + \hat b_\alpha)$, where $c_\alpha$ are the coupling 
parameters and $\hat{v}$ a coupling operator acting in the state-space of the system. 

Within the Floquet-Born-Markov approximation, the rates for the driven system are given by Fermi's golden rule 
\cite{BluemelEtAl91,KohlerEtAl97,BreuerEtAl00,HoneEtAl09},
\begin{equation}
\label{eq:rates_floquet}
R_{ji}=\sum_{m=-\infty}^{\infty}R_{ji}^{(m)},\quad
R_{ji}^{(m)} =\frac{2\pi}{\hbar}|v_{ji}(m)|^2g(\varepsilon_j-\varepsilon_i-m\hbar\omega).
\end{equation}
Here  $v_{ji}(m)=\frac{\omega}{2\pi}\int_0^{2\pi/\omega}\rd t\re^{\ri m\omega t}\la j(t)|\hat{v}|i(t)\ra$ are the Fourier 
coefficients of the coupling matrix elements, where the index $m$ accounts for the absorption or emission of
$|m|$ energy quanta $\hbar\omega$ due to the driving. The quantity 
\be
g(E)=\frac{J(E)}{e^{\beta E}-1}=g(-E)e^{-\beta E}
\ee 
is the bath correlation function, determined by the inverse temperature $\beta=1/T$ (the Boltzmann constant is set to one) 
and the spectral density 
\be
J(E)=\sum_{\alpha} c_{\alpha}^2[\delta(E-\hbar\omega_{\alpha})-\delta(E+\hbar\omega_{\alpha})]=-J(-E).\ee
We will assume Ohmic baths characterized by a spectral density that increases linearly with $E$, $J(E)\propto E$. 

In the autonomous system, Eq.~\eqref{eq:rates_floquet} simplifies to 
\begin{equation}
\label{eq:rates}
R_{ji}=\sum_{b\in\{1,2\}} R_{ji}^{(b)},\quad  R_{ji}^{(b)}= \frac{2\pi}{\hbar}|v_{ji}^{(b)}|^2 g_b(E_j-E_i),
\end{equation}
Here $v_{ji}^{(b)}=\langle j|\hat{v}^{(b)}|i\rangle$ now denote the matrix elements of the coupling operator of heat bath 
$b$ with respect to the eigenstates $|i\ra$ with energy $E_i$. The rate is further characterized by the correlation 
functions $g_b(E)=J_b(E)[\exp(\beta_b E)-1]^{-1}$ of both baths, with spectral density $J_b(E)$ and inverse temperature
$\beta_b$.

Later we will see that the rate-asymmetry matrix 
\begin{equation}
A_{ij}=R_{ij}-R_{ji}
\label{eq:rate_asymmetry}
\end{equation}
plays a major role since many properties of the system depend on this matrix only.
In the time-periodically driven case it reads
\begin{align}\label{eq:Aper}
A_{ij}=&\sum_{m=-\infty}^{\infty}A^{(m)}_{ij},\nonumber\\
A^{(m)}_{ij} =& R^{(m)}_{ij} - R^{(m)}_{ji} =\frac{2\pi}{\hbar}|v_{ji}(m)|^2J(\varepsilon_j-\varepsilon_i-m\hbar\omega)
\end{align}
whereas for the autonomous system one has 
\begin{align}\label{eq:Atwo}
A_{ij}=&\sum_{b\in\{1,2\}}A_{ij}^{(b)},\nonumber\\
	A_{ij}^{(b)}=&R_{ij}^{(b)}-R_{ji}^{(b)}=\frac{2\pi}{\hbar}|v^{(b)}_{ji}|^2J_b(E_j-E_i).
\end{align}
Note that the rate-asymmetry matrix is independent of the bath temperature(s). 

In contrast to equilibrium, a non-equilibrium steady state can retain a constant energy flow through the system.
For the periodically driven system, the transition described by the rate $R_{ji}^{(m)}$ causes a change of the bath
energy by $\epsilon_i-\epsilon_j+m\hbar\omega$. The total energy flow from the system 
to the bath is thus given by
\begin{equation}
\label{eq:heat_flow_driven_single_particle}
Q(t)=\sum_{ijm}
 (\epsilon_i-\epsilon_j+m\hbar\omega)R_{ji}^{(m)}p_i(t).
\end{equation}
Note that also \emph{pseudotransitions} described by rates $R_{ii}^{(m\ne0)}$ contribute to the heat flow
\cite{LangemeyerHolthaus14}. These transitions change the state of the bath, but not that of the system. For the 
autonomous system the energy flow into bath $b$ reads
\begin{equation}
\label{eq:heat_flow_undriven_single_particle}
Q_b(t)=\sum_{ij}(E_i-E_j) R_{ji}^{(b)}p_i(t).
\end{equation}

\subsection{Master equation for the ideal quantum gas}
\label{sec:ideal_gas_rate_equation}
We now generalize the single-particle problem to a gas of $N$ indistinguishable, non-interacting particles.
In our approach we assume the total particle number $N$ to be fixed, like in the canonical ensemble.
For our considerations the canonical description poses the advantage that it contains the single-particle case as the 
natural limit $N=1$, and does not require to define new terms describing the particle exchange with the bath.

The many-body Hilbert space is spanned by Fock states enumerated by the occupation numbers of the $M$ single-particle 
states, $\bn=(n_1,n_2,\ldots,n_M)$. To obtain the many-body rate equations we replace the single-particle jump operators
$\hat L_{ij}=|i \ra \la j|$ in \Eq{eq:EOMrho_Lindblad} by their Fock-space representation
\be\label{eq:mbj}
\hat{L}_{ij}=\aod_i \ao_j.
\ee
Here $\ao_i$ denotes the annihilation operator of a particle, boson or fermion, in the single-particle mode $i$. Quantum jumps still correspond to processes transferring a single particle from one 
mode to another. The validity of the full rotating-wave approximation is, thus, still determined by the {single-particle} 
problem. Moreover, the total particle number $N$ is conserved by the dynamics. 

As before, the dynamics of the many-body occupation probabilities $p_{\bn}=\la \bn|\hat\rho|\bn\ra$ decouple from 
the off-diagonal elements, which decay over time. The corresponding equations of motion are now given by
(see Appendix~\ref{sec:appendix_MPmasterequation} for details)
\be
\dot p_{\bn}(t)
= \sum_{ij} (1+\sigma n_j)n_i  \left[R_{ij} p_{\bn_{ji}}(t)-R_{ji} p_\bn(t)\right],
\label{eq:mpr}
\ee
which is the many-body generalization of the Pauli master equation \eqref{eq:spme}.
Here $\bn_{ji}=(n_1,\ldots,n_i-1,\ldots,n_j+1,\ldots)$ denotes the occupation numbers obtained from
$\bn$ by transferring one particle from $i$ to $j$.
The effective transition rate depends on the quantum statistics via the choice of $\sigma$, with $\sigma=1$ for bosons
(reflecting the enhancement of transitions into occupied states) and $\sigma=-1$ for fermions (reflecting the Pauli 
exclusion principle). The classical case of distinguishable (Boltzmann) particles corresponds to  
$\sigma=0$; here the transition rates are independent of the occupation of the final state.\footnote{The bosonic master equation
(\ref{eq:mpr}) with $\sigma=1$, as well as the corresponding mean-field equation (\ref{eq:EOM_nmean_MF}), also resemble rate 
equations that are used to describe stochastic processes in classical systems, as we mention them already in the 
introduction.}

For the periodically driven ideal gas the energy flow from the system into the bath is given by
\begin{align}
\label{eq:heat_flow_driven_many_particle}
Q(t) =&\sum_{ m}\sum_\bn\sum_{ij} (\epsilon_i-\epsilon_j+m\hbar\omega)R_{ji}^{(m)} (1+\sigma n_j)n_i p_\bn(t)
\nonumber\\
=&\sum_{m}\sum_{ij} (\epsilon_i-\epsilon_j+m\hbar\omega) R_{ji}^{(m)}\big[\la\no_i\ra(t)+\sigma\la\no_i\no_j\ra(t)\big].
\end{align}
Analogously, for the autonomous ideal gas the energy flow into bath $b$ reads
\begin{align}
\label{eq:heat_flow_undriven_many_particle}
Q_b(t)=&\sum_{\bn}\sum_{ij}(E_i-E_j)R_{ji}^{(b)}(1+\sigma n_j)n_i  p_\bn(t)
\nonumber\\
=&\sum_{ij} (E_i-E_j) R_{ji}^{(b)}\big[\la\no_i\ra(t)+\sigma\la\no_i\no_j\ra(t)\big].
\end{align}

\subsection{Non-equilibrium steady state}
\label{sec:steady_state}

In the following we are interested in the properties of the steady state of the ideal quantum gas, whose density operator 
shall simply be denoted by $\hat{\rho}.$\footnote{Whenever we are discussing transient behavior and time-dependent 
quantities (which happens only a few times) this will be indicated by writing out explicitly the time argument. For 
example, $\hat{\rho}(t)$ denotes the time-dependent density operator or $\la\hat{o}\ra(t)$ a time-dependent expectation 
value. Otherwise, i.e.\ when writing $\hat{\rho}$ or $\la\hat{o}\ra$, we are always referring to steady-state quantities.} 
It is diagonal in the occupation number basis,
\begin{equation}\label{eq:ness}
\hat{\rho} = \sum_{\bn} p_\bn |\bn\ra\la\bn|,
\end{equation}
with $p_\bn$ determined by solving Eq.~(\ref{eq:mpr}) for $\dot p_\bn=0$. The uniqueness of the steady state \cite{Schnakenberg76} is inherited 
from the single-particle system, since every Fock state is connected to every other Fock state by a sequence of
allowed single-particle transitions when this is assumed for the single-particle system. 

The steady-state expectation value of an arbitrary observable $\hat{o}$ is denoted by  
\be
\la\hat{o}\ra=\tr(\hat{\rho}\hat{o}).
\ee
Expectation values that we will consider in the following are the mean occupations that we denote by
\begin{equation}
\bar{n}_i = \la \no_i\ra,
\end{equation}
with the number operator $\no_i = \aod_i \ao_i$ and the two-particle correlations 
$\la \no_i \no_j \ra$ or, rather, their non-trivial part
\begin{equation}
\f{i}{j}=\la \no_i \no_j \ra - \avgo{n_i}\avgo{n_j} = \la (\no_i-\bar{n}_i)(\no_j-\bar{n}_j)\ra.
\label{eq:non_trivial_correlations}
\end{equation}

For the scenarios depicted in Fig.~\ref{fig:sys_config} the steady state of the system will be a \emph{non-equilibrium} steady 
state. This can be illustrated already on the level of the single-particle problem (\ref{eq:spme}). Let us first 
recapitulate the case of thermal equilibrium. The transitions induced by a single bath of inverse temperature $\beta$ in 
an autonomous system are described by rates that obey
\begin{equation}\label{eq:db}
\frac{R_{ji}}{R_{ij}} = e^{-\beta(E_j-E_i)}.
\end{equation}
This can be inferred from Eq.~(\ref{eq:rates}) for the case of a single bath. This condition implies that the steady 
state, obtained by solving Eq.~(\ref{eq:spme}) is given by the Gibbs state with $p_i=Z^{-1}e^{-\beta E_i}$ and
$Z=\sum_i e^{-\beta E_i}$. For this equilibrium state, the sum on the right-hand side of Eq.~(\ref{eq:spme}) vanishes 
term by term. Thus, the net probability flux between two states $i$ and $j$ vanishes. 
This is the property of detailed balance, which is characteristic for the thermodynamic equilibrium. 

The rates characterizing the periodically driven system, Eq.~(\ref{eq:rates_floquet}), or the autonomous system coupled 
to two heat baths of different temperature, Eq.~(\ref{eq:rates}), are a sum of rates corresponding to different energy 
changes in the bath or to different bath temperatures, respectively. As a consequence, they do not obey the condition
(\ref{eq:db}) anymore. This implies that, generally, the steady state also does not fulfill detailed balance anymore. 
While the net probability flux into a state $i$, determined by the right-hand-side of Eq.~(\ref{eq:spme}), still has to 
vanish, the probability current from a certain state $j$ to state $i$ can be non-zero, i.e.\ the sum in Eq.~(\ref{eq:spme}) 
does not vanish term by term. The lack of detailed balance characterizes a non-equilibrium steady state. In contrast to 
the equilibrium state, which is determined by the temperature of the bath only, the non-equilibrium steady state 
depends on the very details of the bath(s) (the temperature, the coupling operator, and the spectral density). 
This makes the computation of the many-body non-equilibrium steady state a difficult problem. However, it also offers opportunities to realize states with properties that 
are hard (or impossible) to achieve in equilibrium.

\subsection{Model systems}
\label{sec:models}
\label{sec:model_tb_chain}
\begin{figure}[t]
\centering
\includegraphics[width=1\linewidth]{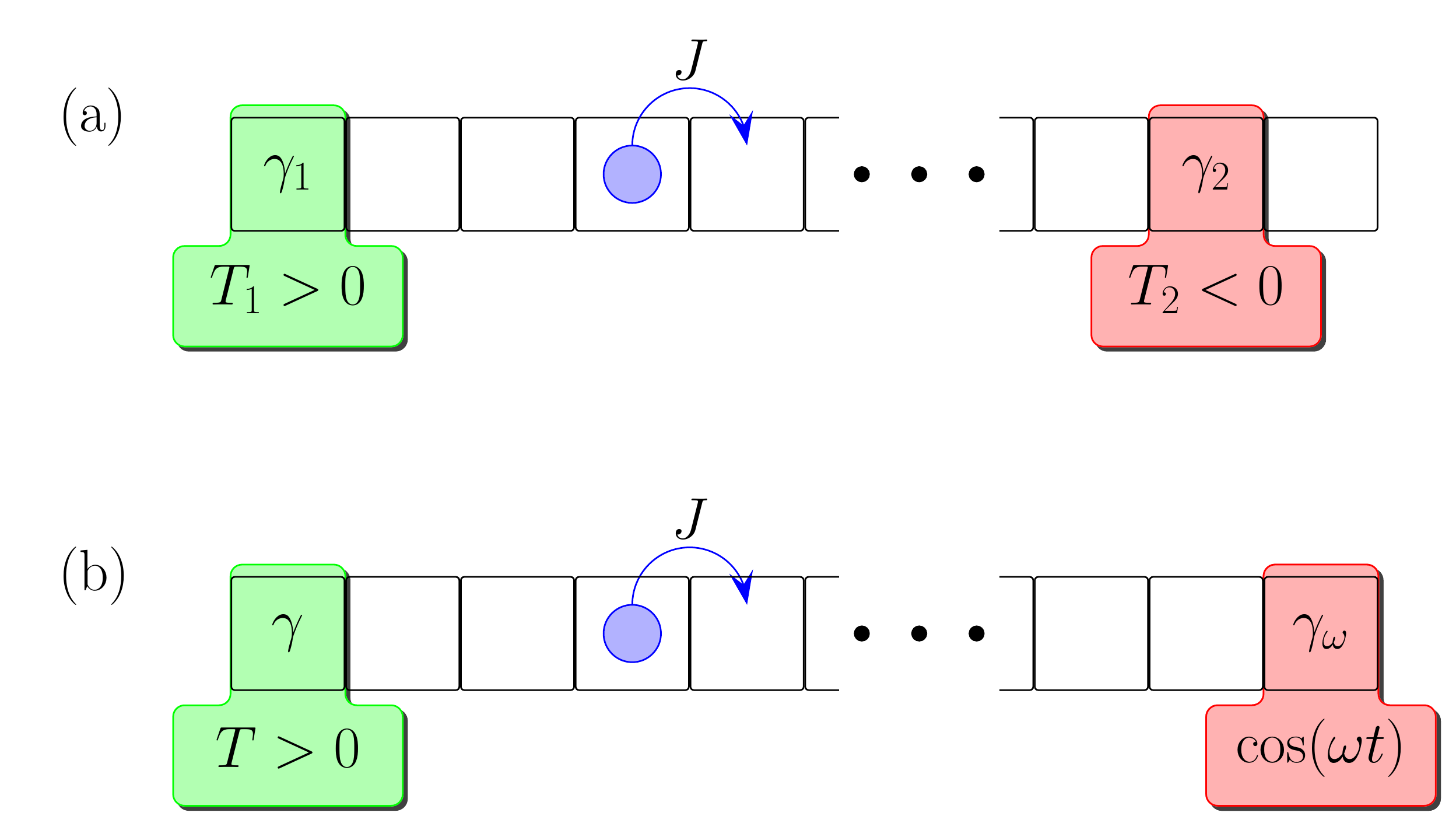}
\caption{(color online) Two model systems. (a) Tight-binding chain coupled to two heat baths of respective temperatures $T_1$ and $T_2$ 
and coupling strengths $\gamma_1$ and $\gamma_2$. (b) Tight-binding chain subjected to a time-periodic potential modulation  at one end with driving strength $\gamma_\omega$ and angular frequency ${\omega}$ and coupled to a heat bath of temperature $T$ at the other end with coupling strength $\gamma$.}
\label{fig:tb_chain}
\end{figure}

Throughout this paper, we will illustrate our findings using three different model systems. Let us 
briefly define them here. Note that our results are not limited to these example systems. 

The first model system is a tight-binding chain of $M$ lattice sites. 
It is described by the Hamiltonian 
\begin{equation}
\hat{H}=-J \sum_{\ell=1}^{M-1} (\hat{c}_\ell^{\dagger} \hat{c}_{\ell+1}+\hc),
\end{equation}
wherein $\hat{c}_{\ell}$ ($\hat{c}_\ell^\dag$) denotes the annihilation (creation) operator for a particle at site $\ell$.
The single-particle eigenstates $|i\ra$, with $i=0,1,\ldots, M-1$, are delocalized. They are described by wave functions
$\la\ell|i\ra\propto\sin(k_i\ell)$, with wave numbers $k_i=(i+1)\pi/(M+1)$ 
and possess energies $E_i=-2J\cos(k_i)$ between $-2J$ and $2J$.
As sketched in Fig.~\ref{fig:tb_chain}(a), the chain is coupled to two baths, on the left and right end of the chain. The 
left (right) bath is locally coupled to the first (next-to-last) site of the chain via the coupling operators
$\hat{v}_1=\gamma_1\hat{c}_1^{\dagger}\hat{c}_1$ and $\hat{v}_2=\gamma_2\hat{c}_{M-1}^{\dagger}\hat{c}_{M-1}$, 
respectively.\footnote{We avoid the choice of coupling the second bath to the last site $M$, since for such a symmetric 
configuration the generic effect of fragmented Bose condensation \cite{VorbergEtAl13} is absent.} This coupling describes 
a bath induced fluctuation of the on-site energy. The steady state will depend on the coupling strength only through 
their relative weight $\gamma_2/\gamma_1$, while their absolute weight determines how fast the system relaxes. The 
temperatures of the baths are different from each other. We will, moreover, mainly focus on the interesting case where 
one of the baths is population inverted. For such a situation the notion of the single-particle ground state becomes 
meaningless, allowing for fragmented Bose condensation with multiple condensates \cite{VorbergEtAl13}, see
Section~\ref{sec:bosons} below. We model the population inverted bath by a negative temperature $T_2<0$ and a spectrum 
that is bounded from above ($\omega_\alpha<0$).

The second model system is also given by a tight-binding chain of $M$ sites. However, instead of coupling it to a 
second bath, the chain is periodically driven in time. Its Hamiltonian is given by
\begin{equation}
H(t)=-J \sum_{\ell=1}^{M-1}\left( \hat{c}_\ell^{\dagger} \hat{c}_{\ell+1} + \hc\right)
			+ \gamma_{\omega}J\cos (\omega t)\hat{c}_M^{\dagger}\hat{c}_M,
\end{equation}
with the dimensionless driving strength $\gamma_{\omega}$ and angular frequency $\omega$. The coupling to a bath of 
inverse temperature $\beta$ is realized via the coupling operator $\hat{v}=\gamma\hat{c}_1^{\dagger}\hat{c}_1$, as 
depicted in Fig.~\ref{fig:tb_chain}(b). The steady state will depend on the dimensionless driving strength $\gamma_\omega$,
which determines the single-particle Floquet modes and the structure of the rate matrix $R_{ij}$. However, the coupling 
strength to the heat bath $\gamma$ has no impact on the steady state, but rather determines how fast the system relaxes. 

Finally, as a third model, we consider a system of $M$ single-particle states with the transition rates $R_{ij}$ given by 
uncorrelated random numbers, independently drawn from an exponential distribution
\begin{equation}
P(R_{ji})=\lambda^{-1}\exp(-\lambda R_{ji}).
\label{eq:random_rate}
\end{equation}
The parameter $\lambda$ controls the time scale of the relaxation, but does not influence the steady state. The diagonal 
elements $R_{ii}$ can be set to 0 as they drop out of all relevant equations (such as Eq.~\eqref{eq:spme}). This choice 
of rates clearly models a non-equilibrium situation, since detailed balance is violated almost surely. It is motivated by 
the rates computed for fully chaotic periodically driven quantum systems coupled to a heat bath \cite{Wustmann10}. A concrete example is given by the kicked rotor coupled to a bath which is discussed for single particles in Ref.~\cite{KetzmerickWustmann10} and for many particles in the supplemental
material of Ref.~\cite{VorbergEtAl13}.

\section{Methods}
\label{sec:methods}
In this paper we are interested in the properties of non-equilibrium steady states (\ref{eq:ness}) of driven dissipative 
ideal quantum gases of $N$ particles, described by the master equation (\ref{eq:spme}) with jump operators (\ref{eq:mbj}) 
or, equivalently, by the rate equation (\ref{eq:mpr}). 
Even though the particles are non-interacting, finding the steady state is a true many-body problem. Unlike in equilibrium, the many-particle solution cannot be obtained from the single-particle solution in a straight-forward manner. 
This is a consequence of the interaction with the bath and reflected in the fact that the right-hand side of the master equation
(\ref{eq:spme}) is quadratic in the jump operators (\ref{eq:mbj}) and, thus, quartic in the bosonic or fermionic field 
operators $\hat{a}_i^{(\dag)}\!.$ As a consequence, equation \eqref{eq:mpr} quickly becomes intractable when the particle 
number is increased. Therefore, it is crucial to develop and apply suitable methods for the approximate treatment of the 
problem. This shall be done in this section.

In the following, we will first describe quantum-jump-type Monte-Carlo simulations based on averaging over random walks 
in the classical space of sharp occupation numbers. This numerical method is quasi exact (the statistical error is 
controlled) and allows for the treatment of moderately large systems. In order to treat even larger systems and to obtain 
an intuitive picture of the dynamics, we will then describe a mean-field theory, which will be based on a description in 
terms of the mean occupations $\bar{n}_i$. Finally, we augment the mean-field theory by taking into account fluctuations 
given by non-trivial two-particle correlations.

\subsection{Monte-Carlo simulations}
\label{sec:MonteCarlo}

Quantum-jump Monte-Carlo simulations \cite{PlenioKnight98,MolmerCastin99} are an efficient method for computing the
time evolution of open quantum systems described by a Markovian master equation of Lindblad form. Instead of 
integrating the time evolution of the full density matrix, the method is based on integrating the time evolution of 
single states (the Monte-Carlo wave function). In doing so, the dissipative effect of the environment is included by 
interrupting the continuous time evolution by a sudden quantum jump, described by one of the jump operators. When such a 
quantum jump occurs, and which one, is drawn from a suitable probability distribution. The time evolution of expectation 
values can then be obtained by averaging over an ensemble of Monte-Carlo wave functions. The error depends on the 
ensemble size and can, in principle, be made arbitrarily small. 

When treating the master equation (\ref{eq:spme}) with jump operators (\ref{eq:mbj}) we encounter a convenient 
situation. The dissipation can be described by jump operators (\ref{eq:mbj}) that transfer a particle from one
single-particle eigenstate (or Floquet state) to another one, i.e.\ between two states of sharp occupation numbers $\bn$.
At the same time, these occupation numbers are conserved by the evolution generated by the system Hamiltonian, since we 
are dealing with a system of non-interacting particles. Therefore, 
the time evolution is exhausted by taking into account quantum jumps. 
This corresponds to a random walk in the classical space spanned by the Fock states $|\bn\ra$ (not their superpositions). 
The Monte-Carlo wave function $|\bn(t)\ra$ jumps between Fock states $|\bn_k\rangle$, in which it resides for time 
intervals of length~$t_k$,
\begin{equation}
\label{eq:mc_jump}
|\bn(t)\ra=|\bn_k\ra \quad\text{with $k$ such that } T_{k-1}\le t<T_k,
\end{equation}
where $T_k=\sum_{l=1}^{k}t_l$.

\begin{figure}[t]
\centering
\includegraphics[width=1\linewidth]{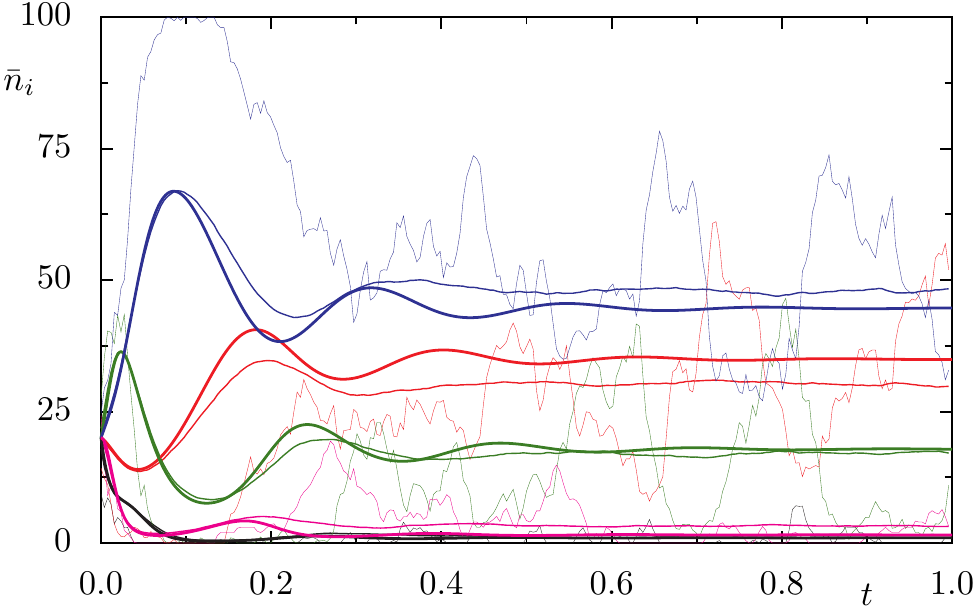}
\caption{(color online) Time evolution of the mean occupations $\avgo{n_i}(t)$ for one realization of the random-rate model for $M=5$ 
states and $N=100$ particles. Time is measured in units of the inverse mean rate $\lambda$ [see Eq.~\eqref{eq:random_rate}]. Initially, each single-particle state is occupied with the same probability. The thin lines 
are obtained from a single Monte-Carlo wave function, the intermediate lines from an ensemble of $L=1000$ Monte-Carlo 
wave functions, and the thick lines from mean-field theory. The mean-field results show small systematic 
deviations from the Monte-Carlo result.}
\label{fig:mc_time}
\end{figure}

We use the Gillespie algorithm \cite{Gillespie76} in order to compute the time evolution. At the beginning, the system is 
prepared according to the chosen initial conditions. Then the algorithm alternates between the following two steps.
(i) The time interval $t_k$ determining how long the system will remain in the current state is drawn randomly from an 
exponential distribution $P(t_k)\propto\exp(-t_k/\bar{t}(\bn_k))$ with mean dwell time 
\begin{equation}
\bar{t}(\bn_k) = \frac{1}{\sum_{i,j}R_{ij}(1+\sigma n_i)n_j}.
\label{eq:bart}
\end{equation}
(ii) The new state with occupation $\bn_{k+1}$ is drawn randomly with branching probability reflecting the many-body transition rates $R_{ji}(1+\sigma n_j)n_i$.
Since only single-particle jumps are involved in Eq.~\eqref{eq:mpr}, the next state is obtained from
the current state by transferring a particle from a randomly drawn departure
state $i$ to the randomly drawn target state $j$. This single-particle jump has the probability
\begin{equation}
P(i\to j,\bn_k)=\bar{t}(\bn_k) R_{ji}(1+\sigma n_j)n_i.
\end{equation}
These two steps are repeated until $T_k=\sum_{l=1}^k t_l$ exceeds the desired evolution time $t_\text{fin}$.

From an ensemble of $L$ Monte-Carlo wave functions $|\bn^{(\alpha)}(t)\ra$ labeled by $\alpha=1,2,\ldots,L$, one can then 
compute the expectation value of an observable $\hat{o}$, 
\begin{align}
\la \hat{o}\ra_{\mathrm{ensemble}}(t)=&\frac{1}{L}\sum_{\alpha=1}^{L}{\la\bn^{(\alpha)}(t)|\hat{o}|\bn^{(\alpha)}(t)\ra}.
\end{align}
Figure~\ref{fig:mc_time} shows the time evolution of the mean occupations $\la\no_i\ra(t)$ for $N=100$ particles on
$M=5$ states, for a single Monte-Carlo wave function (thin lines) and for an ensemble with $L=1000$ (intermediate lines). 
One can clearly observe the relaxation to a steady state reached after a relaxation time of $\tau_r\approx 0.5$. Slight 
temporal fluctuations observed for times $t>\tau_r$ decrease with ensemble size $L$. The mean-field theory (thick lines) described below predicts the occupations rather well, but with small systematic deviations from the Monte-Carlo result. 

When computing steady-state expectation values $\la \hat{o} \ra$, the effect of temporal fluctuations can be reduced by 
combining ensemble averaging with time averaging,
\begin{align}
\la \hat{o} \ra 
=& \frac{1}{L}\sum_{\alpha=1}^L \sum_{k} \frac{\la\bn^{(\alpha)}_{k}|\hat{o}|\bn^{(\alpha)}_{k}\ra}{t_k}.
\end{align}
Here it is useful to constrain the inner sum to $k>k^{(\alpha)}_r$, with $k^{(\alpha)}_r$ such that $T_{k_r^{(\alpha)}}>t_r$, 
in order to exclude the transient relaxation process from the time average. Since we assume that every state is connected 
with all the other states by a sequence of transitions with non-vanishing rates, one can obtain accurate steady-state 
expectation values from a single Monte-Carlo trajectory, provided $t_\text{fin}$ is sufficiently large so that the system 
forgets its initial state after a certain correlation time. Averaging over a long time is, therefore, equivalent to 
averaging over an ensemble. We determine these uncertainties according to the Gelman-Rubin criterion
\cite{gelman1992inference}, generally setting the relative uncertainties below one percent (small enough to make 
statistical fluctuations barely noticeable in any figure). For a bosonic system, this allows us to access particle numbers $N\sim 10^5$ for $M=100$ single-particle states.

\subsection{Mean-field theory}
\label{sec:MeanField}

In order to treat even larger systems and to gain some intuitive understanding of the non-equilibrium steady state of 
ideal quantum gases, it is desirable to use also analytical methods. One of them is a mean-field description of the 
system in terms of the mean occupations $\bar{n}_i$ \cite{VorbergEtAl13}. 

The time evolution of the mean occupations is given by the equations
\begin{align}\label{eq:n_idyn}
 \frac{\diff}{\diff t}  \avgo{n_i}(t)
=& \tr\left( \no_i  \frac{\diff}{\diff t}\hat \rho(t) \right)
\nonumber\\
=&  \sum_{j} R_{ij}\bigg\{\big[\avgo{n_j}(t)+\sigma \la \no_i\no_j\ra(t)\big]
\nonumber\\&\qquad
-R_{ji}\big[\avgo{n_i}(t)+\sigma \la \no_i\no_j\ra(t)\big]\bigg\}
\end{align}
for all $i$ (see  Appendix \ref{sec:appendix_EOM}). Here we encounter the typical hierarchy: The time evolution of 
single-particle correlations (expectation values of operators that are quadratic in the field operators) are 
governed by two-particle correlations (expectation values of operators that are quartic in the field 
operators). The evolution of the latter will in turn be determined by three-particle correlations and so on. 

In order to obtain a closed set of equations in terms of the mean occupations, we employ the factorization approximation 
\be\label{eq:MFtriv}
\la\no_i\no_j\ra(t)= \bar{n}_i(t) \bar{n}_j(t) +\ze_{ij}(t) \approx  \bar{n}_i(t) \bar{n}_j(t)
\ee
for $i\ne j$. 
Here non-trivial correlations are neglected, $\ze_{ij}(t)\approx 0$, so that two-particle correlations are 
approximated by a product of single-particle expectation values as if Wick's theorem was valid. 
In this way we arrive at the set of non-linear mean-field equations
\begin{align}\label{eq:EOM_nmean_MF}
   \frac{\diff}{\diff t} \avgo{n_i}(t)
 \approx&\sum_{j} \bigg\{ R_{ij}\avgo{n_j}(t)\big[1+\sigma \avgo{n_i}(t)\big]
\nonumber\\&\qquad
			-R_{ji}\avgo{n_i}(t)\big[1+\sigma \avgo{n_j}(t)\big]\bigg\}.
\end{align}

In the classical case of distinguishable particles, which can be shown to be captured by $\sigma=0$, the mean-field 
equation is exact. In this case, the equations of motion for the mean occupations $\bar{n}_i(t)$ are of the same form as 
the single-particle master equation (\ref{eq:spme}) for the probabilities $p_i(t)$. Therefore, in the classical system 
the mean occupations are determined by the single-particle problem and read $\bar{n}_i(t)=p_i(t)N$. In contrast, for 
quantum gases of indistinguishable bosons or fermions the dynamics and the steady state will depend in a non-trivial way
on the total particle number. In this case, the classical solution can still be an approximate solution of the quantum system as 
long as $\bar{n}_i\ll1$ for all $i$, so that two-particle correlations $\la\no_i\no_j\ra$ are negligible. However, as 
soon as the quantum degenerate regime is reached, where $\bar{n}_i\gtrsim1$ at least for some $i$, quantum statistics and 
with that the particle number will matter.

The mean-field equations of motion can also be obtained by making a \emph{Gaussian ansatz}, 
\begin{equation}\label{eq:GaussianState}
 \hat \rho_g = \frac{1}{Z} \exp\left[ -\sum_i \eta_i \no_i \right] ,
\end{equation}
with partition function  $Z$
for the many-body density operator. For this ansatz the mean occupations are given by
\begin{eqnarray}
\label{eq:nu_n}
\la\no_i\ra_g= \frac{1}{\re^{\eta_i}- \sigma}.
\label{eq:gamma_vs_n}
\end{eqnarray}
Thus, the $M$ parameters defining the Gaussian state are determined completely by the $M$ mean-occupations, 
$\eta_i=\ln(\la \no_i\ra_g^{-1}+\sigma)$, as they can be obtained by solving the mean-field equations
Eqs.~(\ref{eq:EOM_nmean_MF}). 
Non-trivial correlations vanish and multi-particle correlation functions can be decomposed into products of 
single-particle correlations determined by Wick decomposition. For the two-particle correlations the Gaussian ansatz gives
\cite{Castin04} 
\begin{eqnarray}
 \la \no_i \no_j\ra_g= 
 \left\{\begin{array}{ll}
   \la\no_i\ra_g \big[ (1+\sigma)\la\no_i\ra_g + 1\big] & \mbox{ for } i = j \\
   \la\no_i\ra_g\la\no_j\ra_g   & \mbox{ for } i \neq j .
 \end{array} \right.
\end{eqnarray}
for bosons ($\sigma = 1$) and fermions ($\sigma = -1$). 
For $i\ne j$ we find $\la\no_i\no_j\ra_g=\la\no_i\ra_g\la\no_j\ra_g$. Therefore, starting from
Eq.~(\ref{eq:n_idyn}) and making the Gaussian ansatz for the density operator, we recover the mean-field equations of
motion (\ref{eq:EOM_nmean_MF}) with $\bar{n}_i(t)=\la\no_i\ra_g$. 

With the quantities $\la\no_i^2\ra_g$, the Gaussian ansatz also determines the fluctuations of the occupations $\no_i$ as 
well as of the total particle number $\hat{N}=\sum_i\no_i$. One finds
\begin{align}\label{eq:Deltanig}
\big\la\big(\no_i-\la\no_i\ra_g\big)^2\big\ra_g = \la\no_i^2\ra_g-\la\no_i\ra_g^2 
	= \la\no_i\ra_g +\sigma\la\no_i\ra_g^2
\end{align}
and
\begin{align}\label{eq:DeltaN}
\big\la\big(\No-\la\No\ra_g\big)^2\big\ra_g 
  =&  \sum_i\Big(\la\no_i^2\ra_g - \la\no_i\ra_g^2\Big)
	\nonumber\\&
		+\sum_{i,j\ne i} \Big(\la\no_i\no_j\ra_g -\la\no_i\ra_g\la\no_j\ra_g\Big)
\\\label{eq:DeltaNg}
=& \sum_i\big\la\big(\no_i-\la\no_i\ra_g\big)^2\big\ra_g.
\end{align}
The Gaussian state does not describe a system with a sharp particle number, so that we can only require that the mean 
particle number obeys
\be
\la\hat N\ra_g = N .
\ee

Fluctuations of the total particle number are an immediate consequence of enforcing trivial correlations
$\la\no_i\no_j\ra=\bar{n}_i\bar{n}_j$ for $i\ne j$ (unless also the occupations of the individual states $i$ are 
sharp so that their number fluctuations $\la\no_i^2\ra-\bar{n}_i^2$ vanish). This can be seen from Eq.~(\ref{eq:DeltaN}), 
where we have not yet used the properties of the Gaussian state like 
in Eq.~(\ref{eq:DeltaNg}). It is intuitively clear that a sharp total particle number induces non-trivial correlations 
among the occupations. If the measurement of the occupation $\no_i$ gives a value $n_i$ that is smaller (larger) than the 
expectation value $\bar{n}_i$, a sharp total particle number implies that the number of particles in all other states 
is given by $N-n_i$ and, thus, larger (smaller) than the original expectation value $N-\bar{n}_i$. As a consequence, 
the probability of measuring a certain value $n_j$ of the occupation $\no_j$ with $j\ne i$ will depend on the value $n_i$ 
measured for the occupation $\no_i$.

The role played by fluctuations of the total particle number becomes less and less important in large systems. 
Namely, the variance of the total particle number (\ref{eq:DeltaNg}) is the sum over the variances of the occupations of 
individual modes (\ref{eq:Deltanig}), which are intensive. Thus the fluctuations of the total particle number grow in 
a subextensive fashion like the square root of the system size. That is the relative fluctuations of the total particle 
number vanish in the limit of large systems. This is the mechanism underlying the equivalence of the canonical and the
grand-canonical ensemble. There is one important exception, however. This is the case of Bose-Einstein condensation, 
where in a bosonic system a mode $i$ acquires a macroscopic occupation. If the total particle number is not conserved 
also the number fluctuations of the condensate mode will be as large as the number of condensed particles; in this case
the right-hand side of Eq.~(\ref{eq:Deltanig}) is dominated by the second term. The extensive number fluctuations in the 
condensate mode will then dominate the sum of Eq.~(\ref{eq:DeltaNg}) and give rise to extensive total number fluctuations, 
which are non-negligible in large systems. 
This phenomenon is know as the grand-canonical fluctuation catastrophe \cite{HolthausEtAl98}. 

However, one should note that the dynamics of the mean occupations $\bar{n}_i(t)$ described by Eq.~(\ref{eq:n_idyn}) 
do not depend on the occupation number fluctuations of the modes (the term $j=i$ vanishes so that $\la\no_i^2\ra$ does 
not enter on the right-hand side). The mean-field equations of motion (\ref{eq:EOM_nmean_MF}) can, therefore, provide a 
good approximation to the mean occupations $\bar{n}_i$ also in systems featuring Bose condensation
(see reference \cite{VorbergEtAl13}). This can be seen also in Fig.~\ref{fig:mc_time}, where despite the fact that half 
of the particles occupy a single mode, mean-field theory accurately describes both the transient and the long-time 
behavior of the mean occupations.

The grand-canoncial ensemble of an ideal quantum gas in equilibrium with inverse temperature $\beta$ and chemical 
potential $\mu$ is described by a Gaussian density operator (\ref{eq:GaussianState}) with $\eta_i=\beta (E_i-\mu)$. The  
mean occupations \Eq{eq:gamma_vs_n} follow the Bose-Einstein (Fermi-Dirac) distribution for $\sigma=1$ ($\sigma=-1$). 
The grand-canonical ideal gas is thus described exactly within the mean-field theory. 
This can be seen explicitly by plugging the Gaussian state $p_\bn\propto\prod_i e^{-\beta(E_i-\mu)n_i}$  (solving the
mean-field equation) into the full many-body rate equations (\ref{eq:mpr}). By employing condition (\ref{eq:db}), which 
is fulfilled in an equilibrium situation, one can see that the sum on the right-hand side vanishes term by term. This 
implies also that the equilibrium state obeys detailed balance as it should. Deviations from mean-field theory occur as a 
consequence of two factors, (i) the assumption of a sharp total particle number and (ii) the violation of the 
detailed-balance condition (\ref{eq:db}). 

Both factors (i) and (ii) are independent of each other, as can be illustrated using two examples. The canonical 
equilibrium state with sharp particle number is characterized by the non-Gaussian probabilities
\begin{equation}
p_{\bn}=\begin{cases}\frac{1}{Z_N}\exp\left(-\sum_i\beta E_i n_i\right) &\mbox{ if }\sum_i n_i = N\\
0 &\mbox{ otherwise}
\end{cases}
\label{eq:gaussian_factorized}
\end{equation}
with the partition function $Z_N$.
This state can be obtained by projecting the Gaussian state onto the subspace of sharp total particle number $N$. 
As a consequence of the sharp particle number, it does not solve the mean-field equation, as was discussed above. 
However, it still obeys detailed balance. Namely, plugging it into Eq.~(\ref{eq:mpr}) the sum on the right-hand-side vanishes 
term by term as long as the condition (\ref{eq:db}) is fulfilled.
On the other hand, we can allow the particle number to fluctuate freely, but violate condition (\ref{eq:db}). Then it will 
generally not be possible to find a solution of the mean-field form (\ref{eq:GaussianState}) 
that solves the many-body rate equations (\ref{eq:mpr}), because the number of 
independent equations exceeds the number of parameters $\eta_i$.
In the following, we are interested in the situation, where a system of sharp particle number is driven into a steady state far 
away from equilibrium, so that both factors (i) and (ii) are present. Here, the mean-field theory can still provide a good approximation, as can be checked by comparing it to 
quasi-exact results obtained from Monte-Carlo simulations. 

Within the mean-field approximation, the heat flow for the autonomous system to bath $b$, given by
Eq.~(\ref{eq:heat_flow_undriven_many_particle}), takes the form
\begin{equation}
\label{eq:heat_flow_undriven}
Q^{(b)}(t)=\sum_{i,j\ne i} (E_i-E_j)R_{ji}^{(b)}\avgo{n_i}(t)\big[1+\sigma\avgo{n_j}(t)\big].
\end{equation}
The heat flow from the periodically driven ideal gas into the heat bath (\ref{eq:heat_flow_driven_many_particle}) reads
\begin{align}
\label{eq:heat_flow}
Q(t)=&\sum_m\sum_{i,j\ne i}
(\epsilon_i-\epsilon_j+m\hbar\omega)R_{ji}^{(m)} \avgo{n_i}(t)\big[1+\sigma\avgo{n_j}(t)\big]
\nonumber\\&
+\sum_m \sum_{i} m\hbar\omega R_{ii}^{(m)} \big[\bar{n}_i(t) +\sigma \la\no_i^2\ra(t)\big].
\end{align}
Here the second sum captures the heat flow related to pseudotransitions [see discussion below Eq.~\eqref{eq:heat_flow_driven_single_particle}]. Their contribution depends on $\la\no_i^2\ra$ 
and, thus, on the occupation number fluctuations of the modes. However, as discussed above, in a bosonic system of sharp 
total particle number and where some modes feature macroscopic occupation, the Gaussian expectation value $\la\no_i^2\ra_g =\la\no_i\ra_g \big[ 2\la\no_i\ra_g + 1\big]$ does generally not provide a good approximation for the 
condensate mode(s). Therefore, it might be useful to introduce another approximation for $\la\no_i^2\ra$ in an \emph{ad hoc} fashion. Another possibility is to augment the mean-field theory such that it is able to treat systems with sharp 
particle number and, thus, with non-trivial two-particle correlations. Such a method will be presented in the following 
subsection.

\subsection{Augmented mean-field theory}
\label{sec:augmented}

By construction, the mean-field theory fails to take into account non-trivial two-particle correlations $\ze_{ij}$ as 
they result from having a sharp total particle number and from driving the system out of equilibrium, so that the
detailed-balance condition (\ref{eq:db}) is violated. The effects of a fluctuating total number of particles can be 
assessed by projecting the Gaussian state onto the subspace of $N$-particle states,
$\hat{\rho}_{\text{proj}}\propto \hat{P}_N\hat{\rho}_g\hat{P}_N$ with 
$\hat{P}_N=\sum_{\bn | \sum_i \no_i=N}|\bn\ra\la\bn|$. This introduces non-trivial correlations, which can be obtained from
$\la \no_i \no_j\ra=\tr\left(\rho_{\mathrm{proj}}\no_i \no_j\right)$. However, evaluating this matrix element is an onerous task 
even within efficient algorithms (see Appendix~\ref{sec:app_proj} for an example), since all $N$-particle Fock states 
have to be accounted for. Moreover, such an approach still does not include effects related to the breaking of detailed 
balance. 

In order to include the effects of non-trivial occupation correlations and fluctuations by analytic means, we introduce 
an augmented mean-field theory. This approach includes the two-point correlation functions $\la \no_k \no_i \ra$ into the 
hierarchy of equations of motions. In the original full hierarchy, the corresponding equations of motion take the form
\begin{align}\label{eq:EOM_nncorr}
\nonumber
\frac{\diff}{\diff t} \la \no_k \no_i\ra
 =&  \sum_j\big\{ \sigma(A_{kj} + A_{ij}) \la \no_k \no_i \no_j \ra + R_{kj} \la \no_i \no_j \ra
  \nonumber\\&
 + R_{ij} \la \no_k \no_j \ra   - \left(R_{jk} + R_{ji}\right) \la \no_k \no_i \ra 
\nonumber\\&
+ \delta_{ik} \big[ R_{kj}(\bar{n}_j+\sigma\la \no_k \no_j \ra) 
\nonumber\\&
\qquad + R_{jk}(\bar{n}_k + \sigma\la \no_k \no_j \ra )\big]\big\}
\nonumber\\&
 - R_{ik} (\avgo{n_k}+\sigma \la \no_k \no_i \ra) - R_{ki} (\avgo{n_i}+\sigma \la \no_k \no_i \ra) .
\end{align}
Here, as well as in the rest of this subsection, we suppress time arguments. 
This equation still involves the third-order correlations $\la \no_k \no_i \no_j \ra$.

The hierarchy can be closed by assuming trivial three-particle correlations. For that purpose we separate the 
number operators like 
$\no_i = \avgo{n_i} + \zetao_i$
into their mean values $\bar{n}_i$ and their fluctuations 
\begin{equation}
\zetao_i=\no_i-\bar{n}_i\quad\text{with}\quad \la \zetao_i \ra=0.
\end{equation}
We now approximate
\begin{equation}\label{eq:condition_extendedMF}
 \la \zetao_k \zetao_i \zetao_j \ra = 0,
\end{equation}
while allowing, in contrast to mean-field theory, for non-trivial two-particle correlations $\f{k}{i}=\la \zetao_k\zetao_i \ra$ [Eq.~\eqref{eq:non_trivial_correlations}]. Thus, the equations of motion for the mean occupations are given by
\begin{align}\label{eq:EOM_xmean}
  \frac{\rd \avgo{n_k}}{\rd t}
  = &\sigma \sum_{j} A_{kj} \left[ \avgo{n_k} \avgo{n_j} + \f{k}{j} \right] \nonumber\\
 & +  \sum_{j} \left( R_{kj} \avgo{n_j}  - R_{jk} \avgo{n_k} \right),
\end{align}
which is equivalent to the exact equation~(\ref{eq:n_idyn}). The equations of motion for the non-trivial two-particle 
correlations are obtained from \Eq{eq:EOM_nncorr} by employing the approximation (\ref{eq:condition_extendedMF}). It is
non-linear and  reads (see  Appendix~\ref{sec:appendix_EOMcorrelations} for details)
\begin{align}\label{eq:EOM_zzcorr_MF}
\frac{\diff \f{k}{i}}{\diff t}
\approx& \sum_j\big\{ \sigma \big[ A_{kj} \bar{n}_k \f{i}{j} + A_{ij} \bar{n}_i \f{k}{j}
+(A_{kj} + A_{ij}) \bar{n}_j \f{k}{i} \big] 
\nonumber \\& 
+  R_{kj} \f{i}{j} + R_{ij} \f{k}{j}
- (R_{jk} + R_{ji}) \f{k}{i} 
\nonumber \\& 
+ \sigma    (\delta_{ki}-\delta_{ji})(R_{kj} + R_{jk}) ( \bar{n}_k \bar{n}_j + \f{k}{j}) 
\nonumber \\& 
+ (\delta_{ki}-\delta_{ji})(  R_{kj} \bar{n}_j + R_{jk} \bar{n}_k)\big\}.
\end{align}
The steady state values of $\bar{n}_k$ and $\f{k}{i}$ have to be determined by solving Eqs.~\eqref{eq:EOM_xmean} and 
\eqref{eq:EOM_zzcorr_MF} with the left-hand-side set to zero. 

Within the augmented mean-field theory the state is not only described in terms of the mean occupations $\bar{n}_i$, but 
also in terms of non-trivial two-particle correlations $\ze_{ki}$. As a consequence, we cannot only fix the mean 
total particle number to a value $N$ by requiring
\be
\la\No\ra = \sum_i \bar{n}_i = N.
\ee
Also the fluctuation of the total particle number can be fixed to a value $\Delta N$
\be
\la\No^2\ra-\la\No\ra^2=\sum_{ij}\ze_{ij} = \Delta N^2 .
\ee
This includes the choice
\be
\Delta N = 0
\ee
for a system of sharp particle number. Whereas the mean-field theory was found to be equivalent to a Gaussian ansatz for 
the density operator, we cannot give an analytical expression for the density operator corresponding to the augmented 
theory.

\section{Ideal Bose gases and Bose selection}
\label{sec:bosons}
In this section we discuss in detail the steady state of non-interacting bosonic quantum gases. Let us first recapitulate the case of 
thermodynamic equilibrium. 

\subsection{Equilibrium and Bose condensation}
\label{sec:equilibrium}

Under equilibrium conditions, where the rates obey the condition (\ref{eq:db}), the mean-field equations of motion 
(\ref{eq:EOM_nmean_MF}) with $\sigma = 1$ for bosons are solved by a steady state characterized by the mean occupations 
\be\label{eq:gcni}
\bar{n}_i=\frac{1}{e^{\beta (E_i-\mu)}-1}, 
\ee
corresponding to Eq.~(\ref{eq:nu_n}) with $\eta_i=\beta(E_i-\mu)$. For this solution the right-hand side of
Eq.~(\ref{eq:EOM_nmean_MF}) vanishes term by term, indicating detailed balance. The occupation numbers (\ref{eq:gcni}) 
obtained from the non-number-conserving mean-field theory correspond to the exact grand-canonical mean occupations
\cite{Pathria} and provide a good approximation also for the canonical ensemble with sharp particle number $N$. In the 
latter case, the chemical potential has to be chosen such that 
\be
\sum_i\bar{n}_i = N. 
\ee
Assuming the states of the system to be labeled such that  
\be
E_0<E_1\le E_2\le\cdots,
\ee
meaningful positive occupation numbers correspond to values of the chemical potential below the ground-state energy,
$\mu<E_0$. The chemical potential increases either when $\beta$ is increased at fixed $N$ or when $N$ is increased at 
fixed $\beta$. 

When in a system of finite extent, with discrete energies $E_i$, the particle number $N$ is increased at fixed $\beta$,
the chemical potential will eventually approach the ground-state energy so that $E_0-\mu \ll E_1-E_0$. Once this happens 
at a characteristic particle number $N^*$ specified below, the mean occupations of the excited states can be approximated by 
\be\label{eq:depletion}
\bar{n}_i \simeq \frac{1}{e^{\beta (E_i-E_0)}-1}  \quad\text{ for }\quad i\ge 1.
\ee
Thus, for $N\gg N^*$ the occupations of excited states become independent of $\mu$ (therefore also 
of $N$) and saturate. 
The occupation of the single-particle ground-state still 
depends on the chemical potential; assuming $\beta(E_0-\mu)\ll1$, one finds
\be\label{eq:condensate}
\bar{n}_0 \simeq \frac{1}{\beta (E_0-\mu)} \equiv N_0, 
\ee
with 
\be\label{eq:condensate2}
N_0 \simeq N-\sum_{i\ge 1} \frac{1}{e^{\beta (E_i-E_0)}-1}, 
\ee
such that $\mu\simeq E_0-T/N_0$. All particles that cannot be ``accommodated'' in the excited states will occupy the 
ground state. This is the phenomenon of \emph{Bose-Einstein condensation} (or, strictly speaking, its finite size precursor).

In a finite system Bose-Einstein condensation is a crossover, 
occurring when $N$ becomes comparable to the characteristic value $N^*$, which is directly given by the depletion of the 
condensate, 
\be\label{eq:NcharEq}
N^*=\sum_{i\ge 1} \frac{1}{e^{\beta (E_i-E_0)}-1} .
\ee

\begin{figure*}[t]
\centering
\includegraphics[width=1\linewidth]{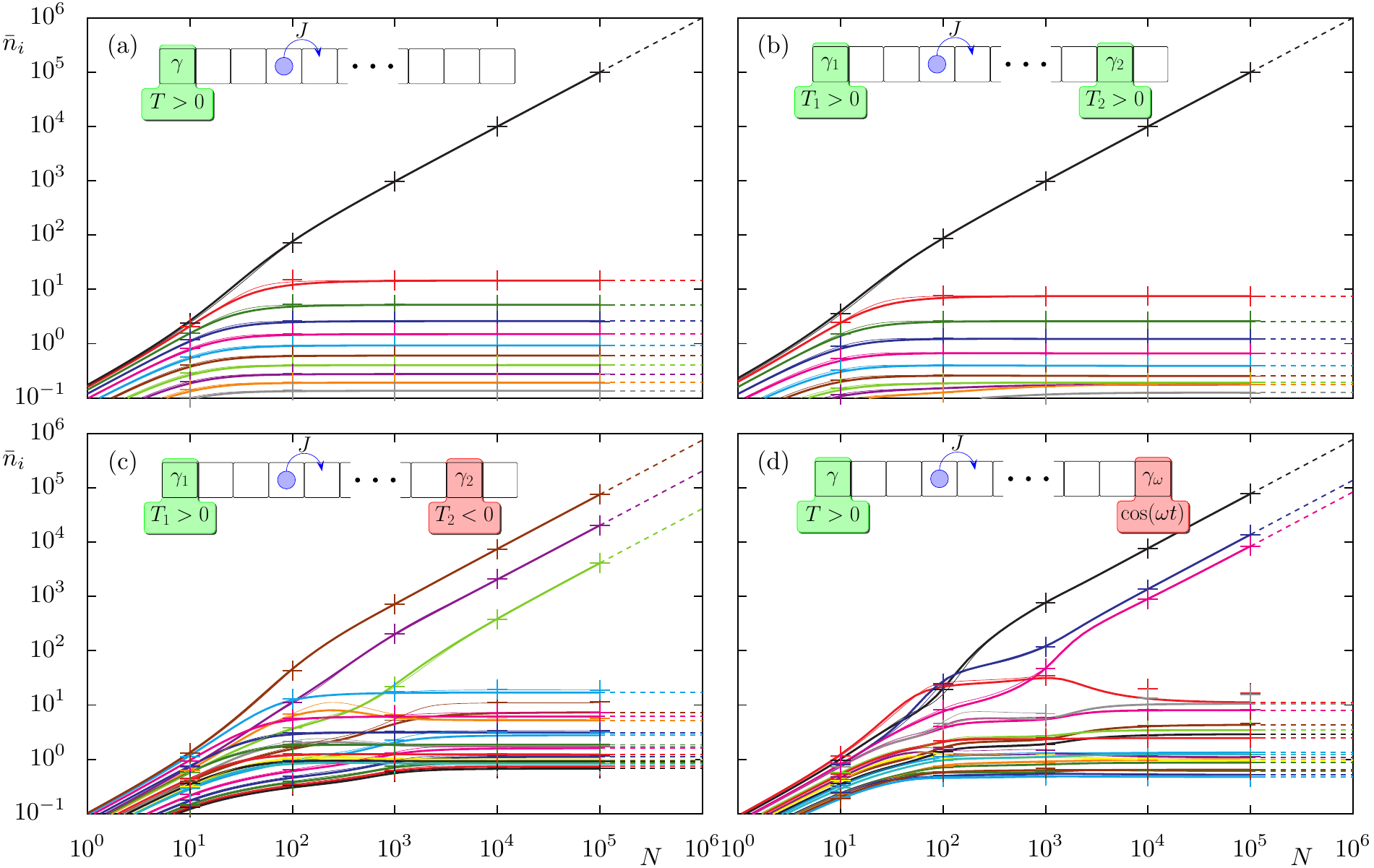}
\caption{(color online) Mean occupations versus total number of bosons for the steady state of a tight-binding chain of $M=20$ sites and 
tunneling parameter $J>0$. The data is obtained from mean-field theory (thick solid lines), asymptotic mean-field theory
(dashed lines), augmented mean-field theory (thin solid lines), and exact Monte-Carlo simulations (crosses). 
(a) Equilibrium situation, the chain is coupled to one bath of temperature $T=1J$. (b) The chain is driven away from 
equilibrium by two heat baths of different positive temperature ($T_1=1J$ and $T_2=0.5J$), coupled to the first and the next to last site with $\gamma_1=\gamma_2$. (c) Same as in (b), but now the second bath is population inverted and 
described by the negative temperature $T_2=-J$. The color code is the same as in panels (a) and (b), where the occupations decrease with increasing energy. (d) The chain is driven away from equilibrium by a periodic potential 
modulation at the last site with amplitude $\gamma_\omega=2.3 J$ and frequency $\hbar\omega=1.5J$. The Floquet states are colored like the stationary states (a-c) from which they evolve adiabatically when the driving is switched on
(see Fig.~\ref{fig:heat_transport_2}).}
\label{fig:BS}
\label{fig:mean_field}
\end{figure*}

In the thermodynamic limit, defined by taking particle number $N$ and volume $V$ to infinity while holding the density
$n=N/V$ at a constant finite value, Bose condensation is a sharp phase transition. At a critical density
$n_c=N^*/V$, the occupation of the ground state becomes macroscopic and the ratio $N_0/N$, the condensate fraction, 
assumes a non-zero value. At the transition $E_0-\mu= T/N_0$ becomes zero. However, Bose condensation does not 
necessarily survive the thermodynamic limit. For a homogeneous Bose gas of spatial dimensionality $D\le2$, the ratio
$N^*/V$ diverges in the thermodynamic limit due to large occupations of low-energy states, so that no phase transition 
exists. In this case Bose condensation can still be observed as a crossover in systems of finite size. This is 
illustrated in Fig.~\ref{fig:BS}(a), where we plot the mean occupations of a bosonic one-dimensional 
tight-binding chain of $M=20$ sites versus the particle number $N$. In this system $M$ plays the role of a 
dimensionless volume $V$ so that the density is given by the dimensionless filling factor $n=N/M$. One can observe a 
sharp crossover: For $N>N^*$ the occupations of the excited states saturate so that newly added particles will all become 
part of the condensate in the ground state as described by Eqs.~(\ref{eq:depletion}), (\ref{eq:condensate}) and
(\ref{eq:condensate2}). 

\subsection{Driven-dissipative Bose gas and Bose selection}
\label{sec:driven_dissipative_bose_gas}
The other panels of Fig.~\ref{fig:BS} show the mean occupations $\bar{n}_i$ versus $N$ for situations where the tight-binding chain is driven into a steady state far from equilibrium, either by coupling it to a 
second bath of different temperature or by time-periodic forcing (see section \ref{sec:models}). In each of these 
panels, we can again identify a sharp crossover. When the particle number $N$ reaches a characteristic value $N^*$,
many occupations saturate as in equilibrium. However, as a striking effect, newly added particles can now occupy a whole 
group of states [Fig.~\ref{fig:BS}(c-d)], with constant relative occupations among these states. These selected 
states take over the role played by the condensate mode in equilibrium. This phenomenon has been termed \emph{Bose
selection} \cite{VorbergEtAl13}. It turns out to be the generic behavior in the ultra degenerate regime of large density 
at fixed finite system size. 

As becomes apparent from Fig.~\ref{fig:BS}, we can distinguish two scenarios. Either a single state becomes selected. 
This includes the case of equilibrium Bose condensation depicted in panel (a), but also the non-equilibrium situation 
shown in panel (b), where a Bose gas is driven out of equilibrium by the coupling to two heat baths of different positive 
temperature. Or multiple states become selected as it can be seen in panel (c) and (d), corresponding to situations where 
a system is driven out of equilibrium by an additional population-inverted bath of negative temperature or by periodic 
forcing. As we will see in the following, the essential difference between both scenarios is that in the situations (a) 
and (b) the notion of the single-particle ground state is still meaningful. In panel (b) both baths favor larger 
occupations in states of lower energy and thus the largest occupation occurs in the ground state. This is not the 
case anymore for the situations (c) and (d). The population-inverted negative temperature bath of the system of panel (c) 
favors larger occupations in states of higher energy counteracting the effect of the positive-temperature bath. For the 
periodically driven system of panel (d), the quasienergies of the single-particle Floquet states are determined modulo
$\hbar\omega$ only, so that a ground state is not even defined.

\begin{figure}[t]
\centering
\includegraphics[width=1\linewidth]{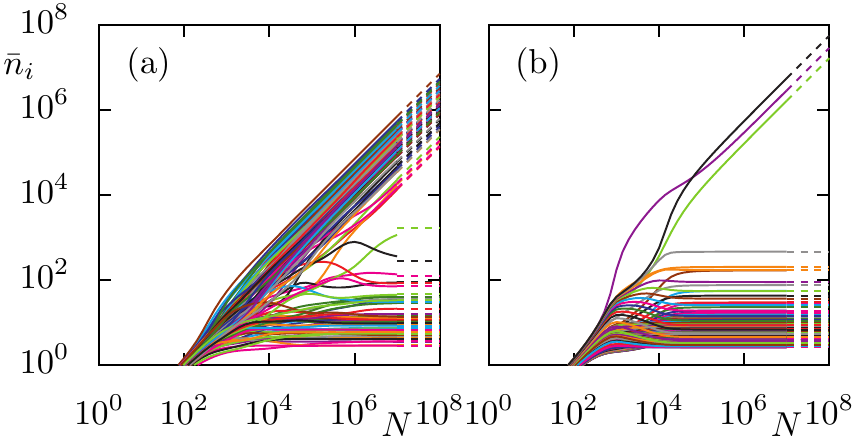}
\caption{(color online) Mean occupations versus total number of bosons for (a) one realization of the random-rate model with $M=100$ states and (b) a tight-binding chain of $M=100$ sites coupled to two heat baths, namely one with temperature $T_1=10J$ at the first site and a population inverted bath described by the negative temperature $T_2=-10J$ at the fifth-last site, with equal coupling strength, $\gamma_1=\gamma_2$ [see inset of Fig.~\ref{fig:BS}(c)].}
\label{fig:largeM}
\end{figure}

Within the scenario of having multiple selected states we can, furthermore, distinguish two possibilities. For that 
purpose we have to consider systems of a large number of states $M$. In Fig.~\ref{fig:largeM} we plot the mean 
occupations for two systems with $M=100$ states. Panel (a) corresponds to one realization of the random-rate model and 
panel (b) is obtained for a tight-binding chain coupled to a second population-inverted bath like in Fig.~\ref{fig:BS}(c). 
For the random-rate model (a) the number of selected states $M_S$ is of the order of the system size $M$, roughly half of 
the states become selected for sufficiently large $N$. This implies that none of the selected states acquires a 
macroscopic occupation of the order of the total particle number. 
For the tight-binding chain (b) we find that the number of selected states $M_S$ is still of the order of one, namely 
three states are selected. As a consequence, each selected state acquires a macroscopic occupation of the order of the 
total particle number and hosts a Bose condensate. This corresponds to fragmented Bose condensation\footnote{Note that 
the system does not feature a single condensate in a state being a coherent superposition of the highly occupied selected 
modes, but independent condensates in each mode. Namely, according to the Penrose-Onsager criterion Bose-Einstein 
condensation is defined by a macroscopic eigenvalue of the single-particle density matrix $\la a_i^\dagger a_j\ra$
\cite{PenroseOnsager56}. In the situation discussed here, the off-diagonal elements of $\la a_i^\dagger a_j\ra$ are 
negligible as a consequence of the weak coupling to the bath. Therefore, each macroscopic mean-occupation
$\bar{n}_i=\la a_i^\dagger a_i\ra$ corresponds to a macroscopic eigenvalue of the single-particle density matrix and an 
independent Bose condensate.}, which is therefore a generic situations for driven Bose gas, unlike in equilibrium where 
this requires a rare ground state degeneracy. 
Thus, all in all we can distinguish three generic types of Bose selection occurring in the ultradegenerate regime of
driven-dissipative ideal Bose gases: standard Bose condensation where a single state acquires a macroscopic occupation, 
fragmented Bose condensation where a small number (of order one) of selected states each acquires macroscopic occupation, 
and the selection of a large number of states with non-extensive individual occupations that together attract most 
particles of the system.  

In the following we will provide a theory for Bose selection based on mean-field theory in the asymptotic 
limit of large $N$. It can be viewed as a generalization of the Eqs.~(\ref{eq:depletion}), (\ref{eq:condensate}) and
(\ref{eq:condensate2}) describing equilibrium Bose condensation to the case of driven-dissipative ideal Bose gases.
Later on, also effects beyond mean-field will be discussed in terms of the augmented mean-field theory.

\subsection{Asymptotic mean-field theory}
\label{sec:Asymptotic}

A theoretical description of Bose selection can be based on mean-field theory, given by Eq.~(\ref{eq:EOM_nmean_MF}) 
with $\sigma=1$. For the steady state this equation reads
\begin{align}\label{eq:mf_steady}
0= \sum_{j} \Big[ R_{ij}\avgo{n_j}\big(1+\avgo{n_i}\big)
			-R_{ji}\avgo{n_i}\big(1+\avgo{n_j}\big)\Big]
\end{align}
for all $i$. Since Bose selection occurs in the asymptotic limit of large densities, it appears natural to approximate 
\be
1+\avgo{n_k}\approx  \avgo{n_k}
\ee
in this equation. One then obtains the equations\footnote{\label{fn:lotka_volterra}It is interesting to note that these equations correspond to the
conservative Lotka-Volterra equations $\dot{\bar{n}}_i = \avgo{n_i}\sum_jA_{ij} \avgo{n_j}$ as they are used to model population dynamics. Indeed, for fully connected rate matrices, the selected states 
correspond directly to those species that will not be extinct, but survive \cite{chawanya2002large,KnebelEtAl13,KnebelEtAl15}. Differences appear, however, for not fully connected rate matrices as will be discussed at the end of subsection \ref{sec:ZeroRates}. }
\begin{equation}
0=\avgo{n_i}\sum_j(R_{ij}-R_{ji}) \avgo{n_j}=\avgo{n_i}\sum_jA_{ij} \avgo{n_j}.
\label{eq:naiv_app}
\end{equation}
One can immediately see that some of the mean occupations $\bar{n}_i$ have to vanish on this level of approximation. 
Namely, if we assume that a subset $\mathcal{S}$ of single-particle states possesses non-zero occupations, these 
states have to obey the linear equations
\begin{equation}
\label{eq:selected_a}
 0 = \sum_{j\in \mathcal{S}} A_{ij}  \avgo{n_j}, \qquad i\in\mathcal{S},
\end{equation}
which directly follow from Eq.~(\ref{eq:naiv_app}). However, without fine-tuning of the skew-symmetric asymmetry matrix
$A_{ij}=-A_{ji}$, these equations have a solution only if $\mathcal{S}$ contains an odd number of states (since a
skew-symmetric matrix generically possesses an eigenvalue zero only when acting in an odd-dimensional space). Moreover, 
even if a formal solution can be found for a certain set $\mathcal{S}$, it is not guaranteed that this solution will 
correspond to physically meaningful solutions, where all occupation numbers are non-negative. Both conditions constrain the set $\mathcal{S}$, so 
that generically it will not contain all states. Those states contained in the (yet to be determined) set $\mathcal{S}$ 
correspond to the Bose selected states. 

In order to compute the occupations of the non-selected states, we have to include another level of approximation. For 
that purpose we use that the occupation of a non-selected state is determined predominantly by transitions from or 
into selected states. The large occupations of the selected states enhances the corresponding rates with respect to the 
rates for transitions from or into other non-selected states. Thus, neglecting transitions among non-selected states and still assuming $n_j+1\approx n_j\ \forall j\in \BES$, from Eqs.~\eqref{eq:mf_steady} for non-selected states $i$ we obtain
\begin{equation}\label{eq:nonselected_a}
 \avgo{n_i} = \frac{1}{g_i -1}
\quad\text{with}\quad
g_i = \frac{\sum_{j\in \BES} R_{ji}\avgo{n_j}}{\sum_{j\in \BES}R_{ij}\avgo{n_j}}, \quad i\notin \BES.
\end{equation}
This approximation is reminiscent of the Bogoliubov approximation \cite{Bogoliubov47} for the weakly interacting Bose gas, 
where interactions among non-condensed particles are neglected. 

The set $\mathcal{S}$ has to be chosen such that 
physically meaningful occupations 
\be\label{eq:condition}
\bar{n}_i\ge 0 
\ee 
are obtained for all $i$, [i.e.\ both for the selected states, whose relative occupations are determined by
Eq.~(\ref{eq:selected_a}), and for the non-selected states, with the occupations given by Eq.~(\ref{eq:nonselected_a})]. 
We will prove in the following subsection \ref{sec:proof} that there exists a unique set $\mathcal{S}$ for which 
condition (\ref{eq:condition}) is fulfilled. Thus, the problem to be solved does not simply consist in solving
Eqs.~(\ref{eq:selected_a}) and (\ref{eq:nonselected_a}) for a given set $\mathcal{S}$. It is rather the task of finding 
both the occupations $\bar{n}_i$ \emph{and} the set $\mathcal{S}$, for which the relations (\ref{eq:selected_a}),
(\ref{eq:nonselected_a}), and (\ref{eq:condition}) are fulfilled. 

By identifying the states of the set $\mathcal{S}$ with the selected states, we can now explain the major features of the 
results presented in Fig.~\ref{fig:BS}. One observation is that for large $N$ the relative occupations among the 
selected states become independent of $N$. This is explained by the fact that these relative occupations are determined 
by the set of linear Eqs.~(\ref{eq:selected_a}), which does not depend on $N$. A second observation is that the 
occupations of the non-selected states saturate in the limit of large $N$. Such a behavior is predicted by
Eq.~(\ref{eq:nonselected_a}), where the $g_i$ are determined by the $N$-independent relative occupations of the selected 
states. This implies also that the total occupation of the selected states, 
\be\label{eq:Ns}
N_S = \sum_{i\in \mathcal{S}}\bar{n}_i = N-\sum_{i\notin\mathcal{S}} \frac{1}{g_i -1},
\ee
grows linearly with $N$. Finally, we can estimate the characteristic particle number $N^*$ at which the crossover to Bose 
selection occurs to be given by the depletion of the selected states, i.e.\ by the total number of particles in
non-selected states, 
\be\label{eq:Nchar}
N^* = \sum_{i\notin\mathcal{S}} \frac{1}{g_i -1} . 
\ee

The set of selected states is determined completely by the rate-asymmetry matrix $A_{ij}$. Namely this matrix determines 
not only the relative occupations among the selected states via Eqs.~(\ref{eq:selected_a}), but also the sign of the 
occupations (\ref{eq:nonselected_a}) of the non-selected states, which have to be positive. The latter can be seen by writing
Eq.~(\ref{eq:nonselected_a}) as
\begin{equation}\label{eq:nonselected_re}
 \bar{n}_i = \frac{1}{g_i-1} = -\frac{\sum_{j\in\BES}R_{ij}\bar{n}_j}{\sum_{j\in\BES}A_{ij}\bar{n}_j},
				\quad i\notin \BES .
\end{equation}
Here the numerator is always positive, since both the rates $R_{ij}$ and the occupations $\bar{n}_j$ are positive, and the 
sign of the denominator is determined by $A_{ij}$, since it depends on the relative occupations among the selected 
states, which are determined by $A_{ij}$ via Eqs.~(\ref{eq:selected_a}). The fact that the rate-asymmetry matrix $A_{ij}$, 
given by Eq.~(\ref{eq:Aper}) or by Eq.~(\ref{eq:Atwo}), does not depend on the bath temperature(s), implies that the 
set of selected states $\BES$ also does not depend on the bath temperature(s). However, the occupations
(\ref{eq:nonselected_a}) of the non-selected states are temperature dependent, as $R_{ij}$ appears on the right-hand 
side of Eq.~(\ref{eq:nonselected_re}). This implies that both the total number of particles in selected states
$N_S$ [Eq.~(\ref{eq:Ns})] as well as the characteristic particle number $N^*$ [Eq.~(\ref{eq:Nchar})] at which Bose 
selection sets in depend on the bath temperature(s). 

Finding the set of selected states $\mathcal{S}$ is generally a non-trivial problem. A brute-force algorithm would go 
through all possible sets containing an odd number of single-particle states, whose number grows exponentially with the 
number of modes $M$, until the desired set $\mathcal{S}$ is found. An efficient algorithm for finding $\mathcal{S}$ will 
be presented in subsection \ref{sec:algorithm} below. Already the question of how many states will be selected is not 
straightforward to answer, apart from the fact that (without fine tuning) it is always an odd number.

A special case is the scenario of having a single selected state $k$, corresponding to standard Bose condensation. Here, 
the occupations of the non-selected states (\ref{eq:nonselected_a}) reduce to the simple expression 
\be\label{eq:noncon}
\avgo{n_i} = \frac{1}{R_{ki}/R_{ik}-1} , \quad i\ne k.   
\ee
The fact that these occupations must be positive reveals that this scenario occurs when the state $k$ is
\emph{ground-state-like} in the sense that for all states $i$ the rate $R_{ki}$ from $i$ to $k$ is always larger than the backward rate $R_{ik}$,
\be\label{eq:gsl}
R_{ki}-R_{ik} = A_{ki} > 0 \quad \forall i \neq k.
\ee
The term ``ground-state-like'' refers to the situation of thermal equilibrium, where the relation
(\ref{eq:db}) implies that the condition (\ref{eq:gsl}) is fulfilled for $k$ being the ground state. These arguments 
reveal why we find a single selected state for the tight-binding chain which is driven between two heat baths of 
different positive temperature [Fig.~\ref{fig:BS}(b)]. In this situation the notion of the single-particle ground state 
still remains meaningful even away from equilibrium. This is generally different when the system is coupled to a 
population-inverted bath described by a negative temperature, like in Fig.~\ref{fig:BS}(c), or in a periodically 
driven system, like in Fig.~\ref{fig:BS}(d). In the former case the condition (\ref{eq:gsl}) cannot be expected to hold 
for $k$ being the ground state and in the latter case the ground state is not even defined (since quasienergies are 
determined modulo $\hbar\omega$ only). 

We can compare our theory to the theory of equilibrium Bose condensation as it was reviewed in subsection
\ref{sec:equilibrium}. First of all, we would like to note that the equilibrium situation is contained in our asymptotic 
mean-field theory as a special case. Namely, the equilibrium expression (\ref{eq:depletion}) for the excited-state 
occupations is reproduced, when the relation (\ref{eq:db}) is plugged into Eq.~(\ref{eq:noncon}). 
Generally, our Eq.~(\ref{eq:nonselected_a}) generalizes Eq.~(\ref{eq:depletion}); likewise Eqs.~(\ref{eq:Ns}) and
(\ref{eq:Nchar}) are generalizations of Eqs.~(\ref{eq:condensate2}) and (\ref{eq:NcharEq}), respectively. However, the 
fact that the relative occupations among the selected states and, even more, also the set $\mathcal{S}$ of selected 
states have to be determined adds an additional layer of complexity to the theory of non-equilibrium Bose selection.

\subsection{Systematic high-density expansion}
In this subsection we show that the asymptotic mean-field theory described in the previous subsection corresponds 
to the leading orders of a systematic expansion in the inverse total particle number $N^{-1}$. This implies that it 
correctly captures the mean-field result in the limit of large $N$. 

Let us expand the mean occupations as a series in powers of the inverse particle number $N^{-1}$
\be
\label{eq:expand_xmean}
  \avgo{n_i} =  N\nu_i + \nu_i^{(1)} +  N^{-1} \nu_i^{(2)} + N^{-2} \nu_i^{(3)} +\cdots
\ee
and require 
\be\label{eq:NuNorm}
\sum_i \nu_i = 1, \qquad \sum_i\nu_i^{(r)} = 0
\ee
for the leading order as well as for the corrections of order $r\ge1$. These requirements ensure that the mean total 
particle number is given by $N$, when the series is truncated after a certain order $r$. Such an expansion is 
equivalent to an expansion in the inverse particle density $n^{-1}=M/N$. We can now plug the ansatz (\ref{eq:expand_xmean}) 
into the mean-field Eqs.~(\ref{eq:mf_steady}), 
\begin{align}
\label{eq:asymptotic_mf}
0=&\nu_i\sum_jA_{ij} \nu_j \nonumber\\& 
+\frac{1}{N}\sum_j\Big[R_{ij}\nu_j-R_{ji}\nu_i+A_{ij}\Big(\nu_i\nu_j^{(1)}+\nu_i^{(1)}\nu_j\Big) \Big]
\nonumber\\&
+\frac{1}{N^2}\sum_j \Big[R_{ij}\nu_j^{(1)}-R_{ji}\nu_i^{(1)}
\nonumber\\&
\quad\quad+ A_{ij}\Big(\nu_i^{(2)}\nu_j+\nu_i^{(1)}\nu_j^{(1)}+\nu_i\nu_j^{(2)}\Big)\Big]
\nonumber\\&
+O\bigg(\frac{1}{N^3}\bigg), 
\end{align}
and ask that all terms that correspond to the same power of $N$ vanish independently. 
In this way we get a hierarchy of equations determining the coefficients of the expansion (\ref{eq:expand_xmean}) order 
by order.

Collecting the terms of the leading order gives rise to a set of equations for the leading coefficients
$\nu_i$. These equations take the form of Eqs.~(\ref{eq:naiv_app}), but with $\bar{n}_i$ replaced by $\nu_i$,
\be\label{eq:selected_orig}
0 =\nu_i\sum_j A_{ij}\nu_j.
\ee
Repeating the arguments of the previous section we see that the leading-order coefficient is non-zero only for a (yet to 
be determined) set of selected states $\BES$, so that
\be\label{eq:NSleading}
\nu_i = 0 ,\quad i\notin\BES, 
\ee
and 
\be
\label{eq:selected}
 0 = \sum_{j\in \mathcal{S}} A_{ij}  \nu_j, \quad i \in\mathcal{S}.
\ee

The next order determines the coefficients $\nu^{(1)}_i$. Thanks to Eq.~(\ref{eq:NSleading}) the coefficients of the
non-selected states are not coupled to each other and depend on the leading-order occupations of the selected states only, so that we arrive at the simple expression 
\be\label{eq:nonselected}
\nu^{(1)}_i = -\frac{\sum_{j\in\BES}R_{ij}\nu_j}{\sum_{j\in\BES}A_{ij}\nu_j},
				\quad i\notin \BES .
\ee
This expression directly corresponds to Eq.~(\ref{eq:nonselected_a}), but with $\bar{n}_{i}$ replaced by $\nu_i$ for the 
selected and by $\nu_i^{(1)}$ for the non-selected states. The leading corrections to the occupations of the selected 
states appear in the same order and can be determined by solving the linear equations  
\begin{align}
\sum_{j\in\BES}A_{ij} \nu_i\nu_j^{(1)}
=& \sum_{j\in\BES}\Big(R_{ji}\nu_i-R_{ij}\nu_j\Big)
\nonumber\\&
+\nu_i\sum_{j\notin\BES}\Big(R_{ji}-A_{ij}\nu_j^{(1)}\Big),
\quad i \in \BES,
\end{align}
where we used $0=\sum_{j\in\BES}A_{ij}\nu_j$ [Eqs.~\eqref{eq:selected_orig} and \eqref{eq:NSleading}], and taking Eq.~\eqref{eq:NuNorm} for $r=1$ into account.
Higher orders in the expansion \eqref{eq:asymptotic_mf} can become relevant when some rates vanish as discussed in
Sec.~\ref{sec:preasymptotic}.

Truncating the $1/N$ expansion after the first order, one obtains
\be\label{eq:FirstOrder}
\bar{n}_i\simeq \left\{\begin{array}{ll} 
	\nu_i N+\nu^{(1)}_i & \text{ for } 	i\in\BES \\
	\nu^{(1)}_i & \text{ for } i\notin\BES .
	\end{array}\right.
\ee
However, asymptotically in the limit of large $N$, it will be sufficient to take into account only the leading 
contributions, so that the mean occupations can be approximated as
\be\label{eq:asymp1}
\bar{n}_i\simeq \left\{\begin{array}{ll} 
	\nu_i N & \text{ for } 	i\in\BES \\
	\nu^{(1)}_i & \text{ for } i\notin\BES .
	\end{array}\right.
\ee
This corresponds to the approximation of the previous subsection, apart from the slight difference that, previously,
we normalized the total occupation of the selected states $N_S$ to the first-order result
$N_s^{(1)}=\sum_{i\in\BES}\big[\nu_iN+\nu_i^{(1)}\big] = N-\sum_{i\ne\BES}\nu_i^{(1)}$. This is implicit in Eq.~(\ref{eq:Ns}) and corresponds to the approximation
\be\label{eq:asymp2}
\bar{n}_i\simeq \left\{\begin{array}{ll} 
	\nu_i N_s^{(1)} & \text{ for } 	i\in\BES \\
	\nu^{(1)}_i & \text{ for } i\notin\BES ,
	\end{array}\right.
\ee
This normalization, which for finite $N$ takes care of the fact that the leading 
contributions to the occupations of the selected and the non-selected states stem from different orders, is thus a compromise 
between Eq.~(\ref{eq:FirstOrder}) and Eq.~(\ref{eq:asymp1}). For large but finite $N$ it is better than
Eq.~(\ref{eq:asymp1}), since it produces the correct total particle number, but it does not require to compute 
corrections $\nu_i^{(1)}$ for the selected states that enter Eq.~(\ref{eq:FirstOrder}). Therefore, we will use
Eq.~(\ref{eq:asymp2}), corresponding to the asymptotic theory as it was presented in the previous subsection, in the 
following. In the asymptotic limit $N\to\infty$ all three expressions (\ref{eq:FirstOrder}), (\ref{eq:asymp1}), and
(\ref{eq:asymp2}) are, of course, equivalent.

The requirement of having a positive particle number in the asymptotic limit of large $N$ is given by
\be\label{eq:positive}
 \left\{\begin{array}{ll} 
	\nu_i >0 & \text{ for } 	i\in\BES \\
	\nu^{(1)}_i > 0 & \text{ for } i\notin\BES .
	\end{array}\right.
\ee
In order to find a compact formulation of finding an asymptotic solution obeying this condition it 
is convenient to introduce the numbers $\mu_i=\sum_j A_{ij}\nu_j$. According to Eq.~(\ref{eq:selected}) they vanish for
$i\in\BES$, while Eq.~(\ref{eq:nonselected}) tells us that they should be negative to ensure positive occupations 
of the non-selected states. 
The problem of finding an asymptotic mean-field solution can, therefore, be reduced to the problem 
of finding a set $\BES$ of selected states and numbers $\nu_i$ and $\mu_i$ such that \cite{VorbergEtAl13}
\begin{equation}
\mu_i=\sum_j A_{ij}\nu_j
\text{ with } \begin{cases}
 \nu_i>0 \text{ and } \mu_i=0 \mbox{ for } i\in \BES,
\\  \nu_i=0 \text{ and } \mu_i<0 \mbox{ for } i\notin \BES.
\end{cases}
\label{eq:allconditions}
\end{equation}
The non-generic situation with $\nu_i=\mu_i=0$ for some $i$ corresponds to transitions, which we discuss in the next 
subsection. Before we prove that a unique set $\mathcal{S}$ obeying the relations (\ref{eq:allconditions}) exists, 
let us point out that these relations are valid only in the case of fully connected rate 
matrices. If we allow for zero rates $R_{ij}=0$, the set of selected states is not determined by the conditions
(\ref{eq:allconditions}) anymore, as we discuss in subsection \ref{sec:ZeroRates} below.

It is interesting to note that the conditions (\ref{eq:allconditions}) that determine the selected states are equivalent to those determining the surviving species under the dynamics of the Lotka-Volterra equations given in footnote \ref{fn:lotka_volterra} \cite{KnebelEtAl13,KnebelEtAl15}.
Differences appear for non fully connected rate matrices (see discussion at the end of Sec.~\ref{sec:ZeroRates}).

\subsection{Existence and uniqueness of the set of selected states}
\label{sec:proof}
In this subsection we provide a proof for the uniqueness and the existence of the set of selected states for fully connected rate matrices (which we repeat for completeness from the supplemental material of Ref.~\cite{VorbergEtAl13}.) In the following we will use the vector and matrix notation, with $\bm\nu$ and $\bm\mu$ denoting the vectors with
elements $\nu_i$ and $\mu_i$, respectively, and $R$ and $A$ denoting the rate matrix and the rate-asymmetry matrix with 
elements $R_{ij}$ and $A_{ij}$, respectively. Let us, furthermore, decompose $A$ like 
\begin{equation}
A=
\left(
\begin{array}{c|c}
 A^{\BES} & A^{\BES\bar\BES}  \\
\hline
 A^{\bar\BES\BES} & A^{\bar\BES} \\
\end{array}
 \right)
\label{eq:decompose_A}
\end{equation}
wherein the submatrix $A^{\BES}=\{A_{ij}\}_{i,j\in \BES}$ denotes the rate-asymmetries among selected states, 
$A^{\bar\BES\BES}=-(A^{\BES\bar\BES})^T=\{A_{ij}\}_{i\notin \BES, j\in \BES}$ the rate-asymmetries among non-selected and 
selected states, and $A^{\bar\BES}=\{A_{ij}\}_{i,j\notin \BES}$ the rate-asymmetries among non-selected states.
The conditions \eqref{eq:allconditions} with $ i\in \BES$ require us to determine $\BES$ such that $A^{\BES}$ has a 
vanishing eigenvalue. Skew-symmetric matrices generically have a vanishing eigenvalue only if their dimension is odd. As 
the square submatrix $A^{\BES}$ of $A$ is still skew-symmetric, we can immediately conclude that the number $M_S$ of Bose 
selected states is odd. 
The conditions \eqref{eq:allconditions} stipulate, furthermore, that the corresponding eigenvector $\nu_i$, $i\in \BES$  
has positive components. Finally, the conditions for $ i\notin \BES$  tell us that this eigenvector should result in 
a vector with non-positive components when it is multiplied with the submatrix $A^{\bar\BES\BES}$. 

We now prove the uniqueness of the set $\BES$. Assume first that there exist two different sets $\BES_1$ and $\BES_2$, both leading to physical solutions $\bm{\nu}_1$ 
and $\bm{\nu}_2$ with $\bm{\mu_1}=A\bm{\nu_1}$ and $\bm{\mu_2}=A\bm{\nu_2}$ obeying Eq.~\eqref{eq:allconditions}. Using
\begin{align}
\bm{\nu}_2^T\bm{\mu}_1&=\bm{\nu}_2^TA\bm{\nu}_1
=(\bm{\nu}_2^TA\bm{\nu}_1)^T=\bm{\nu}_1^TA^T \bm{\nu}_2
=
-\bm{\nu}_1^TA\bm{\nu}_2
\nonumber\\& =- \bm{\nu}_1^T\bm{\mu}_2,
\end{align}
it then follows from Eq.~\eqref{eq:allconditions} that
\begin{equation}
0\ge \bm{\nu}_2^T\bm{\mu}_1 = - \bm{\nu}_1^T\bm{\mu}_2\ge0.
\end{equation}
This requires that both $\bm{\nu}_2^T\bm{\mu}_1 =0$ and $\bm{\nu}_1^T\bm{\mu}_2=0$, such that
$\BES_2\subset \BES_1$ and $\BES_1\subset \BES_2$, leading us to conclude that $\BES_1=\BES_2\equiv \BES$. Given the set 
$\BES$, the homogeneous linear system for $\bm{\nu}$ generically has a single solution only. Therefore, the solution to 
the generic steady-state problem has to be unique.

In order to prove the existence of the set $\BES$, we now restrict $\BES$ to sets comprising an odd number $M_S$ of states, according to the generic conditions described above.
Each choice of $\BES$ gives rise to a (possibly non-physical) solution $\bm{\nu}_{\BES}$ with $\bm{\mu}_{\BES}=A \bm{\nu}_{\BES}$. The vector of signs $\bm{\sigma}$ with 
\begin{equation}
\begin{cases}
\sigma_i=\text{sign}({\nu}_i) \mbox{ if } i\in \BES,\\
\sigma_i=-\text{sign}({\mu}_i) \mbox{ if } i\notin \BES,
\end{cases}
\end{equation}
distinguishes physical solutions ($\sigma_i=1$ for all $i$) from non-physical solutions. Here, we fix an overall sign due to the orientation of the vector $\bm{\nu}_{\BES}$ by the convention $\sigma_1=1$.
Now we observe: (i) Cycling through all odd-numbered subsets $\BES$, each possible vector $\bm{\sigma}$ occurs at most once. Namely, if $\BES_1$ and $\BES_2$ gave rise to the
same vector $\bm{\sigma}$ then the modified rate imbalance matrix $\tilde{A}_{ij}=\sigma_iA_{ij}\sigma_j$ had two physical solutions with different selected sets $\BES_1$ and $\BES_2$, in contradiction to the previously established uniqueness of the solutions. (ii) The number $2^{M-1}$ of possible
vectors $\bm{\sigma}$ equals the number $\sum_{M_S=1,3,\ldots} {M\choose M_S}=2^{M-1}$ of possible sets $\BES$.
Therefore, each vector $\bm{\sigma}$ occurs once. In particular, this includes the vector with $\sigma_i=1$ for all $i$, 
leading to the solution with positive macroscopic and microscopic occupations. This guarantees the existence of a 
physical solution.

\subsection{Transitions}
\label{sec:transition}

\begin{figure}[t]
\centering
\includegraphics[width=1\linewidth]{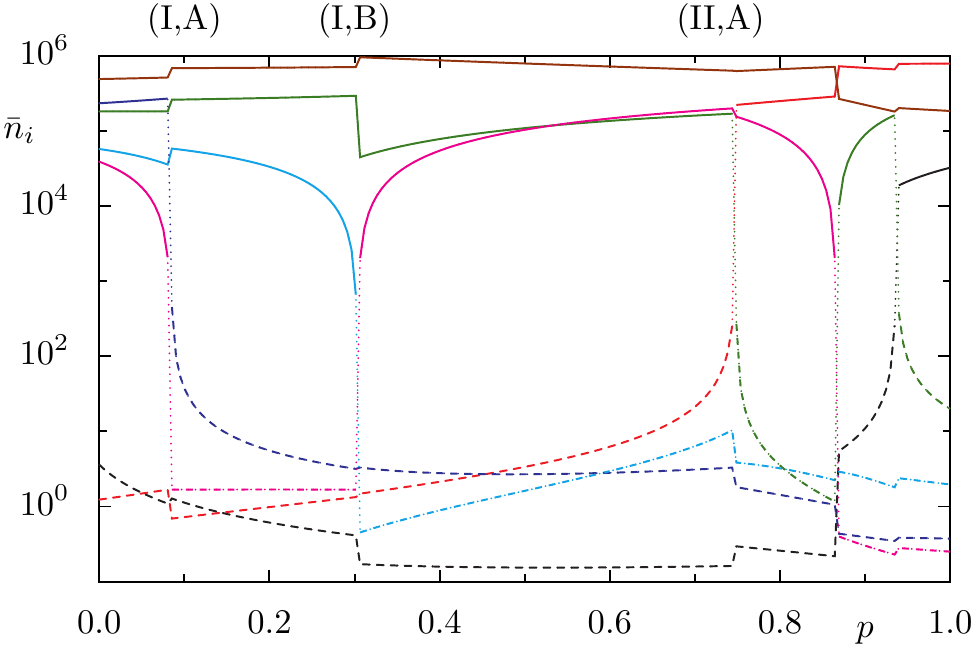}
\caption{(color online) Mean occupations in response to the variation of a dimensionless parameter $p$, for a small system of $N=10^6$ 
bosons on $M=7$ states for the random-rate model. The rate matrix $R(p)$ is a superpostion of two independently drawn rate 
matrices $R^{(1)}$ and $R^{(2)}$, with the relative weight controlled by $p$, $R(p)=(1-p)R^{(1)}+pR^{(2)}$. The results 
are obtained using mean-field theory (dotted lines), asymptotic theory (solid lines for selected states and dashed lines 
for non-selected states). Each color refers to a specific state. At each transition two states are exchanged between the 
sets of selected and non selected states.}
\label{fig:transition_example}
\end{figure}

In this section we will discuss transitions, where the set of selected states $\BES$ changes in response to the variation 
of a parameter $p$. Examples for such transitions can be observed in Fig.~\ref{fig:transition_example}. This figure shows 
the mean occupations versus the parameter $p$ for a model defined by the superposition of two random rate matrices
$R^{(1)}$ and $R^{(2)}$, with the relative weight controlled by $p$, $R(p)=(1-p)R^{(1)}+pR^{(2)}$. One can see that in a 
transition two states are exchanged between the set of selected states and the set of non-selected states, such 
that the number of selected states is odd before and after the transition. Approaching a transition from the left, the 
transition is found to be triggered by a state $i^<$. This state $i^<$ can either be a selected state whose occupation 
drops until it becomes non-selected at the transition (case I) or a non-selected state whose occupation increases until 
it becomes selected at the transition (case II). Furthermore, one can observe that at the transition a second state $i^>$ 
becomes involved abruptly that changes from the selected to non-selected (case A) or vice versa (case B). When 
approaching the transition from the right, the states $i^<$ and $i^>$ change their role, so that the former partner state 
$i^>$ plays the role of the triggering state. 

\begin{figure}[t]
\centering
\includegraphics[width=1\linewidth]{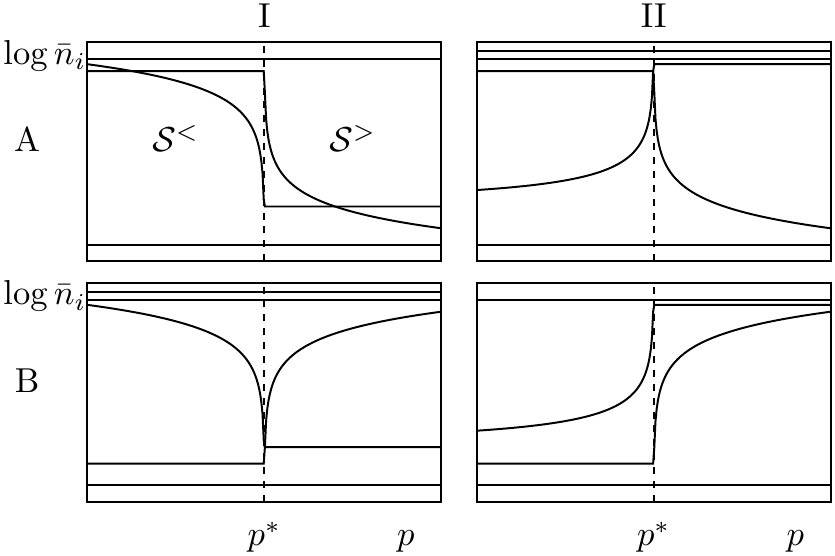}
\caption{(color online) Four generic types of transitions, where the set of selected states changes from $\BES=\BES^<$ to $\BES=\BES^>$ 
when a parameter $p$ reaches a critical value $p^*$. In each transition two states are exchanged between the sets of 
selected and non-selected states, so that the number $M_\BES$ of selected states remains odd. When approaching the 
transition from the left, it is triggered by a state $i^<$, either a selected state whose occupation drops until it 
becomes non-selected at the transition (case I) or a non-selected state whose occupation increases until it becomes 
selected at the transition (case II). A second state $i^>$ becomes involved abruptly at the transition that changes from 
selected to non-selected (case A) or vice versa (case B). This state plays the role of the triggering state when the 
transition is approached from the right. Types (I,A) and (II,B) form one class, since they transform into each other when the transition is passed in opposite direction.}
\label{fig:transition}
\end{figure}

The four combinations of cases I or II and A or B define four generic types of transitions that are depicted in
Fig.~\ref{fig:transition}. Type (I,A) and type (II,B), where the number $M_\BES$ of selected states is lowered or raised 
by two, respectively, transform into each other when the transition is passed in opposite direction. Therefore, they form 
one class. In type (II,A) transitions, which are triggered by non-selected states from both side, and type (I,B) 
transitions, which are triggered from selected states from both sides, the number $M_\BES$ of selected states does not 
change. They define two distinct classes, since they cannot be transformed into each other.

These observations based on Fig.~\ref{fig:transition_example} turn out to be generic. In the following we will describe 
them within the asymptotic mean-field theory. We have already defined the left 
triggering state $i^<$ and its partner state, the right triggering state $i^>$. Let, moreover, $p^*$ be the critical 
parameter at which the transition occurs and $\BES^<$ and $\BES^>$ be the sets of selected state on the left-hand and the 
right-hand side of the transition, respectively [Fig.~\ref{fig:transition}]. Within the asymptotic theory a transition 
must occur when the  occupation $\bar{n}_i$ of a state $i$ would change its sign at a critical parameter $p=p^*$. This 
state $i$ plays the role of the triggering state $i^<$. If $i^<$ is a selected state (before the transition), the 
transition occurs when $\nu_{i^<}$ drops to zero, so that in zeroth order the occupation of this state becomes zero. 
In that case the state $i^<$ can, thus, be viewed as a non-selected state at the transition. If $i^<$ is a non-selected 
state, the transition occurs when $\mu_{i^<}$ becomes zero, so that in first order the occupation of this state diverges. 
In that case the state $i^<$ must, therefore, be viewed as a selected state at the transition. Thus, at the transition
$p=p^*$, which corresponds to a fine-tuned situation, the set of selected states contains an \emph{even} number of states 
and is given by 
\begin{equation}
\BES^*=\begin{cases} \BES^< \cup \{i^<\} &\mbox{ if } i^<\notin \BES^<\\ \BES^<\setminus \{i^<\} &\mbox{ if } i^<\in 
\BES^<.\\
\end{cases}
\end{equation}
As the number of Bose-selected states has to become odd after the transition, one further state $i^{>}$ has to 
be involved. The set $\BES^*$ can also be expressed in terms of this partner state,
\begin{equation}
\BES^*=\begin{cases} \BES^> \cup \{ i^>\} &\mbox{ if }  i^>\notin \BES^>\\ \BES^>\setminus \{i^>\} &\mbox{ if }  i^>\in 
\BES^>.
\end{cases}
\end{equation}
In the following we will describe how to determine this partner state in order to find the set $\BES^>$ of selected states
on the other side of the transition. 

The intricate details of the transition are encoded in the truncated matrix  $A^{\BES^*}$, obtained from $A^*=A(p^*)$ by 
removing all rows and columns corresponding to non-selected states $i\notin \BES^*$ like in \Eq{eq:decompose_A}. 
According to the transition criteria, this matrix has at least one vanishing eigenvalue. As the matrix is
even-dimensional and skew-symmetric, its eigenvalues are imaginary and come in pairs of opposite sign. Thus, one 
eigenvalue of zero implies another one, so that generically the kernel of $A^{\BES^*}$ will be two-dimensional at the 
transition. One vector lying in the kernel of  $A^{S^*}$ is given by the limiting occupations $\nu_i$ as one approaches
$p^*$ from below. We denote this vector by $\bm{\nu}^<$ (note that this is now truncated to the states of $\BES^*$). 
Analogously, there is a second vector $\bm{\nu}^>$ from the limiting occupations as one approaches $p^*$ from the right, 
which also lies in the kernel. We will now establish a relation between both vectors $\bm{\nu}^<$ and $\bm{\nu}^>$. 

For that purpose, we introduce an interpolating vector $\bm{\nu}(a)=a\bm{\nu}^<+(1-a)\bm{\nu}'$, where $\bm{\nu}'$ is the
element of the kernel of $A^*$ which is orthogonal to $\bm{\nu}^<$ while $a$ is an interpolation parameter. 
The occupations of the non-selected states (and their sign) is determined by the vector $\bm{\mu}$ given by
Eq.~\eqref{eq:allconditions}. For the two possible solutions $\bm{\nu}^<$ and $\bm{\nu}^>$, this vector reads  $\bm{\mu}^<=A^{\bar\BES^*\BES^*}\bm{\nu}^<$ and $\bm{\mu}^>=A^{\bar\BES^*\BES^*}\bm{\nu}^>$, respectively.
Both vectors are connected by the interpolation $\bm{\mu}(a)=a\bm{\mu}^<+(1-a)\bm{\mu'}$ with
$\bm{\mu}'=A^{\bar\BES^*\BES^*}\bm{\nu}'$. Herein $A^{\bar\BES^*\BES^*}$ is obtained from $A^*=A(p^*)$ as described by
\Eq{eq:decompose_A}. Due to the selection criterion \Eq{eq:allconditions}, we require physical solutions
${\nu}_i(a)\ge 0\ \forall i\in \BES^*$ and $\mu_i(a)\le0\ \forall i\notin \BES^*$. Choosing the orientation of $\bm{\nu}'$
conveniently, this is fulfilled for the finite interval $0<a<a^>$. The extremal point $a^>$ is determined by ramping up $a$
until either an element of $\bm{\nu}(a)$ or $\bm{\mu}(a)$ becomes zero. The index of this element corresponds to the 
state $i^>$ and the extremal point $a^>$ determines the solution $\bm{\nu}^>$, via $\bm{\nu}(a^>)=\bm{\nu}^>$. 

Exactly at the transition, the interval $0<a<a^>$ corresponds to physically meaningful solutions with positive occupation 
numbers. Its extremal points describe the solutions $\bm{\nu}^<$ and $\bm{\nu}^>$ found when approaching the transition 
from the left and right hand side, respectively. The narrower the interval, i.e.\ the smaller $\Delta a = a^>$, the more 
similar will both solutions $\bm\nu^<$ and $\bm\nu^>$ be. That means the smaller will be the discontinuous changes in the 
occupations of the states $i\notin\{i^<,i^>\}$ that are not directly involved in the transition, as they are visible also 
in Fig.~\ref{fig:transition_example}. The width $\Delta a$ associated with a typical transition must, moreover, be 
expected to shrink with the system size. Namely, each of the $M$ single-particle states of the system provides a 
constraint that potentially limits this interval, since the number of conditions Eq.~\eqref{eq:allconditions} 
proliferates with $M$. So in large systems one cannot only expect more transitions to occur when a parameter is varied, 
but also that the discontinuous jumps, which the non-participating occupations undergo at each transition, become smaller. 

\begin{figure*}[t]
\centering
\includegraphics[width=1\linewidth]{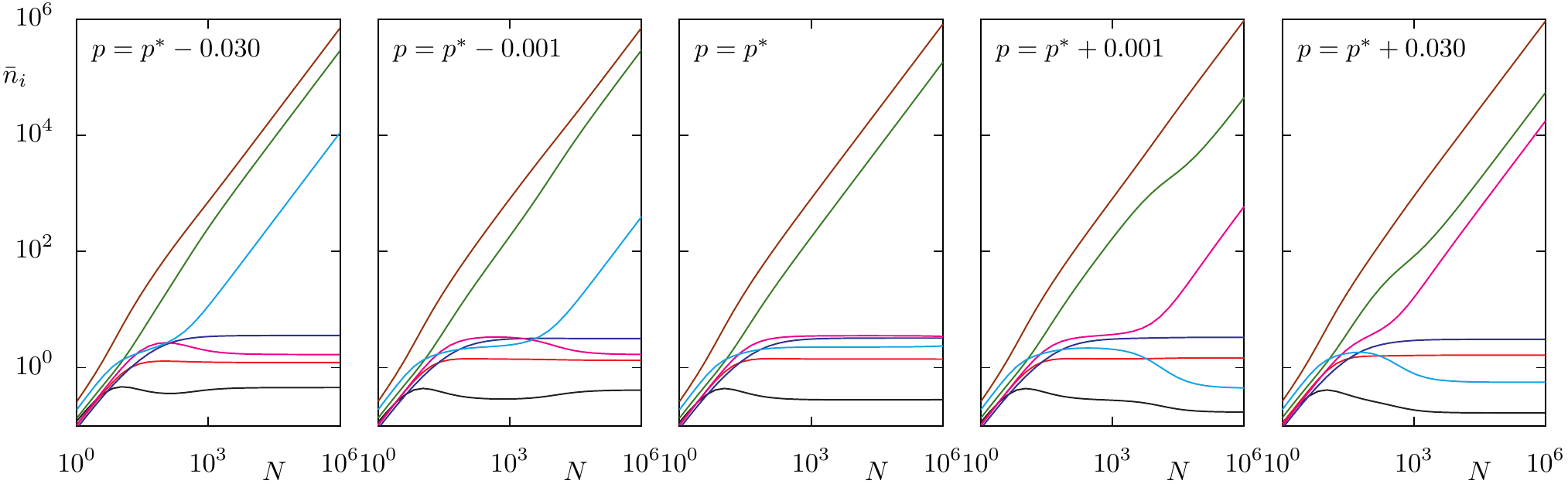}
\caption{(color online) Occupations numbers versus total particle number $N$ close to a transition. The system is described by the same 
rate matrix $R(p)$ as that of Fig.~\ref{fig:transition_example}. The parameters $p$ used in the different panels are chosen to be 
close (or at) the transition labeled (I,B) in Fig.~\ref{fig:transition_example}, with $p^*=0.303179$ denoting the corresponding critical parameter.}
\label{fig:TranFinN}
\end{figure*}

Before moving on, let us briefly discuss the case of finite particle numbers $N$, where the sharp transition becomes a 
crossover of finite width. This can be observed in Fig.~\ref{fig:TranFinN}. Here we plot the mean occupations versus the 
total particle number $N$ for a system described by the same rate matrix $R(p)$ used in Fig.~\ref{fig:transition_example}. The 
five panels of Fig.~\ref{fig:TranFinN} are obtained for parameters $p$ close to (or at) the transition labeled (I,B) in
Fig.~\ref{fig:transition_example}, with the critical parameter denoted by $p^*$. The first panel corresponds to a parameter well 
on the left-hand side of the transition. Here asymptotically three states become selected. When coming closer to the 
transition, but still staying on its left-hand side (second panel), we can observe that a preasymptotic regime appears. Namely, at 
large, but finite $N$ the system approaches a state with two selected 
states, before eventually in the asymptotic limit $N\to\infty$, a third state becomes selected as well. This third state 
corresponds to the triggering state $i^<$. The two states that appear to be selected in this preasymptotic regime 
correspond to those two states that are selected at the transition (middle panel). The fourth panel corresponds to a 
parameter, where the transition has just been passed. Here (roughly) the same preasymptotic state is found, before 
asymptotically for $N\to\infty$ a third state joins the group of selected states. Now the third state is given 
by $i^>$. The fifth panel is, finally, obtained for a parameter well on the right-hand side of the transition. Here again
no preasymptotic regime is found. The emergence of a preasymptotic regime close 
to the transition implies that the fine-tuned rate matrix $R(p^*)$, which gives rise to two selected states, provides an 
accurate description of the system within a finite interval of parameters near the transition.

\subsection{Efficient algorithm for finding the selected states}
\label{sec:algorithm}

In principle, finding the unique set $\BES$ of Bose-selected states requires to sample all possible subsets, whose number 
grows exponentially with $M$, until one succeeds to satisfy the conditions \eqref{eq:allconditions}. Testing all sets by 
brute force quickly becomes unpractical already for moderately large values of $M$.  While the mean-field occupations and 
especially their dependence on the total particle number can provide some guidance, this method also quickly reaches its 
limits when $M$ is further increased. Here we describe an efficient algorithm for finding the set of selected states. It 
uses the theory of transitions that we presented in the previous subsection. 

In order to solve the problem of finding the set of selected states for a given rate-imbalance matrix $A$, we construct the auxiliary rate-imbalance matrix 
\be\label{eq:Atilde}
\tilde{A}_{ij}(p) = A_{ij} +  p B_{ij}
\ee
by adding the real-valued skew-symmetric matrix $B$, weighted with the real parameter $p$, to the original one. The 
problem defined by the new matrix $\tilde{A}(p)$ will be solved by a set $\tilde{\BES}(p)$ of selected states. The  
matrix $B$ is constructed as follows: It shall possess a cross-like structure, with non-zero elements only in the column 
and the line labeled by $k$, 
\be\label{eq:B}
B_{ij} =  \delta_{ik}b_j - \delta_{kj} b_{i}  ,
\ee
so that 
\be\label{eq:AtildeCond}
\tilde{A}_{kj}(1) = A_{kj} + b_j > 0,\qquad\forall j\neq k.
\ee
This condition, which corresponds to the relation (\ref{eq:gsl}), ensures that for $p=1$ only the state $k$ will be 
selected, $\tilde{\BES}(1)=\{k\}$. Relation (\ref{eq:AtildeCond}) can be achieved with minimal effort by setting
\be\label{eq:bj}
b_i = \left\{\begin{array}{ll} 
	+|A_{ki}| + \epsilon_i  >0 & \text{ if } A_{ki}\le 0 \\
	0            & \text{ otherwise } 
	\end{array}\right.
\ee
with arbitrary $\varepsilon_i>0$. Our strategy will now consist in ramping the parameter $p$ down from $p=1$, where the 
solution $\tilde{\BES}(1)=\{k\}$ is known by construction, to $p=0$, where we would like to know the solution
$\tilde{\BES}(0)=\BES$. During this ramp, we will monitor all transitions, i.e.\ changes of the set $\tilde{\BES}(p)$,
that are happening, so that at the end we will arrive at the desired solution. For that purpose it seems favorable
(though not necessarily required) to choose the state $k$ such that a minimum of the elements $b_{j}$ defined like
(\ref{eq:bj}) has to be non-zero, and to choose the $\epsilon_i$ different from each other, 
$\epsilon_i\ne\epsilon_j$ for $i\ne j$, in order to separate the transitions when varying $p$. 

In order to follow the state of the system during the parameter ramp, we take advantage of the specific way the matrix
$\tilde{A}(p)$ depends on the parameter $p$. Namely, the cross structure (\ref{eq:B}) of the matrix $B$ implies that 
the occupations of the system change in a linear fashion unless a transition occurs: If the vector $\tilde{\nu}(p_0)$ solves 
the problem (\ref{eq:allconditions}) for $\tilde{A}(p_0)$, then one has 
\be\label{eq:lin}
\tilde{\bm \nu}(p) = C(p)\Big[ \tilde{\bm\nu}(p_0) + \tilde{\bm\nu}'(p_0) (p-p_0)\Big] \; \text{ for }\; p_a<p<p_b,
\ee
with a global normalization factor $C(p)>0$ such that $\sum_{j\in\tilde{\BES}(p)}\tilde{\bm \nu}(p)=1$. Here the limits
$p_a$ and $p_b$ are given by those values of $p$, where the set of selected state changes away from $\tilde{\BES}(p_0)$ 
in a transition. The proof of this statement is rather technical and delegated to appendix \ref{sec:app_algorithm}, where 
we also describe how to obtain $\tilde{\bm\nu}'(p_0)$. Expression (\ref{eq:lin}) can be employed to predict the positions 
$p_a$ and $p_b$ of the transitions as those points, where either an element $\tilde{\nu}_i(p)$ of $\tilde{\bm\nu}(p)$ or 
an element $\tilde{\mu}_i(p)$ of the associated vector $\tilde{\bm\mu}(p)=\tilde{A}(p)\tilde{\bm\nu}(p)$ would change 
sign. The label $i$ of this state corresponds to the state that triggers the transition. 

With these ingredients, our algorithm works as follows: Start from $p=1$, where $\tilde{\BES}(1)=\{k\}$, and evaluate 
where the next transition occurs when $p$ is lowered and by which state $i^>$ it will be triggered. Next, employ 
the theory of transitions described in the previous subsection to determine the partner state $i^<$, which at the 
transition also changes between the sets of selected and non-selected states. In this way the new set of selected 
states solving $\tilde{A}(p)$ after the transition has been found. Then compute where the next transition occurs
when $p$ is lowered further, iterating this procedure until $p=0$ is reached. The time needed to find the set of selected 
states in this way scales polynomial with the system size $M$. For the random-rate model, which constitutes a rather 
difficult problem since on average half of the states are selected \cite{VorbergEtAl13}, we find this time to scale as $\sim M^\alpha$ with $\alpha\approx 4$.
This allows us to find the set of selected states for systems of up to $M=1000$ states. 
An alternative algorithm for solving the problem
(\ref{eq:allconditions}) has recently been presented in Ref.~\cite{KnebelEtAl15} and is based on linear programming.

\subsection{Small rates and preasymptotic regime}
\label{sec:preasymptotic}

\begin{figure}[t]
\centering
\includegraphics[width=1\linewidth]{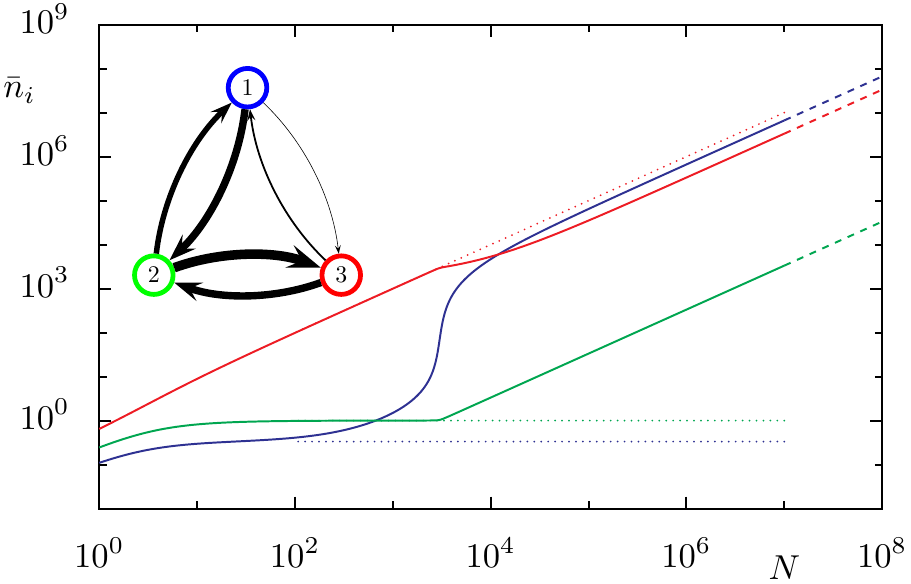}
\caption{(color online) Effect of small rates for a minimal three-state model with rate matrix (\ref{eq:ThreeState}).
Occupations $\bar{n}_i$ versus total particle number $N$ obtained using mean-field theory
(solid lines) and asymptotic theory (dashed lines) for the rate matrix $R$ given by Eq.~\eqref{eq:ThreeState}, which is visualized in the inset (line widths reflect rates). 
Furthermore, the dotted lines show occupations $\bar{n}_i$ obtained
 by the asymptotic theory for the approximate rate matrix $R^a$ 
given by Eq.~(\ref{eq:ThreeStateApprox}), where the small rates have been neglected. Blue, green, and red 
lines describe $\bar{n}_1$, $\bar{n}_2$, 
and $\bar{n}_3$, respectively. 
Near $N\sim 10$ the system approaches a preasymptotic state with a single selected state, 
described by $R^a$, before above $N\sim 10^3$ the true asymptotic state is reached, where all states are 
selected.}
\label{fig:small_rates}
\end{figure}

So far we have assumed strictly positive rates, $R_{ij}>0$, within the asymptotic theory. This assumption is reasonable 
in the sense that exactly vanishing rates, $R_{ij}=0$, can be viewed as a fine-tuned situation. However, obviously, we can 
encounter situations where some rates are much smaller than others, e.g.\
\be
R_{ij} = \left\{\begin{array}{ll}
		O(r) & \text{ for }(i,j)\in G\\
		O(\epsilon r) & \text{ else},
		\end{array}\right.
\ee
with $G$ denoting the subset of pairs $(i,j)$ with large rates of order $r$ and $\epsilon\ll1$ quantifying the 
suppression of small rates of order $\epsilon r$. Such rate matrices can result from a situation where some modes are 
coupled much more weakly to the environment than others. Having such a situation in mind, in the following discussion we will 
consider a rate $R_{ij}$ to be small only when also its backward rate $R_{ji}$ is small too, so that also the corresponding
rate-asymmetry $|A_{ij}|$ is small.\footnote{There can also be small rates without small backward rates, e.g.\ 
between states with a large energy separation. Not considering those rates as small in the below analysis (i.e.\ not 
exploiting the fact that they are small) does not spoil its validity.} 

Having some rates much smaller than others, it appears reasonable to neglect the small rates in an approximation,
\be\label{eq:Rapprox}
R_{ij}\approx R^a_{ij} 
	= \left\{\begin{array}{ll}
		R_{ij} & \text{ for }(i,j)\in G\\
		0 & \text{ else}.
		\end{array}\right.
\ee
As we will argue below, such an approximation will describe the system accurately, provided the total particle number $N$ 
remains below a threshold $N_\text{thr}$ associated with the approximation. Thus, when increasing the particle number $N$,
one might encounter the following scenario: First a preasymptotic state is approached, where the occupations are 
well described by the asymptotic theory based on the approximate rate matrix $R^a_{ij}$, before eventually 
the true asymptotic state of the full rate matrix $R$ is reached above the threshold. This scenario can be observed in 
Fig.~\ref{fig:small_rates}, where we plot the occupations of a minimal three-state model versus $N$. In this model the 
rates are given by 
\begin{align}\label{eq:ThreeState}
&R = r\left(\begin{array}{ccc} 
		0 		& 1 & 2\epsilon\\
		2 		& 0 & 2\\
		1\epsilon 	& 4 & 0
	\end{array}\right), \quad \epsilon = 10^{-3},
\\\label{eq:ThreeStateApprox}
&R^a 
	= r \left(\begin{array}{ccc} 
		0  & 1 & 0\\
		2  & 0 & 2\\
		0  & 4 & 0
	\end{array}\right).
\end{align}

This behavior resembles the preasymptotic behavior found near transitions that we discussed at the end of subsection
\ref{sec:transition}. In both cases the preasymptotic state is described by a fine-tuned rate matrix, either 
characterized by the critical parameter or by setting several matrix 
element to zero. However, since setting several matrix elements to zero corresponds to the fine tuning of several 
parameters, the set of selected states of $R^a$ can be quite different from that of $R$.

The appearance of a preasymptotic regime described by the approximate rate matrix (\ref{eq:ThreeStateApprox}) at intermediate 
particle numbers $N$, as it is visible in Fig.~\ref{fig:small_rates} roughly for $10<N<10^3$, can be explained as 
follows. When applying the asymptotic theory, Sec.~\ref{sec:Asymptotic}, to the approximate rate matrix, where small rates are neglected, we find 
that the selected state 3 acquires an occupation $\sim N$, while the occupations of the non-selected states are $\sim 1$. 
Generally, the fact that the selected state(s) possesses an occupation much larger than the non-selected states justifies 
the $1/N$ expansion (\ref{eq:expand_xmean}), which underlies the asymptotic theory. This explains why the preasymptotic 
regime is reached near $N\sim 10$, when $N\gg1$. However, as soon as the factor $N$ between the occupations of the \
selected and the non-selected states becomes comparable to the inverse suppression factor $\epsilon^{-1}\sim 10^{3}$, the 
weak rates start to spoil the hierarchy of the $1/N$ expansion based on the selected state of $R^a$. Namely the product 
of a small rate with the occupation of a selected state $\sim r\epsilon N$, which was neglected so far, can become 
comparable to the product of a large rate with the occupation of a non-selected state $\sim r$, which has been taken into 
account. This explains why for $N>N_\text{thr}\sim\epsilon^{-1}=10^3$ the system starts to deviate from the solution of 
the approximate rate matrix (describing the preasymptotic state), to approach the true asymptotic state determined by the 
full rate matrix. 

Note that allowing for zero rates, i.e.\ rate matrices that are not fully connected like $R^a$, can have 
several consequences for the asymptotic theory. These are discussed in the following subsection.

\subsection{Zero rates: Not fully connected rate matrices}
\label{sec:ZeroRates}
So far we have assumed fully connected rate matrices within our asymptotic theory. What happens if we allow some rates to 
become zero? This question emerges, e.g., when computing the asymptotic state of an approximate rate matrix $R^{a}$ [Eq.~(\ref{eq:Rapprox})].
First of all, in case the rate matrix is disconnected, so that it is not possible anymore to reach every state $i$ from 
every other state $j$ in a sequence of quantum jumps (and vice versa), then the steady state of the system is 
not unique anymore \cite{Schnakenberg76} and will depend on the initial conditions.\footnote{If, by taking into account 
neglected rates of order $\epsilon r$, the matrix is connected again, then for times longer than $1/(\epsilon r)$ the
non-unique steady states associated with $R^a$ will eventually relax to the unique steady state of the full rate matrix.} 
We will exclude this scenario from the following discussion and focus on situations where the rate matrix is solved by a 
unique steady state. 

In order to discuss the impact of zero rates, let us briefly recapitulate the situation where all states are coupled to 
all other states. In this case the coefficients $\nu_i$ and $\nu^{(r)}_i$ of the $1/N$ expansion (\ref{eq:expand_xmean}) 
are obtained as follows. First the leading coefficients ${\nu_i}$, and with that the set $\BES$ of selected states, have 
to be determined by solving the problem (\ref{eq:allconditions}). Then the sub-leading coefficients $\nu^{(r)}_i$ can be 
obtained iteratively from the hierarchy of equations that results from Eq.~(\ref{eq:asymptotic_mf}) by requiring the 
terms of each power of $N$ to vanish separately. If we denote the terms $\propto N^{-r}$ on the right-hand side of
Eq.~(\ref{eq:asymptotic_mf}) by $I^{(r)}_i$, then this hierarchy of equations reads 
\be\label{eq:hierarchy}
I_i^{(r)}({\bm\nu},{\bm\nu}^{(1)},\ldots,{\bm\nu}^{(r)}) = 0,
\ee
for all $i$ and for $r=0,1,2,\ldots$ and with ${\bm\nu}^{(r)}$ denoting the vector of coefficients ${\nu}^{(r)}_i$. Now the
$\nu^{(1)}_i$ are obtained by solving the set of linear equations $I_i^{(1)}({\bm\nu},{\bm\nu}^{(1)})$, with the 
already determined $\nu_i$ treated as parameters. Then the $\nu^{(2)}_i$ are obtained from the set of linear equations
$I_i^{(2)}({\bm\nu},{\bm\nu}^{(1)},{\bm\nu}^{(2)}) = 0$, with the already determined coefficients $\nu_i$ and
$\nu_i^{(1)}$ entering as parameters, and so on. 

This procedure has to be modified for non fully connected rate matrices. In the following discussion we will assume that 
$R_{ij}=0$ implies $R_{ji}=0$ and, thus, also $A_{ij}=0$, this is analogous to our assumption about the occurrence of 
small rates in the previous section. Let us start with the zeroth-order equation, $I_i^{(0)}=\nu_i\sum_jA_{ij}\nu_j=0$. 
As before, we conclude that the leading coefficients $\nu_i$ are non-zero only for a group of selected states $i\in\BES$, 
\begin{align}\label{eq:NFC0a}
&\sum_{j\in\BES} A_{ij} \nu_j = 0,\qquad &i\in\BES,
\\\label{eq:NFC0b}
&\nu_i = 0, \qquad &i\notin \BES,
\end{align}
The set $\BES$ of selected states has still to be determined from the requirement that the asymptotic occupations of both 
the selected and the non-selected states are positive. It can consist of $K$ uncoupled subsets $\BES_\alpha$,
\be
\BES= \BES_1\cup\BES_2\cup\cdots\cup\BES_K,
\ee
with
\be
 R_{ij}=0 \qquad \text{ for }\qquad i\in\BES_\alpha, j\in\BES_\beta, \alpha\ne\beta. 
\ee
such that each subset $\BES_{\alpha}$ fulfills Eqs.~\eqref{eq:NFC0a} individually,
\begin{align}
\label{eq:sel_subset}
\sum_{j\in \BES_{\alpha}} A_{ij}\nu_j=0,\qquad \forall i \in\BES_{\alpha}.
\end{align}
Without fine-tuning, a solution of $\sum_{j\in\BES} A_{ij} \nu_j = 0$ is guaranteed as long as the number of states in 
each of the subsets $\BES_\alpha$ is odd. However, the total number of selected states $M_S$ can now also be even. It is 
even (odd), if the number $K$ of uncoupled subsets $\BES_\alpha$ is even (odd). In the case of fully connected rate 
matrices, the coefficients $\nu_i$ were determined uniquely by the set $\BES$, Eqs.~(\ref{eq:NFC0a}) and (\ref{eq:NFC0b}),
as well as by the normalization condition (\ref{eq:NuNorm}). For $K>1$ this is not the case anymore. Here the relative 
occupation of a subset $\BES_\alpha$, defined by $\nu_{\BES_\alpha}=\sum_{i\in\BES_\alpha}\nu_i$, is not fixed, since
$\nu_{\BES_\alpha}/\nu_{\BES_\beta}$ with $\alpha\ne\beta$ is not determined by Eq.~(\ref{eq:NFC0a}). Thus, one has $K-1$ 
parameters $\nu_{\BES_\alpha}$ that yet have to be determined from the equations of higher order. 

In order to investigate the first-order equations $I_i^{(1)}=0$, [see Eq.~\eqref{eq:asymptotic_mf}] it is useful to define two groups of non-selected 
states, 
\be
\bar{\BES}=\bar{\BES}'\cup \bar{\BES}'',
\ee
such that states that are directly coupled to selected states via non-zero rates form the set $\bar{\BES}'$ and
states that are not coupled directly to any selected state form the set $\bar{\BES}''$. 
For $i\in\bar{\BES}'$ they lead to the familiar result 
\begin{align}\label{eq:NFC1a}
&\nu_i^{(1)} 
= -\frac{\sum_{j\in\BES} R_{ij}\nu_j }{\sum_{j\in\BES} A_{ij}\nu_j }
,\quad  &i\in\bar{\BES}'.
\end{align}
Note that for states $i\in\bar\BES'$ that are coupled to selected states belonging to two subsets $\BES_\alpha$ and
$\BES_\beta$ (or more), the right-hand side of Eq.~(\ref{eq:NFC1a}) depends on the ratio $\nu_{\BES_\alpha}/\nu_{\BES_\beta}$, 
which is not determined yet. In that case the ratio $\nu_{\BES_\alpha}/\nu_{\BES_\beta}$ can be obtained from
Eqs.~(\ref{eq:sel_relative}) below. The coefficients $\nu_i^{(1)}$ with $i\in\bar{\BES}''$ drop out of the first-order 
equations ($I_i^{(1)}=0$ is fulfilled trivially) and must be determined from the second-order Eqs.~(\ref{eq:NFC2b}) below. 
The first-order equations for the selected states $i\in\BES_{\alpha}$ of a subset $\BES_{\alpha}$ simplify
[with Eq.~\eqref{eq:sel_subset}] to 
\begin{align}
0=&\sum_{j\in \BES_{\alpha}}\Big[R_{ij}\nu_j-R_{ji}\nu_i+A_{ij}\nu_i\nu_j^{(1)} \Big]
\nonumber\\
&+\sum_{j\in \bar{\BES}'}\Big[-R_{ji}\nu_i+A_{ij}\nu_i\nu_j^{(1)} \Big],\qquad i\in \BES_{\alpha}.
\label{eq:NFC1b1}
\end{align}
These equations determine the coefficients $\nu_i^{(1)}$ of the selected states $i$.

Further information can be obtained by summing Eqs.~(\ref{eq:NFC1b1}) over all states $i\in\BES_{\alpha}$. This gives
$0=\sum_{j\in\bar{\BES}'}\sum_{i\in\BES_{\alpha}}\nu_i\Big(R_{ji}+A_{ji}\nu_j^{(1)}\Big)$. Here, all
non-selected states $j\in\bar{\BES}'$ that couple only to selected states of the subset $\BES_{\alpha}$ do not contribute 
to the sum, since according to Eq.~\eqref{eq:NFC1a} their occupations are given by
$\nu_j^{(1)}=-(\sum_{i\in\BES_{\alpha}}R_{ji}\nu_i)/(\sum_{i\in\BES_{\alpha}}A_{ji}\nu_i)$. 
Thus, we obtain
\begin{align}\label{eq:sel_relative}
0=\sum_{j\in\bar{\BES}_{\alpha+}}\sum_{i\in\BES_{\alpha}}\nu_i\Big(R_{ji}+A_{ji}\nu_j^{(1)}\Big),\qquad \forall \alpha
\end{align}
where $\bar{\BES}_{\alpha+}$ denotes the set of non-selected states that couple to the subset $\BES_{\alpha}$ and at 
least to one more selected state of a different subset $\BES_{\beta}$ with $\beta\ne\alpha$. 
If this set $\bar{\BES}_{\alpha+}$ is not empty, Eq.~(\ref{eq:sel_relative}) can be used to determine missing 
relative occupations $\nu_{\BES_{\alpha}}/\nu_{\BES_{\beta}}$. 
We will argue below that in fact all subsets of selected states must form a connected cluster, where two subsets $\BES_{\alpha}$ 
and $\BES_{\beta}$ are defined to be connected if they are coupled directly (via a single quantum jump of non-zero rate) to the same
non-selected state(s). This guarantees that all relative occupations $\nu_{\BES_{\alpha}}/\nu_{\BES_{\beta}}$ can be determined from
Eqs.~(\ref{eq:sel_relative}) and (\ref{eq:NFC1a}), so that the $\nu_i$ can be determined completely. 

From the second-order equations $I_i^{(2)}=0$, we obtain 
\begin{align}\label{eq:NFC2}
0=\sum_j\bigg[&R_{ij}\nu_j^{(1)} - R_{ji}\nu_i^{(1)} 
\nonumber\\& + A_{ij}\Big(\nu_i^{(2)}\nu_j+\nu_i^{(1)}\nu_j^{(1)}
		+\nu_i\nu_j^{(2)}\Big) \bigg],\quad  \forall i.
\end{align}
These equations determine all the coefficients $\nu_i^{(1)}$ that have not been obtained yet, since all $\nu^{(1)}_i$ are 
coupled to each other (at least indirectly). For the missing coefficients $\nu_i^{(1)}$ of states $i\in\bar\BES''$ 
they simplify further to
\be\label{eq:NFC2b}
0=\sum_{j\in\bar\BES}\bigg(R_{ij}\nu_j^{(1)} - R_{ji}\nu_i^{(1)} 
+ A_{ij}\nu_i^{(1)}\nu_j^{(1)} \bigg),\quad  \forall i \in\bar\BES'',
\ee
since the states $i\in\bar{S}''$ couple to non-selected states only. In these equations the coefficients $\nu_j^{(1)}$ for the 
states $j\in\bar\BES'$ are determined already by Eqs.~(\ref{eq:NFC1a}).

The statement that all subsets of selected states must form a single connected cluster (in the sense described 
above) can now be shown by noting that the assumption of several mutually unconnected clusters $A$, $B$, $C$, \ldots leads to a 
contradiction. Let us denote the set of non-selected states directly coupled to the selected states of cluster $X$ by
$\bar\BES'_X$ and note that the mean particle current from one subset of non-selected states $\bar\BES_1$ to another 
one $\bar\BES_2$ is in leading order given by
$J_{\bar\BES_2\bar\BES_1}=\sum_{i\in\bar\BES_2}\sum_{j\in\bar\BES_1} \big(A_{ij}\nu_i^{(1)}\nu_j^{(1)}+R_{ij}\nu_j^{(1)}
-R_{ji}\nu_i^{(1)}\big)$. The total current into $\bar\BES''$ then reads $J_{\bar\BES''}=J_{\bar\BES''\bar\BES'}=
J_{\bar\BES''\bar\BES_A'}+J_{\bar\BES''\bar\BES_B'}+\cdots$. It is directly given by summing the right-hand-sides of
Eqs.~(\ref{eq:NFC2b}). Consequently, it vanishes in the steady state as it should, $J_{\bar\BES''}=0$. The total current into 
cluster $A$ reads $J_{\bar\BES_A'}=J_{\bar\BES_A'\bar\BES''}+J_{\bar\BES_A'\bar\BES_B'}+J_{\bar\BES_A'\bar\BES_C'}+\cdots$. 
Obviously, it should also vanish in the steady state. However, generically this is is not possible for more than a single 
cluster. Namely, (without fine tuning) the individual terms $J_{\bar\BES_A'\bar\BES''}$, $J_{\bar\BES_A'\bar\BES_B'}$, \ldots, 
containing the coefficients $\nu_i^{(1)}$ determined from Eqs.~(\ref{eq:NFC1a}) and (\ref{eq:NFC2b}), can neither be expected to 
vanish individually nor to cancel each other. In contrast, for a single cluster, one has
$J_{\bar\BES_A'}=J_{\bar\BES_A'\bar\BES''}=-J_{\bar\BES''}=0$, as required.

From the rather technical discussion of the preceding paragraphs, we can now draw several important conclusions. 
First of all, Eqs.~(\ref{eq:NFC0a}) and (\ref{eq:NFC0b}) imply that Bose selection 
is still predicted to occur, i.e.\ only a subset $\BES$ of the single-particle states have occupations that grow with the 
total particle number
\be
\bar{n}_i=\nu_i N.
\ee
Second, the asymptotic occupations of the non-selected states are still determined by the 
first-order coefficient $\nu_i^{(1)}$, so that their occupations saturate for large $N$. (In contrast, if $\nu^{(2)}_i$ 
would describe the leading contribution to the occupations of a state $i$, it would become unpopulated in the limit of 
large particle numbers). This is true also for states contained in $\bar{\BES}''$ that are not directly coupled to a 
selected state. Both conclusions, Bose selection and saturation, are confirmed by the preasymptotic state that can be 
observed in Fig.~\ref{fig:small_rates} for $10^1\lesssim N\lesssim 10^3$, which is approximately given by the asymptotic 
state of the rate matrix $R^a$ [Eq.~\ref{eq:ThreeStateApprox}]. 

Finally, a third conclusion is that for rate matrices that are not fully connected the set of selected states $\BES$ is 
not determined by the conditions (\ref{eq:allconditions}) anymore. Namely a negative $\mu_i$ guarantees a positive 
asymptotic occupation $\nu^{(1)}_1$ of a non-selected state $i\in\bar{\BES}'$, but not for a non-selected state
$i\in\bar{\BES}''$. This implies that we cannot apply the efficient algorithm presented in subsection \ref{sec:algorithm}
in order to find the set of selected states (neither can the algorithm of reference \cite{KnebelEtAl15} be used, which is also 
based on the conditions (\ref{eq:allconditions})).
It seems likely that the set 
of selected states of the mean-field equations is still unique and determined by the requirement of having positive occupations, as the full
many-body master equation possesses a unique steady state. However, unlike in the case of fully connected rate matrices, 
we have no proof for this statement.

Let us illustrate the above reasoning using the minimal example given by the rate matrix $R^a$ defined in
Eq.~(\ref{eq:ThreeStateApprox}) of the previous subsection. The corresponding rate-asymmetry matrix reads
 \begin{align}
A^a =  r\left(\begin{array}{ccc} 
		0  & -1 & 0\\
		1 & 0 & -2\\
		0  & 2 &  0
	\end{array}\right). 
\end{align}
Thus, if we were allowed to solve the problem (\ref{eq:allconditions}) to find the set of selected states and the 
asymptotic occupations, we would find two disconnected clusters of selected states given by $\BES_1=\{1\}$ and 
$\BES_2=\{3\}$. Namely,
\be
{\bm \mu} = 
r\left(\begin{array}{ccc} 
		0  & -1 & 0\\
		1 & 0 & -2\\
		0  & 2 &  0
	\end{array}\right)
	\left(\begin{array}{c} 
		\nu_1\\
		0\\
		\nu_3
	\end{array}\right)
	= 
	\left(\begin{array}{c} 
		0\\
		r(\nu_1-2\nu_3)\\
		0
	\end{array}\right)
\label{eq:sel_example}
\ee
solves problem (\ref{eq:allconditions}) non-uniquely for $0<\nu_1<2/3$ and $\nu_3=1-\nu_1$. However, this is \emph{not} the true solution. 
Namely Eq.~\eqref{eq:sel_relative} for $\alpha=1$ simplifies to $0=\nu_1(R_{21}+A_{21}\nu_2^{(1)})=\nu_1(2+\nu_2^{(1)})$ from which $\nu_1=0$ follows  in contradiction to Eqs.~\eqref{eq:allconditions}. This demonstrates that 
Eq.~(\ref{eq:allconditions}) cannot be used in order to determine the selected states in the case of non fully connected rate matrices. 

From Fig.~\ref{fig:small_rates}, where $R^a$ describes the preasymptotic regime
($10^1\lesssim N\lesssim 10^3$), one can infer that only state 3 will be selected. Let us, therefore, solve
Eqs.~(\ref{eq:NFC0a}), (\ref{eq:NFC0b}), (\ref{eq:NFC1a}), and (\ref{eq:NFC2}) for the ansatz
\be
\BES=\{3\} .
\label{eq:sel_state_3}
\ee 
The zeroth order equations (\ref{eq:NFC0a}) and (\ref{eq:NFC0b}) are solved trivially by
\be
\nu_3=1,\qquad \nu_1=\nu_2 = 0. 
\ee
Then $\nu_2^{(1)}$ is obtained from Eq.~(\ref{eq:NFC1a}) and reads
\be
\nu_2^{(1)} = - \frac{R^a_{23}}{A^a_{23}} = 1
\ee
while Eq.~\eqref{eq:sel_relative} is trivially fulfilled since $\bar{\BES}_{\alpha+}$ is empty. Finally, $\nu_1^{(1)}$ results from Eq.~(\ref{eq:NFC2}) for $i=1$, 
\be
\nu_1^{(1)} = \frac{R^a_{12}\nu_2^{(1)}}{R^a_{21}-A^a_{12}\nu^{(1)}_2} 
		= \frac{1}{3},
\ee
We can see that the initial assumption $\BES=\{3\}$ is confirmed by the fact that we obtained meaningful positive 
occupation numbers. The just-obtained asymptotic occupations for the rate matrix $R^{a}$ are plotted as dotted lines in 
Fig.~\ref{fig:small_rates} and provide a good description of the preasymptotic state.

\begin{figure}[t]
\centering
\includegraphics[width=1\linewidth]{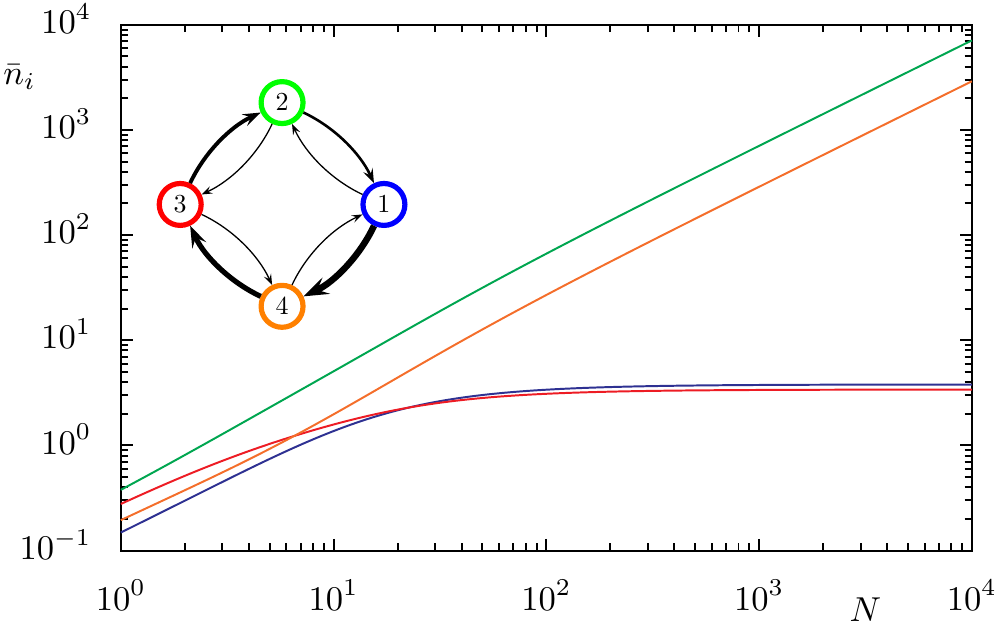}
\caption{(color online) Example for Bose selection of two uncoupled states. Mean occupations obtained from mean-field theory ($\bar{n}_1$ blue, $\bar{n}_2$ green, $\bar{n}_3$ red, $\bar{n}_4$ 
orange) vs the total particle number $N$ for the rate matrix (\ref{eq:FourState}), which is visualized in the inset (line widths reflect rates). The two selected states 2 and 4 are not coupled directly.}
\label{fig:TwoSelected}
\end{figure}

For completeness, we will finally present a simple example for a situation where the set of selected states consists of 
two uncoupled subsets. It is given by a model of four states with rate matrix 
\begin{align}\label{eq:FourState}
R =  r\left(\begin{array}{cccc} 
		0  & 2 & 0 & 1\\
		1  & 0 & 3 & 0\\
		0  & 1 & 0 & 4\\
		5  & 0 & 1 & 0\\
	\end{array}\right).
\end{align}
The occupations plotted in Fig.~\ref{fig:TwoSelected} show that the set of selected states contains the two 
uncoupled states 2 and 4, 
\be
\BES=\BES_1\cup\BES_2, \quad\text{with}\quad \BES_1=\{2\} \text{ and } \BES_2=\{4\}.
\ee

\begin{figure}[t]
\centering
\includegraphics[width=1\linewidth]{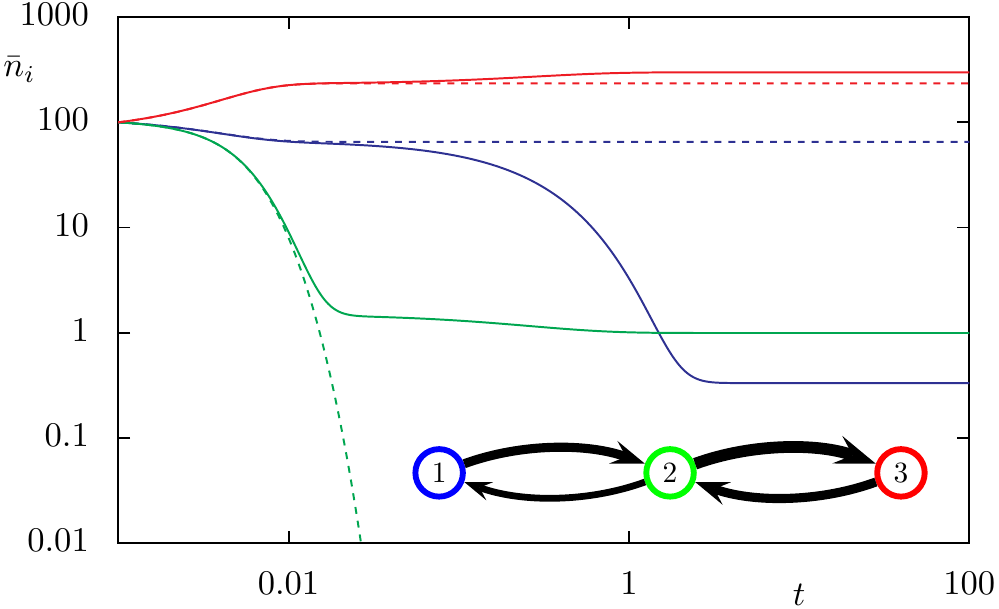}
\caption{(color online) Comparison between the dynamics of the mean-field equations \eqref{eq:EOM_nmean_MF} [solid lines] and the Lotka-Volterra equations \eqref{eq:Knebel} [dashed lines] for the rate matrix \eqref{eq:ThreeStateApprox}, which is visualized in the inset (line widths reflect rates), for $N=300$ particles initially uniformly distributed. While in both cases the population in state 2 (green) decays on an intermediate time scale, the population in state 1 (blue) decays on a longer time scale in the mean-field equations only leading to a single condensate in state 3 (red).} 
\label{fig:LV_vs_MF}
\end{figure}

The case of zero rates has recently also been discussed by Knebel \emph{et al.\ } for the Lotka-Volterra equations of motion \cite{KnebelEtAl13,KnebelEtAl15},
\begin{equation}\label{eq:Knebel}
\dot{\bar{n}}_i=\bar{n}_i\sum_jA_{ij}\bar{n}_j.
\end{equation}
These equations correspond to the leading-order high-density approximation (\ref{eq:naiv_app}) of the mean-field equation (\ref{eq:EOM_nmean_MF}), with $\sigma=1$ for bosons. These leading-order equations describe the dynamics of the Bose gas on an intermediate time scale, before, eventually the sub-leading terms of Eq.~\eqref{eq:EOM_nmean_MF}, which are linear in the occupations, become relevant and determine the steady state. Knebel \emph{et al.\ } show that under the evolution described by Eq.~(\ref{eq:Knebel}) the occupations of some states $i$ die out exponentially fast, while the other states retain non-zero occupations. Interestingly, those states retaining non-zero occupations are determined by the very same condition (\ref{eq:allconditions}) that we found to determine the selected  states for fully connected rate matrices. That means in the case of fully connected rate matrices the selected states are determined already by the leading-order equation (\ref{eq:Knebel}). Note the general difference between the mean-field equation on the one hand and the Lotka-Volterra equation on the other. While in the first case the non-selected states retain a small but non-zero occupation, they die out completely in the latter case.

In the case of non fully connected rate matrices Eqs.~(\ref{eq:allconditions}) still determine uniquely which occupations 
die out under the dynamics of Eq.~(\ref{eq:Knebel}) \cite{KnebelEtAl15}. Thus, the conditions (\ref{eq:allconditions}) 
still describe a dynamical selection mechanism happening on an intermediate time scale. However, in order to compute the 
(true) steady state approached in the long-time limit, also higher-order equations [Eqs.~\eqref{eq:NFC1a},
(\ref{eq:NFC1b1}) and \eqref{eq:NFC2}] have to be taken into account. As a result, the set of selected states in the 
steady state can be different from that obtained from conditions (\ref{eq:allconditions}). 

Let us illustrate the above reasoning using the example of the not fully connected rate matrix \eqref{eq:ThreeStateApprox}.
Fig.~\ref{fig:LV_vs_MF} shows the different dynamics of this system for both the full mean-field equations (\ref{eq:EOM_nmean_MF}) and for the Lotka-Volterra equations of motion (\ref{eq:naiv_app}). In the limit of large $N$ the population of state 2 decays on the intermediate time scale, because the conditions \eqref{eq:allconditions} predict a extinction of occupations $n_2$ on the level of the Lotka-Volterra equations [see Eq.~\eqref{eq:sel_example}]. Eventually, however, when higher-order terms become relevant in the full mean-field equations of motion, also the occupation of state 1 decays so that only state 3 is selected as predicted in Eq.~\eqref{eq:sel_state_3}. This is contrasted by the Lotka-Volterra system, which remains in the situation with two condensates in the states 1 and 3. 

\subsection{Asymptotic theory beyond mean field}
\label{sec:beyond_mf}
Our theoretical description of Bose selection has so far been based on mean-field theory. The data presented in
Fig.~\ref{fig:BS} for a tight-binding chain in and out of equilibrium suggests that mean-field theory provides a rather 
good approximation to the mean occupations. Namely, deviations between the mean-field results (thick solid lines) and the 
exact Monte Carlo data (crosses) are visible only for non-selected states. And where visible deviations occur they are 
still rather small and captured by the augmented mean-field theory (thin solid lines) introduced in Section
\ref{sec:augmented}. Such good agreement can generally not be expected for the number fluctuations of macroscopically 
occupied selected modes, since mean-field theory does not comply with the conservation of the total particle number. 

In this subsection we will investigate corrections to mean-field theory in the asymptotic limit of large total particle 
number $N$, as they are described by the augmented mean-field theory. For simplicity, we will consider the case of fully 
connected rate matrices. We will explain why mean-field theory accurately describes the occupations of the selected 
states and that their correlations, such as number fluctuations, deviate from mean-field theory in a universal fashion. 
Moreover, we will argue that the set of selected states is well described by a Gaussian state projected to the space of 
sharp particle number $N$.

Within the augmented mean-field theory (Section \ref{sec:augmented}) the state of the system is described not only by 
the mean occupations $\bar{n}_i=\la\no_i\ra$, like in mean-field theory, but also in terms of the non-trivial
two-particle correlations $\ze_{ij}=\la\no_i\no_j\ra-\la\no_i\ra\la\no_j\ra$. In order to derive an augmented mean-field 
theory for the asymptotic limit of large total particle numbers $N$, we do not only expand the mean occupations with 
respect to the inverse particle number, but at the same time also the non-trivial two-particle correlations,
\begin{align}
\label{eq:expand_n}
 & \avgo{n_i} =  N\nu_i + \nu_i^{(1)} +  N^{-1} \nu_i^{(2)} + N^{-2} \nu_i^{(3)} +\cdots,
\\
\label{eq:expand_zz}
  &{\f{k}{i}} =N^2 \xi_{ki} + N \xi_{ki}^{(1)} + \xi_{ki}^{(2)} +N^{-1} \xi_{ki}^{(3)} \cdots .
\end{align}
Moreover, we choose again the normalization conditions
\be
\sum_i \nu_i = 1, \qquad \sum_i\nu_i^{(r)}=0, 
\ee
which fix the total particle number $N=\sum_i\bar{n}_i$ in leading order, as well as the conditions
\be\label{eq:ZeroDeltaN}
\sum_{ij} \xi_{ij} =0, \qquad \sum_{ij} \xi_{ij}^{(r)} = 0 ,
\ee
ensuring that the fluctuations of the total particle number $\Delta N=\sum_{ij}\ze_{ij}$ vanish.

We now insert the expansions (\ref{eq:expand_n}) and (\ref{eq:expand_zz}) into the augmented mean-field equations
(\ref{eq:EOM_xmean}) and (\ref{eq:EOM_zzcorr_MF}), with $\sigma=1$ for bosons and with the left-hand side set to zero in 
order to obtain the steady state. In the resulting equations we ask that all terms belonging to a certain power of $N$ 
vanish independently. In this way, we obtain the set of coupled non-linear equations
\begin{align}
\label{eq:aug_asym_n}
 &0 = \sum_j A_{ij} \left[ \nu_i \nu_j + \xi_{ij}\right] , \\ 
\label{eq:aug_asym_kor}
 &0 = \sum_j\left[  A_{kj} \nu_{k} \xi_{ij}  + A_{ij} \nu_{i} \xi_{kj} 
+ (A_{kj} + A_{ij}) \nu_{j} \xi_{ki}\right].
\end{align}
for the leading order. 

Remarkably, we can solve these equations by making the simple ansatz 
\be\label{eq:XiAnsatz}
\xi_{ki} = x(\delta_{ki}\nu_k - \nu_k\nu_i)
\ee
for the leading non-trivial correlations $\xi_{ij}$, with $x$ being a free parameter. The relative weight of both terms 
in the bracket is chosen such that the condition (\ref{eq:ZeroDeltaN}) is obeyed. Entering the ansatz (\ref{eq:XiAnsatz}) 
into Eqs.~(\ref{eq:aug_asym_n}) and (\ref{eq:aug_asym_kor}) reduces these equations to the much simpler conditions
\be
\nu_i \sum_j A_{ij}  \nu_j = 0 .
\ee
These equations are identical to the leading-order conditions (\ref{eq:selected_orig}) of the asymptotic mean-field theory. 
Using the same arguments as in the conventional asymptotic mean-field theory, we have to conclude that the solution must 
be of the form
\begin{align}\label{eq:NuAugAsym1}
&\sum_{j\in\BES} A_{ij} \nu_j = 0,\qquad &i\in\BES,
\\\label{eq:NuAugAsym2}
&\nu_i = 0, \qquad &i\notin \BES.
\end{align}

Equations~(\ref{eq:NuAugAsym1}) and (\ref{eq:NuAugAsym2}) imply Bose selection. Only a subset $\BES$ of selected states 
have non-vanishing occupations in leading order. The set $\BES$ has to be determined by the requirement to have positive 
occupations both for selected and non-selected states. The asymptotic occupations of the latter are given by $\nu_i^{(1)}$
and have to be determined in the next order. Note that the set $\BES$ obtained within the augmented theory can be 
different from the one obtained within mean-field theory. Namely, the occupations of the non-selected states differ in 
both theories, so that in the augmented theory, e.g., a transition where $\BES$ changes might be shifted away from the 
critical mean-field parameter. However, as long as the set of selected states is the same in both theories, the
mean-field result for the asymptotic occupations of the selected states is not corrected anymore. This explains the 
excellent agreement between mean-field theory, augmented mean-field theory, and Monte-Carlo results for the
selected-state occupations in Fig.~\ref{fig:BS}.

According to the ansatz (\ref{eq:XiAnsatz}), we find the asymptotic correlations among the selected states to be given by
\be\label{eq:AsympCorrAug}
\la\no_i\no_j\ra = (1-x) \bar{n}_i\bar{n}_j + x \bar{n}_i\delta_{ij} , \quad i,j\in\BES .
\ee
This is an intriguing result. It implies that the correlations and fluctuations are determined solely by the
mean occupations and a \emph{single parameter} $x$. The scaled two-particle correlations for particles in different 
selected states,
\be\label{eq:gij}
g_{ij} = \frac{\la\no_i\no_j\ra}{\la\no_i\ra\la\no_j\ra} = 1-x, \quad i,j\in\BES, i\ne j 
\ee
asymptotically approach all the same value, which is reduced by $x$ with respect to the mean-field result. 

This very same parameter $x$ also determines the asymptotic number fluctuations of the Bose selected modes, 
\be\label{eq:Delta_n_aug}
\Delta n_i^2 \equiv \ze_{ii}
		= x N^2 (1-\nu_i)\nu_i= x (N-\bar{n}_i)\bar{n}_i ,\, 
 i\in\BES.
\ee
This equation implies that (in leading order) the number fluctuations vanish if we have a single condensate in the state
$i=k$, so that $\nu_k=1$. This is a consequence of the conservation of the total particle number that is incorporated in 
the augmented mean-field theory. It contrasts with the Gaussian result (\ref{eq:Deltanig}) obtained within the
non-number-conserving mean-field theory, which for bosons ($\sigma=1$) reads
$\Delta n_i^2=\bar{n}_i+\bar{n}_i^2=N^2(\nu_i+1/N)\nu_i$. Note, however, that as soon as a system features several 
condensates (macroscopically occupied selected states), their number fluctuations (\ref{eq:Delta_n_aug}) will typically 
be of the order of the total particle number. This reflects the fact that each condensate is effectively in contact with 
a particle reservoir given by the other ones. 

The requirement $\Delta n_i^2>0$ tells us that $x$ is positive. Moreover, it is reasonable to assume that the number 
fluctuations will not be much larger than those obtained within the non-number-conserving mean-field theory, so that
$\Delta n_i^2\lesssim \bar{n}_i^2$ for all $i\in\BES$. Thus, an estimate for an upper bound for $x$ is determined by the 
selected state $i$ with the smallest occupation $\bar{n}_i=\nu_iN$. Therefore,  
\be\label{eq:estimate}
0 < x \lesssim \frac{\nu_\text{min}}{1-\nu_\text{min}}, \qquad \nu_\text{min}=\min_{i\in\BES} \nu_i.
\ee
The precise value of $x$ has to be obtained, however, from the first-order equations. These equations are rather involved 
and we will not discuss them here. They also describe small beyond-mean-field corrections for the asymptotic occupations, 
correlations, and fluctuations of the non-selected states.

\begin{figure*}[t]
\centering
\includegraphics[width=1\linewidth]{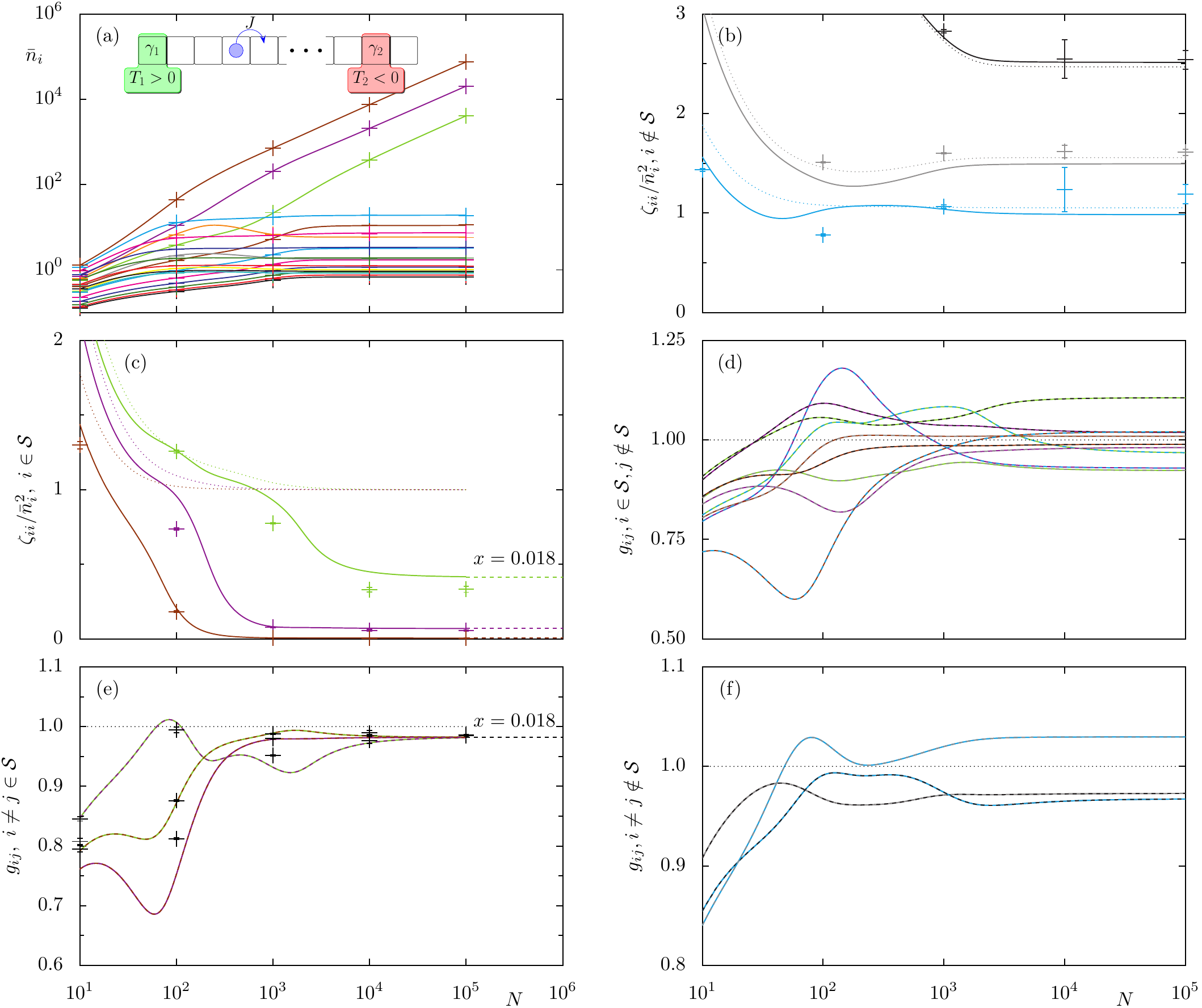}
\caption{(color online) Augmented mean-field theory (solid lines) versus mean-field theory (dotted lines) and Monte-Carlo simulations
(crosses with error bars) for the tight-binding chain with parameters as in Fig.~\ref{fig:BS}(c). All colors are 
consistent with panel (a) [and also Fig.~\ref{fig:BS}], lines describing correlations between two states have alternating color. 
(a) Mean occupations [corresponding to thin solid lines in Fig.~\ref{fig:BS}(c)]. 
(b) Relative number fluctuations $\Delta n_i^2/\bar{n}_i^2=\ze_{ii}/\bar{n}_i^2$ for three exemplary non-selected states. 
(c) Relative number fluctuations $\Delta n_i^2/\bar{n}_i^2=\ze_{ii}/\bar{n}_i^2$ of the selected states. 
(d) Correlations $g_{ij}$ between the selected and three exemplary non-selected states. 
(e) Correlations $g_{ij}$ among selected states $i\ne j$. 
(f) Correlations $g_{ij}$ among exemplary non-selected states $i\ne j$.} \label{fig:MFvsAMF}
\end{figure*}

In Fig.~\ref{fig:MFvsAMF} we compare the augmented theory (solid lines) with Monte-Carlo results (crosses with error bars),
ordinary mean-field theory (dotted lines), and the asymptotic prediction (\ref{eq:AsympCorrAug}) for the selected states
(dashed lines), using the model system of Fig.~\ref{fig:BS}(c). The comparison with the Monte-Carlo data shows that the 
augmented mean-field theory provides an excellent approximation for the mean occupations $\bar{n}_i$ [panel (a)], where
the ordinary mean-field theory shows small deviations for the non-selected states [see Fig.~\ref{fig:BS}(c)]. For 
the two-particle correlations $\la\no_i\no_j\ra$ shown in Fig.~\ref{fig:MFvsAMF}(b)-(f)], the augmented mean-field theory 
still provides a rather good description, though small systematic deviations with respect to the exact Monte-Carlo 
results are now visible, while mean-field theory is not reliable anymore. 

The relative number fluctuations $\Delta n_i^2/\bar{n}_i^2=\ze_{ii}/\bar{n}_i^2$ for the selected states [panel (c)] show 
strong deviations from mean-field theory, once Bose selection sets in near $N=10^2$ [see panel (a)] so that the selected 
modes acquire ``extensive'' occupations. This agrees with our expectation that mean-field theory is not able to describe 
the condensate fluctuations for a system with sharp particle number. The condensate fluctuations are found to be 
consistent with the asymptotic prediction (\ref{eq:Delta_n_aug}) for $x\approx0.018$. Note that the selected state with 
the smallest occupation (roughly 4\%) has asymptotic number fluctuations that are only half as large as the mean-field 
prediction, even though the other two condensates are large enough to serve as a reservoir. Thus $x$ is roughly given by
$\nu_\text{min}/2$ in agreement with the estimate (\ref{eq:estimate}). 

The other quantities displayed in Fig.~\ref{fig:MFvsAMF} are not expected to exhibit such drastic deviations of orders of 
magnitude from mean-field theory, as we observed them for the condensate fluctuations.
Panel (e) shows the scaled correlations (\ref{eq:gij}) among the selected states. The augmented theory asymptotically 
approaches the universal value $1-x$, with $x\approx0.018$. Noticeable deviations of up to $30\%$ occur before reaching 
the asymptotic regime, whereas the deviation from the mean-field result 1 become rather small asymptotically since
$x\ll1$. Similar behavior, i.e.\ larger deviations of up to a few tens of percent for small particle numbers that are 
reduced slightly in the asymptotic regime, can be observed also in the remaining plots of the figure. Panel (b) displays 
the relative number fluctuations $\Delta n_i^2/\bar{n}_i^2$ for three exemplary non-selected states. Relative 
correlations $g_{ij}$ between selected states and exemplary non-selected states as well as among exemplary non-selected states 
are plotted in panel (d) and (f) respectively.

A deeper understanding of the findings presented so far in this subsection, can be gained by noting that the selected 
states are asymptotically described by a projected Gaussian state. This can be seen as follows. For bosons in the steady 
state, the full many-body rate equation (\ref{eq:mpr}) takes the form  
\be
0 = \sum_{ij} (1+n_j)n_i\Big[R_{ij}p_{\bn_{ji}}-R_{ji}p_{\bn}\Big],
\ee
where $p_\bn$ is the full occupation-number distribution. Let us accept that there is a group of selected states, 
whose occupations will grow with the total particle number $N$ while all other occupations saturate. Asymptotically for 
$N\to\infty$, we can then neglect all non-selected states and safely approximate $(1+n_j)\approx n_j$, so that
\be\label{eq:ABC}
0 = \sum_{i,j\in\BES} n_j n_i\Big[R_{ij}p_{\bn_{ji}}-R_{ji}p_{\bn}\Big].
\ee
We can now show that this equation is solved by the projected Gaussian state (\ref{eq:gaussian_factorized}). For this 
state one finds that $p_{\bn_{ji}}=e^{\eta_i-\eta_j}p_\bn$ for the probability of finding the system in the Fock state
$|\bn_{ji}\ra$ obtained from $|\bn\ra$ by transferring one particle from $i$ to $j$. 
Moreover, according to Eq.~(\ref{eq:gamma_vs_n}) one has $e^{\eta_i}=1-1/\bar{n}_i\simeq 1$. Here we have used that,
asymptotically, the mean occupations of the projected Gaussian state become identical to that of the non-projected 
Gaussian state. This implies the intuitive statement that for the projected Gaussian state, the probabilities for finding 
the system in the almost identical Fock states $|\bn_{ji}\ra$ and $|\bn\ra$ asymptotically become identical,
$p_{\bn_{ji}}\simeq p_\bn$. Thus, plugging the projected Gaussian state into the right-hand side of Eq.~(\ref{eq:ABC}), 
we obtain 
\be
p_\bn\sum_{i,j\in\BES} n_j n_i\Big[R_{ij}e^{\eta_i-\eta_j}-R_{ji}\Big]
\simeq p_{\bn}\sum_{i,j\in\BES} n_j n_i A_{ij} = 0 ,
\ee
since $A_{ij}=-A_{ji}$. We have shown that asymptotically in the limit $N\to\infty$ the \emph{full} number distribution 
of the selected states is given by a projected Gaussian state. An important consequence is that mean-field theory 
provides the exact asymptotic mean occupations of the selected states. Another consequence is that correlations
$\la\no_i\no_j\ra$ with $i,j\in\BES$ and, therefore, also the parameter $x$, must be determined completely by the 
asymptotic mean occupations $N\nu_i$ of the selected states.

\subsection{Heat flow through the system: the role of fragmented condensation and pseudotransitions}
\label{sec:heat_transport}
Non-equilibrium steady states of a driven-dissipative quantum system typically feature a steady heat flow between the 
system and its bath(s). This heat flow is described by Eq.~(\ref{eq:heat_flow_undriven_single_particle}) in the case of 
an autonomous system and by Eq.~(\ref{eq:heat_flow_driven_single_particle}) for a periodically driven system. For bosons
($\sigma=1$) in a steady state, these equations read
\begin{align}\label{eq:Qauto}
Q_b=&\sum_{ij} (E_i-E_j) R_{ji}^{(b)}\big[\la\no_i\ra+\la\no_i\no_j\ra\big]
\end{align}
for the heat flow from an autonomous system into bath $b$ and 
\begin{align}\label{eq:QFloquet}
Q =&\sum_{m}\sum_{ij} (\epsilon_i-\epsilon_j-m\hbar\omega) R_{ji}^{(m)}\big[\la\no_i\ra+\la\no_i\no_j\ra\big]
\end{align}
for the heat flow from a Floquet system into a bath. In this subsection, we will investigate such heat flow in the regime 
of Bose selection. The dominant processes contributing to the heat flow will be identified. They are found to be given 
by transitions between different selected states and, for the Floquet system, also by pseudotransitions [corresponding to 
terms with $i=j$ and $m\ne0$ in Eq.(\ref{eq:QFloquet})] associated with a selected state.

\begin{figure}[t]
\centering
\includegraphics[width=1\linewidth]{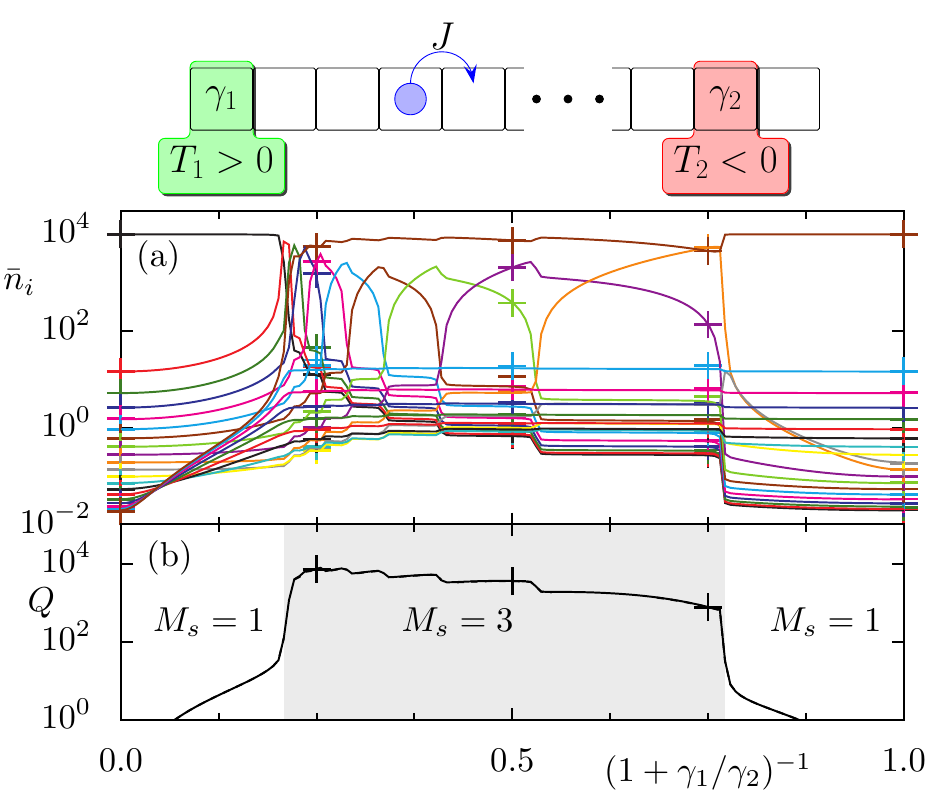}
\caption{(color online) Tight-binding chain with $M=20$ sites coupled to two heat baths. Parameters as in Fig.~\ref{fig:BS}(c), but 
for fixed $N=10^4$ and versus relative coupling strength $\gamma_2/\gamma_1$. (a) Mean occupations obtained from
mean-field theory (solid lines), augmented mean-field theory (dashed line, indistinguishable from mean-field result) and 
Monte-Carlo simulations (crosses). Color code like in Fig.~\ref{fig:BS}, on the left-hand (right-hand) side the occupation decreases (increases) with energy. (b) Heat flow through the system from the hotter negative-temperature bath into the positive-temperature bath.}
\label{fig:heat_transport_1}
\end{figure}

In Fig.~\ref{fig:heat_transport_1} we present data obtained for a tight-binding chain that is driven between two heat 
baths, one of positive temperature and a population-inverted one modeled by a negative temperature. This system 
corresponds to the one of Fig.~\ref{fig:BS}(c), but with the particle number fixed and with the relative coupling between 
both baths, $\gamma_2/\gamma_1$, varied. In panel (a) we plot the mean occupations versus the parameter
$p=(1+\gamma_1/\gamma_2)^{-1}$, which increases with $\gamma_2/\gamma_1$. One can observe several transitions. For $p=0$, 
where the system is only coupled to bath $b=1$, a single state (the ground state) is selected as indicated by a large 
occupation. This corresponds to equilibrium Bose condensation. At a critical coupling to the second bath, near
$p=0.2$, three states become selected. Increasing the coupling to the second bath further, various transitions occur,  
where the set of selected states changes. Eventually, roughly from $p=0.75$ on only the most excited state will be 
selected, corresponding to the equilibrium situation at $p=1$, where the system is coupled to the population-inverted 
bath 2 only. In panel (b) we plot the heat flow from the hotter population-inverted bath through the system into the 
cooler positive temperature bath versus $p$. We can clearly see that the heat flow increases dramatically (by more than 
two orders of magnitude), when fragmented Bose condensation with more than just one selected state occurs.  

This effect, which has been reported already in reference \cite{VorbergEtAl13}, can be understood intuitively. Namely, in 
order to exchange energy with the system, the bath has to drive transitions between states $i$ and $j$ in the system. The 
larger the occupations of $i$ and $j$, the larger will be the rate of the corresponding transition. Therefore, the most 
effective way of exchanging energy with the system is to drive transitions between two largely occupied states. And this 
is possible only if more than just one state is selected. This effect might be employed to control the heat conductivity of a bosonic system by switching between one and three selected states.

\begin{figure}[t]
\centering
\includegraphics[width=1\linewidth]{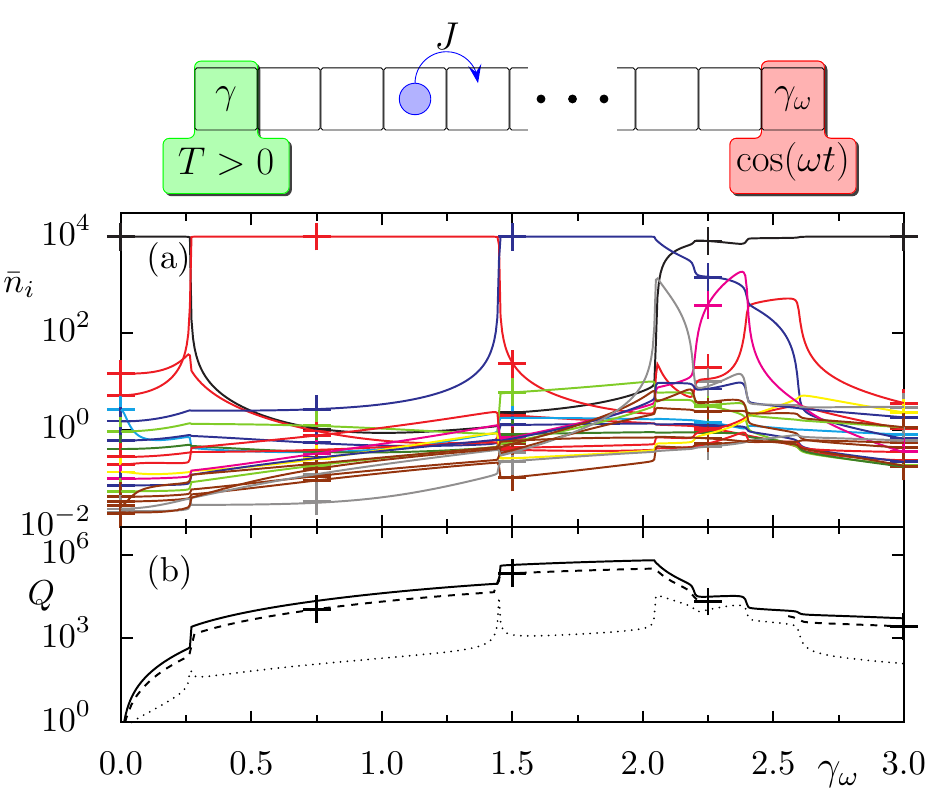}
\caption{(color online) Periodically driven tight-binding chain with $M=20$ sites coupled to a heat bath. Parameters as in
Fig.~\ref{fig:BS}(d), but for fixed $N=10^4$ and versus dimensionless driving strength $\gamma_\omega$. (a) Mean 
occupations obtained from mean-field theory (solid lines) and Monte-Carlo simulations (crosses). Color code like in
Fig.~\ref{fig:BS}, on the left-hand side the occupation decreases (increases) with energy. (b) Heat flow from the 
driven system into the bath obtained from mean-field theory (solid line), augmented mean-field theory (dashed line), and 
Monte-Carlo simulations (crosses). The dotted line is the mean-field heat flow without the contribution from 
pseudotransitions.}
\label{fig:heat_transport_2}
\end{figure}

In Fig.~\ref{fig:heat_transport_2} we show results for a periodically driven tight-binding chain coupled to a heat bath. 
This system corresponds to the one of Fig.~\ref{fig:BS}(d), but with the particle number fixed and with the dimensionless 
driving strength $\gamma_\omega$ varied. From the mean occupations plotted in panel (a), we can observe that for small
$\gamma_\omega$ a single-particle Floquet state is selected, which is connected adiabatically to the ground state 
of the undriven system with $\gamma_\omega=0$. Roughly at $\gamma_\omega=0.25$ and $\gamma_\omega=1.5$ the selected state 
changes in transitions, but still only a single state is selected. Only for a driving strength of about $\gamma_\omega=2$,
a parameter window is reached, where three states become selected and acquire large occupations.  

The heat flow from the system into the bath is plotted in panel (b) of Fig.~\ref{fig:heat_transport_2}. In contrast to the 
autonomous chain, we can observe that the heat flow grows strongly, despite the fact that we have only a single selected 
state. This effect can be attributed to pseudotransitions \cite{LangemeyerHolthaus14} associated with rates $R^{(m)}_{ij}$
with $i=j$ and $m\ne0$. In these processes the bath energy changes by $m\hbar\omega$, while the system's state is not 
altered. Thus, the bath can effectively exchange energy with the system by driving pseudotransitions for a single 
strongly occupied (Bose selected) Floquet mode. This interpretation is supported by the dotted line, showing the share
$Q'$ of the heat flow not related to pseudotransitions,
\begin{align}\label{eq:QFloquet_pseudo}
Q' =&\sum_{m}\sum_{i,j (i\ne j)} (\epsilon_i-\epsilon_j+m\hbar\omega) R_{ji}^{(m)}\big[\la\no_i\ra+\la\no_i\no_j\ra\big].
\end{align}
Away from the undriven limit $\gamma_\omega=0$ and as long as only one Floquet mode $i$ acquires a large occupation,
$Q'$ is typically two orders of magnitude smaller than the full heat flow and, thus, negligible. That means that 
practically all the heat flow is based on pseudotransitions, the double sum in Eq.~(\ref{eq:QFloquet}) is dominated by 
the terms with $i=j$. $Q'$ becomes significant only when several states have a large occupation. As one can clearly 
observe in Fig.~\ref{fig:heat_transport_2}(b), this happens both near transitions, where two states are selected (see 
subsection \ref{sec:transition}), and for $2\lesssim\gamma_\omega\lesssim 2.6$, where three states are selected. 
Here an efficient heat exchange with the bath can be achieved by driving transitions between these largely occupied 
states, like for the autonomous system. 

In  Fig.~\ref{fig:heat_transport_2}(b), we can also observe a noticeable difference between the heat flow obtained from 
mean-field theory (solid line) and augmented mean-field theory (dashed line), in contrast to the autonomous system where 
both theories show very good agreement [on the logarithmic scale of Fig.~\ref{fig:heat_transport_2}(b) both lines overlap].
This is also a consequence of the strong impact of pseudotransitions in the condensate mode, which are determined by the 
condensate fluctuations, a quantity that is overestimated by mean-field theory. This confirms our conclusion that, thanks 
to pseudotransitions not present in autonomous systems, a bosonic Floquet system can be a good heat conductor even when 
most of its particles form a single Bose condensate.

In conclusion, departing from equilibrium offers interesting possibilities to control the heat conductivity of a bosonic 
quantum system, which might be relevant for technological applications.

\section{Ideal Fermi gases}
\label{sec:fermions}

\begin{figure}[t]
\centering
\includegraphics[width=8cm]{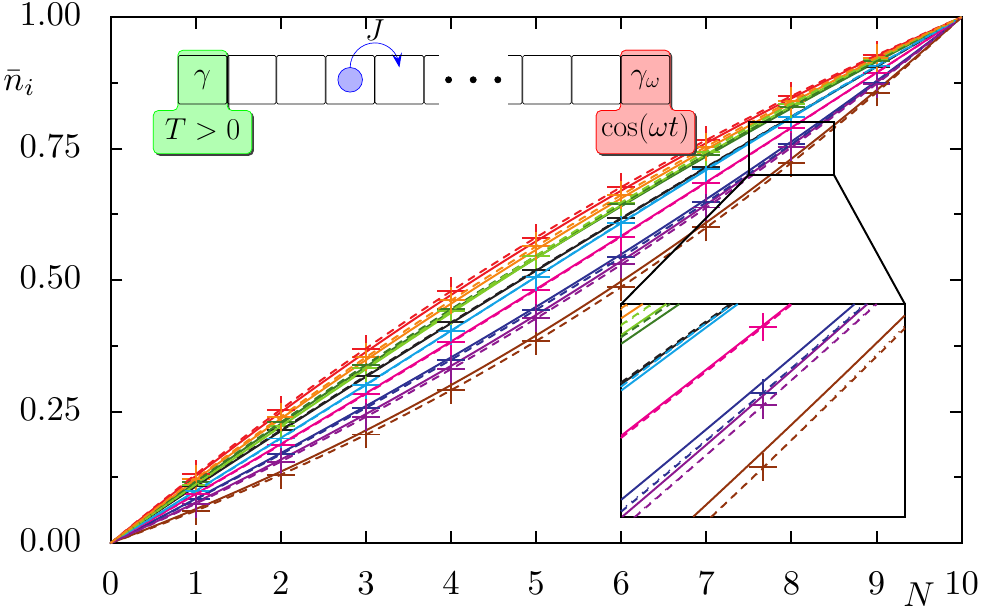}
\caption{(color online) Mean occupations versus total number of fermions $N$ for a driven tight-binding chain with tunneling parameter
$J$ and $M=10$ sites. The chain is coupled to a heat bath of temperature $T=J$ at the first site and it is driven away 
from equilibrium by a time-periodic potential modulation at the last site of frequency $\hbar\omega = 1.5 J$ and driving 
strength $\gamma_\omega=2.3$. Data obtained from mean-field theory (solid lines), augmented mean-field theory
(dashed lines), and exact solution of the many-body rate equation (crosses).}
\label{fig:FermiDriven}
\end{figure}

\begin{figure*}[t]
\centering
\includegraphics[width=1\linewidth]{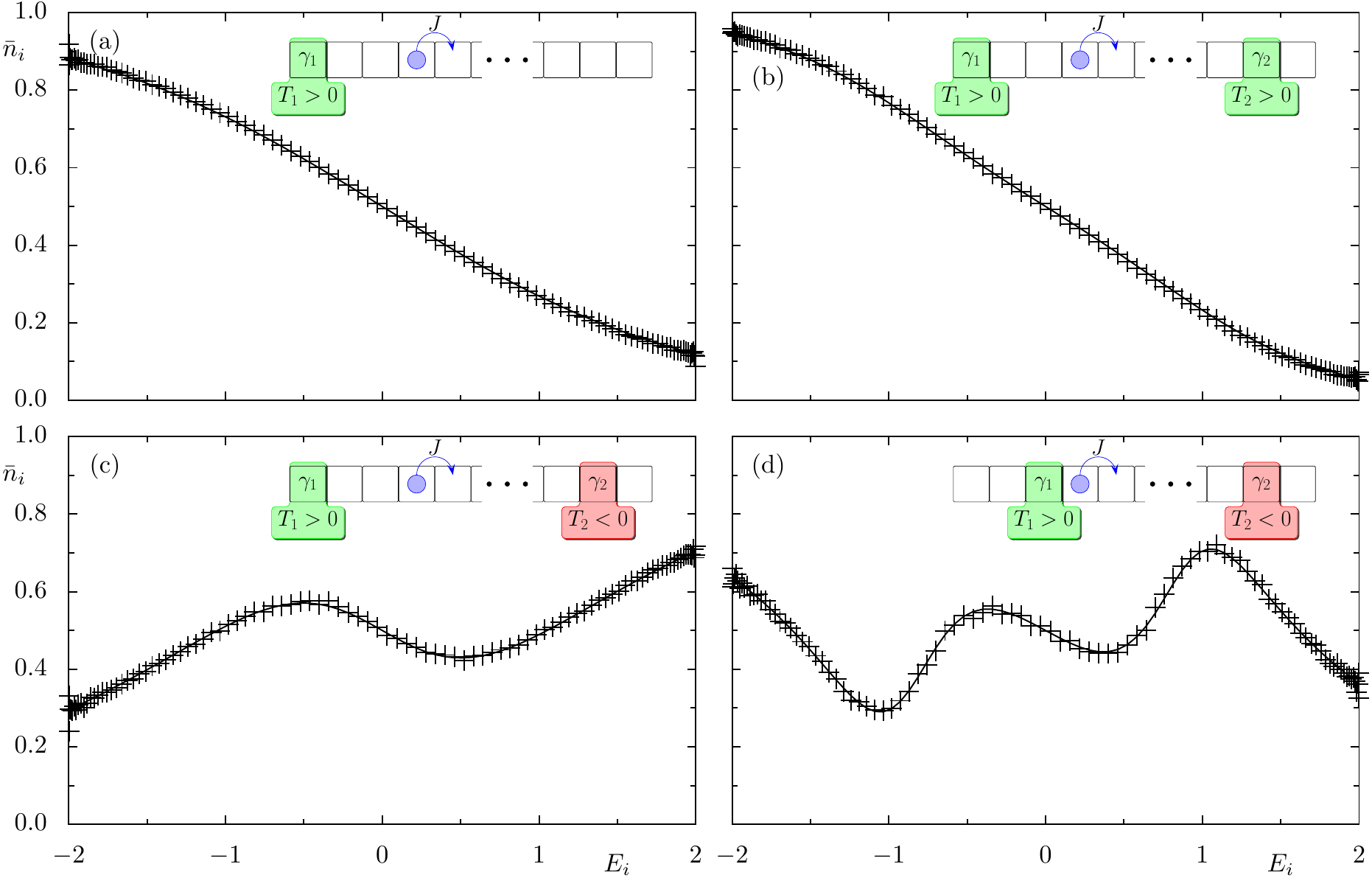}
\caption{(color online) Mean-occupations of the single-particle energy eigenstates in a tight-binding chain of $M=100$ sites occupied 
by $N=M/2$ spinless (i.e.\ spin-polarized) fermions versus the energy (in units of the tunneling parameter $J$).
Data obtained from mean-field theory (solid lines) and exact Monte-Carlo simulations (crosses). 
(a) Equilibrium situation where the chain is coupled to one bath of temperature $T=1J$. 
(b) The chain is driven away from equilibrium by two heat baths of different positive temperature ($T_1=J$ and $T_2=0.5J$), coupled to the 
first and the next to last site with $\gamma_1=\gamma_2$. 
(c) Same as in (b), but now the second bath is population inverted and described by the negative temperature $T_2=-J$. 
(d) Like in (c), but now the first bath is coupled to the third site.}
\label{fig:FermiChain}
\end{figure*}

In this section, we will briefly demonstrate that the theory of section \ref{sec:quantum_gas} and the methods presented in 
section \ref{sec:methods} can also be employed to describe the properties of ideal Fermi gases. As a motivation, we note 
that the physics of such driven-dissipative Fermi gases will have to play an important role, for example, for the 
realization of Floquet topological insulators. These systems are based on lattice potentials that are forced periodically 
in time such that they possess a topologically non-trivial quasienergy band structure giving rise to a quantized (spin) 
Hall conductivity, when one band is filled completely. Proposals for Floquet topological insulators consider irradiated 
electronic systems like graphene \cite{OkaAoki09} and semiconductor heterostructures \cite{LindnerEtAl11}; conceptually 
different schemes for the Floquet engineering of topological band structures have been, moreover, proposed in the context 
of ultracold atomic quantum gases in optical lattices \cite{Kolovsky11,HaukeEtAl12b}. First experimental evidence of a 
(quantized) Hall conductivity in such systems has been observed with ultracold atoms in optical lattices
\cite{JotzuEtAl14,AidelsburgerEtAl15}. These systems are well isolated from their environment. However, achieving this 
goal in an electronic solid-state systems, which cannot be viewed as isolated, is rather challenging. Namely, it cannot 
be expected that the periodically driven system in contact with the heat bath (given among others by phonons) will simply 
form a band-insulating state with one band filled completely. Thus, one either has to resort to bath engineering in order 
to enforce a band insulating state \cite{IadecolaEtA15,SeetharamEtAl15} or explore novel opportunities of tailoring 
interesting system properties related to non-thermal occupations of (quasi)energy bands. In this section we will not 
address the issue of Floquet topological insulators, but present simple examples that show how the general formalism of 
sections \ref{sec:quantum_gas} and \ref{sec:methods} can be applied to compute non-equilibrium steady states of
driven-dissipative Fermi gases. 

In Fig.~\ref{fig:FermiDriven} we plot the mean occupations of a periodically driven tight-binding chain of $M=10$ states 
that is coupled to a heat bath and occupied by $N$ spinless (i.e.\ spin-polarized) non-interacting fermions. The state is trivial not only for zero filling ($N/M=0$), but as a consequence of Pauli 
exclusion also for unit filling ($N/M=1$), corresponding to zero filling of holes. For intermediate filling $N/M$ we find 
occupation numbers whose exact values [obtained from solving the many-body rate equation (\ref{eq:mpr})] are well 
described by mean-field theory. Residual deviations of the mean-field theory are cured within the augmented mean-field 
theory (Section \ref{sec:augmented}).

As another example, we have computed steady states of a fermionic tight-binding chain of $M=100$ sites (see section
\ref{sec:models}) and half filling (N=M/2). In Fig.~\ref{fig:FermiChain} we plot the mean occupations of the
single-particle states $i$ of the chain versus their energy $E_i=-2J\cos(k_i)$, where $k_i$ is the wave number of 
state $i$. In panel (a) the equilibrium situation is shown, where the system is coupled to a single heat bath of 
intermediate temperature $T=J$. The non-equilibrium system coupled to two baths of different positive temperature
$T_1=J$ and $T_2=0.5J$ shows qualitatively similar behavior, as can be seen from panel (b). In both situations (a) and (b)
the occupations decrease with increasing energy. In striking contrast, the occupations depend in a non-monotonous 
fashion on the energy, when the second heat bath is population inverted and described by a negative temperature. This can 
be seen in panel (c) and (d). Moreover, the distribution of occupations depends sensitively on the structure of the 
system-bath coupling. Depending on whether bath 1 is coupled to the first site [panel (c)] or to the third site [panel (d)] 
the occupation of the ground state assumes either a local minimum or a local maximum. Thus, like in the bosonic case, 
already the ideal Fermi gas offers many possibilities of dissipative state engineering far from equilibrium. Exploring 
these possibilities is, however, beyond the scope of the present manuscript.

\section{Conclusions and outlook}
\label{sec:summary}

In this paper, we describe several aspects of non-equilibrium steady states of driven-dissipative ideal quantum 
gases. We focus on systems of sharp particle number that are driven away from equilibrium either by the coupling to 
two heat baths of different temperature or by time-periodic driving in combination with the coupling to a heat bath. 
We describe analytical and numerical methods for treating these systems within the framework of (Floquet-)Born-Markov 
theory and apply them both to bosonic and fermionic quantum gases. On that basis, we work out a theory of Bose 
selection, a non-equilibrium generalization of Bose condensation, where multiple states can acquire large occupations. 
Also the possibility of bath engineering in a fermionic lattice system is pointed out. 
Our results demonstrate that already ideal quantum gases give rise to intriguing and unexpected behavior, 
when they are driven into a steady state far from equilibrium. In the future it will be interesting to find applications 
for dissipative quantum engineering, e.g., in order to control the heat conductivity of a system in a robust 
fashion. On a theoretical level, it will be interesting to extend the formalism to systems exchanging particles with their environment and to include the effect of interactions.

\begin{acknowledgments}
We thank Erwin Frey, Johannes Knebel, Alexander Leymann, Markus Weber, Jan Wiersig and, in particular, Alexander Schnell for valuable discussions. Support through DFG Forschergruppe 760 ``Scattering Systems with Complex Dynamics'' is acknowledged. D.V. is grateful for 
the support from the Studienstiftung des Deutschen Volkes.
\end{acknowledgments}

\begin{appendix}

\section{Many-body rate equation from Lindblad master equation}
\label{sec:appendix_MPmasterequation}
Here we derive the equations of motion for the many-body occupation probabilities $p_{\bn}=\la \bn|\rho|\bn\ra$,
based on the Markovian master equation with the Liouvillian, \Eq{eq:EOMrho_Lindblad}.
Replacing the single-particle operators $|i \ra \la j|$ by their representation in Fock space $\aod_i \ao_j$
the equations of motion for the diagonal elements of the density operator take the form,
\begin{align}
\dot p_{\bn}(t) =& \la\bn|\hat{\rho}(t)|\bn\ra
\nonumber\\
=& 
\sum_{i,j=1}^M R_{ij}\Big(
\la\bn| \aod_i \ao_j \hat{\rho}(t) \aod_j \ao_i |\bn\ra
\nonumber\\&\qquad
-\frac{1}{2} \la\bn| \{ \hat{\rho}(t), \aod_j \ao_i \aod_i \ao_j \}|\bn\ra \Big).
\end{align}
For $i=j$ both terms inside the bracket cancel each other. For $i \neq j$, 
we have $\aod_j \ao_i | \bn \ra = \sqrt{ n_i (1 \pm n_j)} | \bn_{ji} \ra$
and $\aod_j \ao_i \aod_i \ao_j | \bn \ra = n_j (1\pm n_i) | \bn \ra $, where the upper (lower) sign applies to bosons (fermions). Thus, the master equation simplifies to
\begin{align}
\dot p_{\bn}(t) =& \sum_{i,j=1}^M R_{ij} \left[  n_i (1 \pm n_j) p_{\bn_{ji}}(t) - n_j (1\pm n_i) p_\bn(t) \right],
\nonumber\\ =& 
\sum_{i,j=1}^M (1\pm n_j)n_i  \left[R_{ij} p_{\bn_{ji}}(t) - R_{ji} p_\bn(t)\right].
\end{align}
wherein $\bn_{ji}=(n_1,\ldots,n_i-1,\ldots,n_j+1,\ldots)$ denotes the occupation numbers obtained from
$\bn$ by transferring one particle from $i$ to $j$.
We have not explicitly excluded the $i=j$ terms, since they still cancel. The second line was obtained by 
exchanging $i$ and $j$ in the second term.

\section{Equations of motion for mean occupations}
\label{sec:appendix_EOM}
The equations of motion for the mean occupations read 
\begin{align}\label{eq:EOM_nmean1}
 \frac{\diff}{\diff t} \bar{n}_k (t)
 =& \tr \left( \no_k  \frac{\diff}{\diff t} {\hat \rho}(t) \right)
\nonumber\\
=& \sum_{i,j} R_{ij} \tr \Big( \no_k \aod_i \ao_j \hat \rho (t) \aod_j \ao_i
\nonumber\\&\qquad
- \frac{1}{2} \no_k \left\{ \aod_j \ao_i \aod_i \ao_j, \hat \rho (t) \right\} \Big),
\end{align}
where we have employed Eq.~(\ref{eq:EOMrho_Lindblad}) with the jump operators given by Eq.~(\ref{eq:mbj}).
The first term of the sum can be written like 
\begin{align}
 \tr\left( \no_k \aod_i \ao_j \hat \rho (t)\aod_j \ao_i \right)
 =& \tr\left( \no_k \aod_j \ao_i \aod_i \ao_j \hat \rho (t) \right) 
\nonumber\\
 & +\; (\delta_{ik}-\delta_{jk}) \tr\left( \aod_j \ao_i \aod_i \ao_j \hat \rho(t) \right).
\end{align}
Here we have used the invariance of cyclic permutations under the trace as well as the relation 
\be\label{eq:commutator1}
 \left[\aod_j \ao_i, \no_k\right] = \aod_j \ao_i \left( \delta_{ik} - \delta_{jk} \right).
\ee
This relation is valid for particles of either statistics, as it can be obtained both by employing either the commutation 
relation $[\ao_i, \aod_j] = \delta_{ij}$ for bosons or the anticommutation relation $\{\ao_i, \aod_j\} = \delta_{ij}$ for 
fermions. We can now use
\begin{equation}\label{eq:operatorrelation1}
 \aod_j \ao_i \aod_i \ao_j = \no_j (1 \pm \no_i) \mp \delta_{ij} \no_i,
\end{equation}
with the upper (lower) sign referring to bosons (fermions), to arrive at
\begin{align}
  \frac{\diff}{\diff t} \bar{n}_k (t)
 =& \sum_{i,j} R_{ij}
   (\delta_{ik}-\delta_{jk}) \tr \big( \no_j (1 \pm \no_i) \hat \rho(t)\big) 
\nonumber\\
    =&\sum_{j=1}^M \bigg\{ R_{kj}\big[\bar{n}_j(t)\pm \la \no_k\no_j\ra(t)\big]
\nonumber\\&\qquad
-R_{jk}\big[\bar{n}_k(t)\pm \la \no_k\no_j\ra(t)\big]\bigg\} .
\label{eq:EOM_nmean_app}
\end{align}

\section{Mean occupation and correlation in projected Gaussian state}
\label{sec:app_proj}
Calculating expectation values, like mean occupation or second order correlations, for the projected Gaussian state 
\be
\hat{\rho}_{\text{proj}}\propto \hat{P}_N\hat{\rho}_g\hat{P}_N,
\ee 
with 
\be
\hat{P}_N=\sum_{\bn | \sum_i \no_i=N}|\bn\ra\la\bn|
\ee 
is a non-trivial problem. This is why, already in equilibrium it is typically much easier to treat a system in the
grand-canonical rather than in the canonical ensemble. 
In this appendix we describe a method for computing expectation values
\begin{equation}
  \la \hat{A}\ra_{N}=\tr\left(\hat{A}\hat{\rho}_{\text{proj}}\right)
=\frac{1}{Z}\sum_{\bn}\!^N \la \bn | \hat A |\bn \ra
  \re^{-\sum_k \eta_k n_k}
\end{equation}
for projected Gaussian states numerically. Here the sum $\sum_\bn^N$ is constrained to Fock states of total particle 
number~$N$ and $Z=\sum_{\bn}^N\exp(-\sum_k \eta_k n_k)$ denotes the partition function. 

We will focus on the mean occupations
\begin{equation}
 \la\no_i\ra_N=\frac{1}{Z}\sum_{\bn}^N  n_i \re^{-\sum_k \eta_k n_k}
\end{equation}
and the two-particle correlations 
\begin{equation}
\la\no_i \no_j\ra_N
=\frac{1}{Z}\sum_{\bn}^N n_i n_j  \re^{-\sum_k \eta_k n_k}.
\end{equation}
The first expectation value can be written as
\begin{align}
 \la \no_i\ra_N=
&\frac{1}{Z}\sum_{n_i} n_i \re^{-\eta_i n_i} Z_{N-n_i}^{\backslash\{i\}}
\end{align}
wherein 
\begin{align}
Z_{N_R}^{\backslash S_R}= \sum_{\{n_k\}_{k\notin S_R}}\mkern-16mu\!^{N_R} \exp\left(\textstyle{-\sum_{l\notin S_R} \eta_l n_l}\right)
\end{align}
is the partition function of fictitious system obtained by the original one by removing the states $S_R$ and filling it with $N_R$ particles only.

The second expectation value reads
\begin{align}
 \la \no_i\no_j \ra_N=
&\frac{1}{Z} \sum_{n_i=0}^N\sum_{n_j=0}^{N-n_i} n_in_j 
\re^{-\eta_i n_i-\eta_j n_j}   Z_{N-n_i-n_j}^{\backslash\{i,j\}}.
\end{align}
The remaining partition functions can be calculated by exploiting the recursion formula \cite{Borrmann93}
\begin{equation}
  Z_N=\frac{1}{N}\sum_{k+1}^{N}(\pm 1)^{k+1} Z_{N-k}
\end{equation}
This enables the numerical treatment of systems with several thousands particles on $M=10$ states.

\section{Equations of motion for two-particle correlations}
\label{sec:appendix_EOMcorrelations}
In this appendix we derive the equations of motion for the two-particle correlations $\la \no_k \no_i \ra(t)$ [Eq.~\eqref{eq:EOM_nncorr}] and rewrite this as equations of motion for the non-trivial correlations $\zeta_{ki} = \la \zetao_k \zetao_i \ra = \la \no_k \no_i \ra - \bar{n}_k\bar{n}_i$ [Eq.~\eqref{eq:EOM_zzcorr_MF}].
Together with the equations of motion for the mean occupations $\bar{n}_i(t)$, Eqs.~\eqref{eq:EOM_xmean}, they build the set of equation for the augmented mean-field theory described in Sec.~\ref{sec:augmented}. Hereby we 
close the hierarchy of equation by assuming the three-particle correlations to be trivial. For the sake of a simple notation we will suppress the time argument in the following. 

The exact equations of motion for $\la\no_k \no_i\ra$ are obtained from the many-body master equation in Lindblad form \Eq{eq:EOMrho_Lindblad} by multiplying it by $\no_k \no_i$ from the left and taking the trace,
\begin{align}\label{eq:EOM_nncorr1}
 \frac{\diff}{\diff t} \la \no_k \no_i\ra 
=& \tr\left( \no_k \no_i \dot{\hat \rho} \right)
\nonumber\\
=& \sum_{j,l} R_{lj} \tr \Big( \no_k \no_i \aod_l \ao_j \hat \rho \aod_j \ao_l
\nonumber\\&
\phantom{\sum}-\frac{1}{2} \no_k \no_i \left\{ \hat \rho, \aod_j \ao_l \aod_l \ao_j \right\} \Big).
\end{align}
Invoking cyclic permutation under the trace and using \Eq{eq:commutator1} we regroup the operators as
\begin{align}
 \aod_j \ao_l \no_k \no_i
 =&\no_k \no_i \aod_j \ao_l +  
\nonumber\\&
\big[   (\delta_{li}-\delta_{ji}) \no_k + (\delta_{lk}-\delta_{jk}) \no_i 
\nonumber\\
& + (\delta_{li}-\delta_{ji})(\delta_{lk}-\delta_{jk}) \big] \aod_j \ao_l.
\end{align}
Here, the first  term and the anticommutator in \Eq{eq:EOM_nncorr1} form  
a commutator, which vanishes under the trace, $\tr(\rho [ \no_k \no_i, \aod_j \ao_l \aod_l \ao_j])=0$.
Applying also the operator relation \Eq{eq:operatorrelation1} we arrive at
\begin{align}
  \frac{\diff}{\diff t} \la \no_k \no_i\ra
 =& \sum_{j,l} R_{lj} \tr\big\{
   \big[ (\delta_{li}-\delta_{ji}) \no_k +  (\delta_{lk}-\delta_{jk}) \no_i 
\nonumber\\&
+ (\delta_{li}-\delta_{ji})(\delta_{lk}-\delta_{jk})\big]
    \big[ \no_j (1 \pm \no_l) \mp \delta_{jl} \no_j \big] \hat \rho \big\}.
\end{align}
The term $\delta_{jl} \no_j$ vanishes in combination with each of the $\delta$-prefactors, leaving
\begin{align}
   \frac{\diff}{\diff t} \la \no_k \no_i\ra
 =&  \sum_{j,l} R_{lj} \Big[
  (\delta_{li}-\delta_{ji}) \left( \la \no_k \no_j \ra \pm \la \no_k \no_j \no_l \ra \right)
\nonumber  \\
&+ (\delta_{lk}-\delta_{jk}) \left( \la \no_i \no_j \ra \pm \la \no_i \no_j \no_l \ra \right)
\nonumber  \\
&+  (\delta_{li}-\delta_{ji})(\delta_{lk}-\delta_{jk}) \left( \avgo{n_j} \pm \la \no_j \no_l \ra \right)
 \Big] .
\end{align}
Evaluating one of the two sums we arrive at
\begin{widetext}
\begin{eqnarray}\label{eq:EOM_nncorrB}
 \frac{\diff}{\diff t} \la \no_k \no_i\ra
 &=&
\pm \sum_j (A_{kj} + A_{ij}) \la \no_k \no_i \no_j \ra 
\nonumber\\
 &&+ \sum_j \left( R_{kj} \la \no_i \no_j \ra  - R_{jk} \la \no_i \no_k \ra 
   +   R_{ij} \la \no_k \no_j \ra  - R_{ji} \la \no_k \no_i \ra \right) 
\nonumber \\
 &&+ \delta_{ik} \sum_j\left( R_{kj} \left( \avgo{n_j} \pm \la \no_j \no_k \ra \right)
   +  R_{jk} \left( \avgo{n_k} \pm \la \no_k \no_j \ra \right) \right)
\nonumber\\
 &&- R_{ik}  \left(  \avgo{n_k}  \pm \la \no_k \no_i \ra \right) - R_{ki}  \left( \avgo{n_i} \pm \la \no_i \no_k \ra \right),
\end{eqnarray}
\end{widetext}
which is identical to \Eq{eq:EOM_nncorr}. 
We separate the number operators $\no_k$  into their mean part $\bar{n}_k$ and their fluctuations $\zetao_k=\bar{n}_k-\no_k$. 
With that, the correlations read  
$\la \no_k \no_i \ra = \bar{n}_k\bar{n}_i + \zeta_{ki}$
with the non-trivial correlation $\zeta_{ki} = \la \zetao_k \zetao_i \ra$ and 
$ \la \no_k \no_i \no_j \ra =   \la \zetao_k \zetao_i \zetao_j \ra + \bar{n}_k \zeta_{ij}+ \bar{n}_i \zeta_{kj}  + \bar{n}_j \zeta_{ki} + \bar{n}_k \bar{n}_i \bar{n}_j.$
Now Eq.~\eqref{eq:EOM_nncorrB} can be rewritten as
\begin{widetext}
\begin{eqnarray}\label{eq:rhs}
\frac{\diff}{\diff t} \la \no_k \no_i\ra=
&\pm&  \sum_j (A_{kj}+A_{ij}) \left[ \la \zetao_i \zetao_k \zetao_j \ra
 + \bar{n}_i \zeta_{kj}  + \bar{n}_k \zeta_{ij} + \bar{n}_j \zeta_{ik}
 + \bar{n}_k \bar{n}_i \bar{n}_j  \right] 
\nonumber\\
 &+&   \sum_{j}\left( R_{kj} \left[ \bar{n}_i \bar{n}_j + \zeta_{ij} \right]
              - R_{ji} \left[ \bar{n}_i \bar{n}_k +  \zeta_{ik} \right]
	      + R_{ij} \left[ \bar{n}_k \bar{n}_j +  \zeta_{kj} \right]
              - R_{ji} \left[ \bar{n}_k \bar{n}_i + \zeta_{ki} \right]\right)  
\nonumber\\
&+&  \delta_{ki} \sum_j \left(\pm(R_{jk} + R_{kj}) \left[ \bar{n}_j \bar{n}_k +  \zeta_{jk} \right]   +(R_{kj} \bar{n}_j + R_{jk} \bar{n}_k ) \right)
\nonumber\\
& \mp& (R_{ki} + R_{ik}) \left[ \bar{n}_k \bar{n}_i + \zeta_{ki} \right] -(R_{ik} \bar{n}_k + R_{ki}  \bar{n}_i ).
\end{eqnarray}
\end{widetext}
To obtain the equations of motion for the non-trivial correlations $\zeta_{ki}$, we subtract 
\begin{align}
\frac{\diff}{\diff t}(\bar{n}_k\bar{n}_i) =& \phantom{+}\bar{n}_k \sum_j \big(R_{ij}(\bar{n}_j(1\pm \bar{n}_i)\pm\zeta_{ij})
\nonumber\\&
\phantom{+\sum }-R_{ji}(\bar{n}_i(1\pm\bar{n}_j)\pm\zeta_{ij})\big)
\nonumber\\&
+ \bar{n}_i \sum_j \big(R_{kj}(\bar{n}_j(1\pm\bar{n}_k)\pm\zeta_{kj})
\nonumber\\&
\phantom{+\sum}-R_{jk}(\bar{n}_k(1\pm \bar{n}_j)\pm\zeta_{kj})\big)
\end{align}
from Eq.~\eqref{eq:rhs}, to obtain 
\begin{eqnarray}\label{eq:EOM_zzcorr}
 \frac{\diff \zeta_{ki}}{\diff t}
&=& \pm \sum_j \Big[(A_{kj}+A_{ij}) ( \la \zetao_i \zetao_k \zetao_j \ra
   + \bar{n}_j \zeta_{ik} ) 
\nonumber\\
&&\phantom{\pm \sum_j}+ A_{kj}\bar{n}_k\zeta_{ij}+A_{ij}\bar{n}_i\zeta_{kj}\Big]
\nonumber\\
&& +  \sum_j \left[ R_{kj} \zeta_{ji} - R_{jk} \zeta_{ki} + R_{ij} \zeta_{kj} - R_{ji} \zeta_{ki} \right]
  \nonumber \\
&& +\delta_{ki} \sum_j \big[\pm(R_{jk} + R_{kj}) \left( \bar{n}_j \bar{n}_k + \zeta_{jk} \right)  
\nonumber\\
&&\phantom{+\delta_{ki} \sum_j}+ (R_{kj} \bar{n}_j + R_{jk} \bar{n}_k)\big]
\nonumber\\
&&\mp (R_{ki} + R_{ik}) \left( \bar{n}_k \bar{n}_i + \zeta_{ki}  \right)  
\nonumber\\
&&- (R_{ik}  \bar{n}_k + R_{ki}  \bar{n}_i).
\end{eqnarray}
Finally, neglecting non-trivial three-particle correlations, $\la \zetao_k \zetao_i \zetao_j \ra \approx 0$, one arrives at the non-linear 
set of equations (\ref{eq:EOM_zzcorr_MF}), which defines together with Eqs.~\eqref{eq:EOM_xmean} the augmented mean-field theory. 

\section{Parameter-dependent solution of the auxiliary matrix $\tilde{A}(p)$}
\label{sec:app_algorithm}

For the auxiliary rate-asymmetry matrix $\tilde{A}(p)$ given by Eqs.~(\ref{eq:Atilde}) and (\ref{eq:B}) the problem
(\ref{eq:allconditions}) takes the form 
\begin{align}
\tilde{\mu}_i(p)=&\sum_j \big(A_{ij}+pB_{ij}\big)\tilde{\nu}_j(p)
\nonumber\\&
\text{ with } \begin{cases}
 \tilde{\nu}_i>0 \text{ and } \tilde{\mu}_i=0 \mbox{ for } i\in \tilde{\BES}(p),
\\  \tilde{\nu}_i=0 \text{ and } \tilde{\mu}_i<0 \mbox{ for } i\notin \tilde{\BES}(p).
\end{cases}
\end{align}
Together with Eq.~(\ref{eq:B}) restricting $B$ to have a cross-like structure, this implies 
\begin{align}\label{eq:ntilde}
\sum_{j\in \tilde{\BES}(p)}\big(A_{ij}+p\delta_{ik}b_j - p \delta_{kj}b_i\big)\tilde{\nu}_j(p)= 0 ,  \qquad i\in\tilde{\BES}(p).
\end{align}
Let us now show that, unless a transition occurs where the set of selected states $\tilde{\BES}(p)$ changes, the solution 
$\tilde{\nu}(p)$ varies, apart from a normalization factor, linearly with $p$ as written in Eq.~(\ref{eq:lin}). 

For that purpose we decompose the solution $\tilde{\nu}_i(p)$ like
\be
\tilde{\nu}_i(p) = \tilde{\nu}^{(0)}_i + \Delta\tilde{\nu}_i(p)  ,  \qquad i\in\tilde{\BES}(p),
\ee
where $\tilde{\nu}^{(0)}_i$ is defined to solve 
\be\label{eq:ntilde0}
\sum_{i\in\tilde{\BES}(p)} A_{ij} \tilde{\nu}^{(0)}_i =0  ,  \qquad i\in\tilde{\BES}(p).
\ee
These equations possess a solution, since $A_{ij}$ is a skew symmetric matrix acting in the odd-dimensional subspace 
spanned by the selected states. However, the $\tilde{\nu}^{(0)}_i$ can be negative, as $\tilde{\BES}(p)$ contains the 
selected states for the matrix $\tilde{A}(p)$ and not for $A$. 

We can now distinguish two cases. If the state $k$ is not contained in the set of selected states,
$k\notin\tilde{\BES}(k)$,
Eqs.~(\ref{eq:ntilde}) simply reduces to Eq.~(\ref{eq:ntilde0}), so that we find the trivial parameter dependence 
\be
\tilde{\nu}_i(p) = \tilde{\nu}^{(0)}_i   \qquad i\in\tilde{\BES}(p),
\ee
which complies with Eq.~(\ref{eq:lin}). If the state $k$ is contained in the set of selected states,
$k\in\tilde{\BES}(k)$, it is convenient to discard the normalization condition 
$\sum_{i\in\tilde{\BES}(p)} \tilde{\nu}_i(p)=1$ for the moment, in favor of requiring 
\be
\tilde{\nu}_k(p)=\tilde{\nu}^{(0)}_k,
\ee
i.e.\ 
\be
\Delta\tilde{\nu}_k(p) = 0. 
\ee
Note that this requires also to fix the solution of the homogeneous equations (\ref{eq:ntilde0}) such that
$\tilde{\nu}^{(0)}_k>0$, which we can always do. With that, all other states in $\tilde{\BES}(p)$ obey
\be
\sum_{j\in\tilde{\BES}(p)\backslash\{k\}} A_{ij}\Delta\tilde{\nu}_j(p)  = p \tilde{\nu}^{(0)}_k b_i, 
			\qquad i\in \tilde{\BES}(p)\backslash\{k\}.
\ee
This set of inhomogeneous equations possesses a solution, since $A_{ij}$ is a skew-symmetric matrix acting in the
even-dimensional subspace spanned by the states of $\tilde{\BES}(p)\backslash\{k\}$, which has no eigenvalue zero without 
fine tuning. The solution $\Delta\tilde{\nu}_j(p)$ will depend linearly on the parameter $p$. Therefore, one finds
that the $\tilde{\nu}_i(p)$ depend linearly on the parameter $p$,
\be
\tilde{\nu}_i(p) = \tilde{\nu}^{(0)}_i + c_i p  ,  \qquad i\in\tilde{\BES}(p).
\ee  
In order to restore the normalization condition $\sum_{i\in\tilde{\BES}(p)} \tilde{\nu}_i(p)=1$, we can now re-define
\be\label{eq:nutildenorm}
\tilde{\nu}_i(p) = \tilde{C}(p)\Big[\tilde{\nu}^{(0)}_i + c_i p \Big] ,  \qquad i\in\tilde{\BES}(p).
\ee
with normalization constant $\tilde{C}(p)>0$. One finds
\be
\tilde{C}^{-1}(p) = \sum_{i\in\tilde{\BES}(p)}\Big[\tilde{\nu}^{(0)}_i + c_i p \Big] = 1+p\sum_{i\in\tilde{\BES}(p)}c_i,
\ee
where the second equality holds if we choose $\sum_{i\in\tilde{\BES}(p)}\tilde{\nu}^{(0)}_i =1$, which we always can.
Eq.~(\ref{eq:nutildenorm}) implies that Eq.~(\ref{eq:lin}) is fulfilled also if $k\in\tilde{\BES}(k)$.

\end{appendix}

\bibliography{refs}

\end{document}